\documentclass{aa}
\usepackage{lineno}
\usepackage{color}
\usepackage{amssymb}
\usepackage{amsmath}
\usepackage{graphicx}
\usepackage{txfonts}
\usepackage{natbib}
\usepackage{lscape}
\usepackage{arydshln}
\usepackage{stfloats}
\usepackage{longtable}
\bibpunct{(}{)}{;}{a}{}{,} 

\begin{document}
\title{GRB\,120711A: an intense \emph{INTEGRAL} burst with long-lasting soft $\gamma$-ray emission and a powerful optical flash}

\author{A. Martin-Carrillo
          \inst{1}
          \and
          L. Hanlon\inst{1}
          \and
          M. Topinka\inst{1}
          \and
          A. P. LaCluyz\'e\inst{2}
          \and
          V. Savchenko\inst{3}
          \and
          D. A. Kann\inst{4, 5, 6}
          \and
          A. S. Trotter\inst{2,7}
          \and
          S. Covino\inst{8}
          \and
          T. Kr\"uhler\inst{9,10}
          \and
          J. Greiner\inst{5}
          \and
          S. McGlynn\inst{1,4}
          \and
          D. Murphy\inst{1}
          \and 
          P. Tisdall\inst{1}
          \and
          S. Meehan\inst{1}
          \and
          C. Wade\inst{1}
          \and
          B. McBreen\inst{1}
          \and
          D. E. Reichart\inst{2}
          \and
          D. Fugazza\inst{8} 
          \and
          J. B. Haislip\inst{2}
          \and
          A. Rossi\inst{6,11}
          \and
          P. Schady\inst{5}
          \and
          J. Elliott\inst{5}
          \and S. Klose\inst{6}
          }

   \institute{Space Science Group, School of Physics, University College Dublin, Belfield, Dublin 4, Ireland\\
              \email{antonio.martin-carrillo@ucd.ie}
   \and University of North Carolina at Chapel Hill, Campus Box 3255, Chapel Hill, NC 27599-3255, USA
   \and APC, Universite Paris Diderot, CNRS/IN2P3, CEA/DSM, Obs. Paris, 13 rue Watt, 75205 Paris Cedex 13, France
   \and Universe Cluster, Technische Universit\"at M\"unchen, Boltzmannstra{\ss}e
2, 85748 Garching, Germany
   \and Max-Planck-Institut f\"ur extraterrestrische Physik, Giessenbachstrasse 1, 85748 Garching, Germany
      \and Th\"uringer Landessternwarte Tautenburg, Sternwarte 5, 07778 Tautenburg, Germany
   \and Department of Physics, North Carolina A\&T State University, 1601 E Market St, Greensboro, NC 27411, USA
   \and INAF, Osservatorio Astronomico di Brera, Via E. Bianchi 46, I-23807 Merate, Italy
   \and Dark Cosmology Centre, Niels Bohr Institute, University of Copenhagen, Juliane Maries Vej 30, 2100 Copenhagen, Denmark
   \and
   European Southern Observatory, Alonso de C\'{o}rdova 3107, Vitacura, Casilla 19001, Santiago 19, Chile
   \and
   INAF-IASF Bologna, Area della Ricerca CNR, via Gobetti 101, I--40129 Bologna, Italy
\\
             }
 \date{Received xxxx ; accepted xxxx}

\abstract
{A long and intense $\gamma$-ray burst (GRB) was detected by \emph{INTEGRAL} on July 11 2012 with a duration of $\sim$\,115\,s and fluence of 2.8\,$\times$\,10$^{-4}$\,erg\,cm$^{-2}$ in the 20\,keV\,–-\,8\,MeV energy range. GRB\,120711A was at $z$\,$\sim$\,1.405 and produced soft $\gamma$-ray emission ($>$20\,keV) for at least $\sim$\,10\,ks after the trigger. The GRB was observed by several ground-based telescopes that detected a powerful optical flash peaking at an $R$-band brightness of $\sim$\,11.5\,mag at $\sim$\,126\,s after the trigger, or $\sim$\,9th magnitude when corrected for the host galaxy extinction ($A_{\rm V}$\,$\sim$\,0.85). The X-ray afterglow was monitored by the \emph{Swift}, \emph{XMM-Newton}, and \emph{Chandra} observatories from 8\,ks to 7\,Ms and provides evidence for a jet break at $\sim$\,0.9\,Ms. We present a comprehensive temporal and spectral analysis of the long-lasting soft $\gamma$-ray emission detected in the 20\,--\,200\,keV band with \emph{INTEGRAL}/IBIS, the \emph{Fermi}/LAT post-GRB detection above 100\,MeV, the soft X-ray afterglow and the optical/NIR detections from \emph{Watcher}, \emph{Skynet}/PROMPT, GROND, and REM. The prompt emission had a very hard spectrum ($E_{\rm peak}$\,$\sim$\,1\,MeV) and yields an $E_{\gamma,\rm iso}$\,$\sim$\,10$^{54}$\,erg (1\,keV\,--\,10\,MeV rest frame), making GRB\,120711A one of the most energetic GRBs detected so far. We modelled the long-lasting soft $\gamma$-ray emission using the standard afterglow scenario, which indicates a forward shock origin. The combination of data extending from the NIR to GeV energies suggest that the emission is produced by a broken power-law spectrum consistent with synchrotron radiation. The afterglow is well modelled using a stratified wind-like environment with a density profile $k$\,$\sim$\,1.2, suggesting a massive star progenitor (i.e. Wolf-Rayet) with a mass-loss rate between $\sim$\,10$^{-5}$\,--\,10$^{-6}$\,$M_{\odot}$\,yr$^{-1}$ depending on the value of the radiative efficiency ($\eta_{\gamma}$\,=\,0.2 or 0.5). The analysis of the reverse and forward shock emission reveals an initial Lorentz factor of $\sim$\,120\,--\,340, a jet half-opening angle of $\sim$\,2$^{\circ}$\,--\,5$^{\circ}$, and a baryon load of $\sim$\,10$^{-5}$\,--\,10$^{-6}$\,$M_{\odot}$ consistent with the expectations of the fireball model when the emission is highly relativistic. Long-lasting soft $\gamma$-ray emission from other \emph{INTEGRAL} GRBs with high peak fluxes, such as GRB\,041219A, was not detected, suggesting that a combination of high Lorentz factor, emission above 100\,MeV, and possibly a powerful reverse shock are required. Similar long-lasting soft $\gamma$-ray emission has recently been observed from the nearby and extremely bright \emph{Fermi}/LAT burst GRB\,130427A.}

\titlerunning{Multi-wavelength observations of GRB\,120711A}
   \maketitle
%
\section{Introduction}
\label{sec:intro}
Gamma-ray bursts (GRBs) are the most luminous explosions in the Universe and have central engines that drive the outbursts in highly relativistic jets \citep[e.g.][]{Meszaros:2006p261}. The durations of GRBs vary from milliseconds to hundreds of seconds with a few rare cases lasting more than 1000\,s \cite[e.g.][]{2002ApJ...570..573G}. Long-lasting hard X-ray/$\gamma$-ray ($>$\,10\,keV) emission from GRBs has been observed on several occasions with detections of up to $\sim$\,4000\,s \citep[e.g.][]{{2002ApJ...567.1028C},{2002ApJ...570..573G}}. Evidence of emission up to $\sim$\,1000\,s after the trigger has been revealed with BATSE \citep{2002ApJ...567.1028C} and \emph{Fermi}/GBM \citep{2012grb..confE..24F} by stacking a large sample of long GRBs. Recently, \emph{Swift} has observed several very long GRBs with prompt emission that lasted thousands of seconds, which seem to be part of a new population of ultra-long GRBs that might arise from progenitors different from those of standard GRBs \citep{2013arXiv1302.2352L}. This population includes GRB\,101225A \citep{{2011Natur.480...72T},{2013arXiv1302.2352L}}, which lasted $>$\,1.7\,ks in the 15\,--\,150 keV band; GRB\,111209A \citep{2012arXiv1212.2392G}, which was active for about 25\,ks in the 0.3\,--\,10\,keV energy band and is considered the longest GRB ever observed; GRB\,121027A \citep{{2013arXiv1302.4876P},{2013arXiv1302.4878W},{2013arXiv1302.2352L}}, whose X-ray flaring activity lasted $\sim$\,10$^{4}$\,s in the 0.3\,--\,10\,keV band; and GRB\,130925A, which showed highly variable $\gamma$- and X-ray emission extending over $\sim$\,20\,ks \citep{2014arXiv1403.4079E}. These GRBs can be interpreted as the tail of the distribution of long GRBs \citep{2013ApJ...778...54V}.

Since its launch, \emph{Swift} \citep{Gehrels:2004p1618} has helped to unravel the properties of early GRB afterglows in the X-ray and optical bands by providing fast localisations and extensive monitoring of X-ray afterglows up to $\sim$\,10$^{7}$\,s after the burst trigger \citep[e.g.][]{{Nousek:2006p277},{2007ApJ...662..443G},{2010ApJ...711.1008G}}. X-ray light curves are normally well sampled in the 0.3\,--\,10\,keV energy range with a canonical shape consisting of 4-5 segments with different decay slopes \citep[e.g.][]{{Nousek:2006p277},{Zhang:2006p369}}. During the early parts ($<$\,100\,--\,1000\,s post trigger), the X-ray light curves are characterised by a fast-decaying phase commonly associated with high-latitude emission \citep[e.g.][]{{Zhang:2006p369},{2009MNRAS.399.1328G}}. This may be followed by a plateau phase that can be interpreted as late energy injection either from long-lived activity of the central engine \citep[e.g.][]{Zhang:2006p369} or from the reverse shock \citep[e.g.][]{{2007ApJ...665L..93U},{2007MNRAS.381..732G},{Leventis:2014kx}}. During these two phases and especially in the initial fast-decaying segment \citep{{2010MNRAS.406.2113C},{2011A&A...526A..27B},{2011MNRAS.417.2144M}}, many GRBs have soft X-ray flares \citep[e.g.][]{Burrows:2005p2248}. The spectral lag-luminosity relationship observed in these X-ray flares may provide a link to the prompt emission \citep[e.g.][]{2010MNRAS.406.2149M}. At the end of the plateau phase the X-ray GRB afterglow decay steepens in a manner consistent with forward shock emission that becomes dominant, signalling the end of the prompt emission. The afterglow emission is then well described by synchrotron emission from the interaction of the forward shock with the surrounding medium \citep[e.g.][]{{Sari:1998p1670},{2000ApJ...536..195C},{2002ApJ...568..820G},{Zhang:2006p369}}. At late times (a few days), some GRBs show additional steepening that is consistent with either a jet break \citep[e.g.][]{{Frail:2001p1785},{2009ApJ...698...43R}} or shocks that no longer have sufficient energy to accelerate electrons that radiate in the 0.3\,--\,10\,keV energy band \citep{2012ApJ...749...80S}.

Long-lived soft $\gamma$-ray emission (above 10\,keV) from individual GRBs has been reported in only a few cases. For example, \cite{{1999A&A...344L..53B},{1999A&AS..138..443B}} observed emission for $\sim$\,1000\,s in the 35\,--\,300\,keV energy range from GRB\,920723 using the \emph{GRANAT}/SIGMA observatory. Emission up to 60\,keV was also detected from GRB\,990123 using BeppoSAX \citep{2005A&A...438..821M}. Using the \emph{INTEGRAL} observatory, \cite{Grebenev:2007p2363} observed emission for about 20\,s after the prompt emission had ended for GRB\,060428C using IBIS (20\,--\,200\,keV). In the cases of GRB\,980923 \citep{1999ApJ...524L..47G} and GRB\,110918A \citep{2013ApJ...779..151F}, spectral evidence was found that supports an external/forward shock origin for these $\gamma$-ray emission tails. Recently, \cite{2013arXiv1302.0560M} reported emission of up to $\sim$\,270\,s after the end of the prompt emission using \emph{INTEGRAL}/JEM-X (hereafter JEM-X) in several GRBs of the \emph{INTEGRAL} sample.

Before the launch of \emph{Fermi} \citep{DeAngelis:2001p1637}, long-lasting GeV emission was observed by EGRET from GRB\,940217, with no corresponding emission at lower energies \citep{Hurley:1994p2071}. However, \emph{Fermi} has now detected high-energy emission from several GRBs up to $\sim$\,1000\,s after the trigger, which is well modelled by an external shock mechanism at late times (100\,--\,1000\,s post-trigger) \citep[e.g.][]{{2009MNRAS.400L..75K},{2010MNRAS.403..926G}}. These GRBs are characterised by photon indices $\sim$\,2, and E$_{\rm iso}$ between $\sim$\,2\,$\times$\,10$^{52}$\,erg for GRBs with redshift $<$\,1.2  and $\sim>$\,10$^{54}$\,erg for GRBs at higher redshifts \citep{collaboration:2013tp}.

In this paper, the characteristics of the long-lasting emission in GRB\,120711A are described. This is the first GRB observed to have both long-lasting soft $\gamma$-ray emission and to be detected at GeV energies. The characteristics of GRB\,120711A, the observations, and the data analysis are described in section~\ref{sec:observations}. The temporal and spectral results are given in section~\ref{sec:results} and are described using afterglow models in section~\ref{sec:model}. The physical parameters derived from the best-fit model are given in section~\ref{sec:parameters}. In section~\ref{sec:disc}, these results are compared with those obtained from other GRB observations, and a summary and conclusions are presented in section~\ref{sec:conclusions}. 

The dependence of the flux on frequency, $\nu$, and time, $t$, is described throughout by $F_{\nu}\propto\nu^{-\beta}t^{-\alpha}$, where $\beta$ is the spectral energy index that is related to the photon index, $\Gamma$, by $\Gamma=\beta+1$. Thus, in this paper, a negative power-law index corresponds to a rising slope. All errors are quoted at the 1$\sigma$ level for one parameter of interest. The cosmological parameters considered are $H_{0}$\,=\,70\,km\,s$^{-1}$\,Mpc$^{-1}$, $\Omega_{\rm M}$\,=\,0.27, and $\Omega_{\Lambda}$\,=\,0.73. When appropriate, the notation  $Q_{x}$ is equivalent to $Q\times10^{x}$.

\section{GRB\,120711A: observations and data analysis}
\label{sec:observations}
At 02:44:48 UT (T$_{0}$) on July 11 2012, an extremely bright and long GRB was detected by \emph{INTEGRAL} \citep[RA\,=\,06h18m48.7s, Dec\,=\,-71$^{\circ}$00$'$04$''$,][]{2012GCN..13434...1G}. Most unusually, the burst also had long-lasting emission up to $\sim$\,1200\,s after the trigger that was detected by both \emph{INTEGRAL}/IBIS-ISGRI (hereafter IBIS) and \emph{INTEGRAL}/SPI (hereafter SPI) in the 20\,--\,50\,keV energy range \citep[][respectively]{{2012GCN..13435...1B},{2012GCN..13468...1H}}. The burst was rapidly followed by several telescopes. \emph{Fermi}/LAT observations started $\sim$\,300\,s after the trigger and emission was detected up to 2\,GeV \citep{{2012GCN..13444...1T},{2012GCN..13452...1K}}. Robotic optical telescopes detected a rapidly brightening optical counterpart, peaking at magnitude $\sim$\,12 in the $R$ and $V$ bands \citep[][]{2012GCN..13430...1L} while the burst was still in progress. Spectroscopic observations of the optical afterglow using the \emph{Gemini-S} telescope derived a redshift of 1.405 for GRB\,120711A \citep{2012GCN..13441...1T}. A 3$\sigma$ radio upper limit of 96\,$\mu$Jy at 34\,GHz using ATCA was reported at 3.78 days after the GRB by \cite{2012GCN..13485...1H}. No significant emission or absorption lines were found in the X-ray spectrum obtained with \emph{XMM-Newton} 20 hours after the GRB trigger \citep{2013arXiv1307.8345G}.

\subsection{\emph{INTEGRAL} data}
The ESA \emph{INTEGRAL} observatory \citep{Winkler:2003p1629} contains three high-energy instruments: IBIS, sensitive from $\sim$\,20\,keV to $\sim$\,1\,MeV \citep{Lebrun:2003p249}; a high-resolution spectrometer SPI, sensitive in the 20\,keV\,--\,8\,MeV energy range \citep{Vedrenne:2003p343}; and two X-ray monitors, JEM-X, operating in the 3\,--\,35\,keV energy range \citep{Lund:2003p250}. All three high-energy instruments have coded masks and operate simultaneously with the same pointing axis. However, the fields of view (FoV) are different for all instruments, with JEM-X having the lowest value of $\sim$\,5$^{\circ}$. \emph{INTEGRAL} observations normally consist of a series of pointing observations (science windows) of duration $\sim$\,3.5\,ks ($\sim$\,1 hour). Data from the region of GRB\,120711A are available from one hour before the trigger to 12 hours after the burst. The high intensity of the burst resulted in many telemetry gaps during the prompt emission phase in the IBIS data. Therefore we only used data from the SPI instrument for the analysis of the prompt emission. The large off-axis angle ($9.5^{\circ}$) of the GRB at which the burst was observed precluded observations with JEM-X until $\sim$\,T$_{0}$+6\,ks, where T$_{0}$ is the trigger time. All \emph{INTEGRAL} analysis was performed with the Offline Science Analysis (OSAv10)\footnote{\texttt{http://www.isdc.unige.ch/integral/analysis}}. 

\subsection{Swift observations}
\begin{table*}[bt]
\caption{X-ray observations of GRB\,120711A}
\label{tab:xrayobs}
\centering
\begin{tabular}{llccccc}
\noalign{\smallskip}
\hline
\hline
\noalign{\smallskip}
Satellite & ObsID & Start time & Start &Exposure & Net counts & Observed flux \\
\noalign{\smallskip}
& & (UTC)& (ks since T$_{0}$) & (ks)  &($\times$\,10$^{-2}$\,counts/s) & (0.3\,--\,10\,keV, erg\,cm$^{-2}$\,s$^{-1}$) \\
            \hline
\noalign{\smallskip}
\emph{Swift} & 00020223001 & 2012-07-11T05:03:27 & 8.34 & 0.3 & 857\,$\pm$\,18 & (3.7\,$\pm$\,0.1)\,10$^{-10}$ \\
\emph{Swift} & 00020223002 & 2012-07-11T06:49:05 & 14.64 & 1.0 & 87\,$\pm$\,3 & (1.25\,$\pm$\,0.07)\,10$^{-10}$ \\
\emph{Swift} & 00020223003 & 2012-07-11T11:06:23 & 30.12 & 8.0  & 35\,$\pm$\,1 & (1.9\,$\pm$\,0.1)\,10$^{-11}$ \\
\emph{Swift} & 00020223004 & 2012-07-11T17:31:07 & 53.16&9.2 & 6.2\,$\pm$\,0.3 & (3.0\,$\pm$\,0.2)\,10$^{-12}$\\
\emph{XMM-Newton}& 0658401001 & 2012-07-12T00:53:35 & 79.74 & 40.0 & 105.8\,$\pm$\,0.3 & (5.90\,$\pm$\,0.05)\,10$^{-12}$\\
\emph{Chandra}& 13794 & 2012-07-24T03:33:18 & 1126.14 & 9.9 & 1.21\,$\pm$\,0.12 & (1.07\,$\pm$\,0.20)\,10$^{-13}$\\
\emph{XMM-Newton}& 0700380701& 2012-07-28T20:42:08 & 1533.42 & 28.0 & 1.6\,$\pm$\,0.1 & (5.1\,$\pm$\,0.4)\,10$^{-14}$\\
\emph{Chandra}& 14469 & 2012-07-31T10:15:09 & 1755.0 & 19.8 & 0.65\,$\pm$\,0.06 & (4.4\,$\pm$\,0.5)\,10$^{-14}$\\
\emph{XMM-Newton}& 0700380801 & 2012-08-15T08:36:19 & 3045.12 & 38.0 & 0.33\,$\pm$\,0.04 & (1.0\,$\pm$\,0.1)\,10$^{-14}$\\
\emph{Chandra}& 15541 & 2012-09-25T14:33:27 & 6608.94 & 40.5 & 0.06\,$\pm$\,0.02 & (4.2\,$\pm$\,1.5)\,10$^{-15}$\\
\emph{Chandra}& 13795 & 2012-09-30T21:55:40 & 7067.46 & 40.5 & 0.07\,$\pm$\,0.02 & (3.4\,$\pm$\,1.5)\,10$^{-15}$\\
\noalign{\smallskip}
            \hline
\end{tabular}
\end{table*}

\emph{Swift} started observing GRB\,120711A at $\sim$\,T$_{0}$+8\,ks, with the XRT instrument \citep{Burrows:2005p1620}. Data were collected in the window-timing mode for the first package and in photon-counting mode for the remaining observations. The data were analysed using standard procedures through the \texttt{xrtpipeline} and the latest version of the calibration files \citep{Burrows:2005p1620}, and following the procedure described in \cite{{Evans:2007p208},{Evans:2009p4}}. Because of orbital constraints, \emph{Swift} was unable to monitor the source after $\sim$\,T$_{0}$+200\,ks. The list of all XRT observations of GRB\,120711A considered in this analysis is shown in Table~\ref{tab:xrayobs}.

\subsection{XMM-Newton observations}
A target of opportunity was activated with \emph{XMM-Newton} \citep{Jansen:2001p1572} $\sim$\,22\,h after the trigger for a duration of $\sim$\,50\,ks. The first 10\,ks of this observation was not considered because of the high background level. Two more \emph{XMM-Newton} observations were granted (PI. Martin-Carrillo) to observe the late afterglow emission ($>$\,T$_{0}$+1\,Ms). The data from all EPIC cameras, one pn \citep{Struder:2001p1573} and two MOS \citep{Turner:2001p1574}, were analysed using the Science Analysis Software, SASv13\footnote{\texttt{http://xmm.esac.esa.int/sas/}} including the latest version of the calibration files. To help constrain the spectral parameters, the information from all instruments was combined for each observation. In all cases, source and background regions of 20$''$ were taken to extract the source data. Standard filtering and screening criteria were then applied to create the final products. Table~\ref{tab:xrayobs} summarises the key observational parameters.

\subsection{Chandra observations}
Four late follow-up observations ($>$\,T$_{0}$+1\,Ms) performed by \emph{Chandra} \citep{2002PASP..114....1W} are publicly available (Table~\ref{tab:xrayobs}). All observations were performed with the ACIS-S CCD \citep{Garmire:2003p1650} in timed exposured/faint (TE-F) mode. The data were analysed using the latest version of the \emph{Chandra} Interactive Analysis of Observations, CIAOv4.5\footnote{\texttt{http://cxc.harvard.edu/ciao/}} and the calibration database CALDBv4.5.9. The source and background regions were extracted in all the cases from a 3$''$ circle centred on the GRB coordinates and close-by source-free regions.

\subsection{Fermi/LAT observations}
The \emph{Fermi}/LAT detector \citep{2009ApJ...697.1071A} observed the field of view of the GRB from $\sim$\,T$_{0}$+300\,s until $\sim$\,T$_{0}$+1.1\,ks, when the position of the burst was occulted by Earth. A second set of observations was made after the occultation phase between $\sim$\,T$_{0}$+2.5\,ks and $\sim$\,T$_{0}$+7.2\,ks. Because of background constraints, only photons with energies above 100\,MeV were used in this analysis.

\subsection{Watcher observations}
The \emph{Watcher} robotic telescope is located at Boyden observatory in South Africa \citep{2004AIPC..727..741F}. The 40 cm telescope is equipped with an Electron Multiplying CCD (EMCCD) Andor iXon camera with a field of view of 8$'$\,$\times$\,8$'$. The EMCCD camera can operate in a high-gain mode as well as in conventional mode. In the high-gain mode, photoelectrons go through an avalanche process with a gain of up to a factor of 255, permitting frames to be co-added without a read noise penalty.

\emph{Watcher} began observing the field of GRB 120711A with a clear filter 60\,s after the GRB trigger using 5\,s exposures with EMCCD mode on and subsequently 30\,s exposures with EMCCD mode off. The data were analysed using the standard photometry pipeline developed for this telescope \citep{2010AdAst2010E..36F}. After $\sim$\,T$_{0}$+200\,s, the images were stacked until a 3$\sigma$ detection was obtained. The source was calibrated in the $R$-band using the same nearby standard stars as PROMPT (Sect.~\ref{sec:skynet}). Magnitudes were taken from the USNO-B catalogue. The difference between the GRB afterglow spectrum and the reference star spectrum in the transformation from the clear filter to $R$-band typically results in a shift of $\sim$\,+0.15 magnitudes. The resulting error in magnitude includes the error in the calculation of the zero-point, sky background, and read-noise of the camera. In the EMCCD mode, the contribution from read-noise is negligible. An additional correction of +0.35 mag was applied to the calculated $R$-band \emph{Watcher} magnitudes to match the PROMPT $R$-band data. This final adjustment was to simultaneously fit the light curve with both data sets. The shift arises from applying different transformations (clear to $R$-band in the case of \emph{Watcher} and r$^{\prime}$ to $R$-band in the case of PROMPT) and the use of different catalogues in the photometry pipeline of each telescope (USNO-B and APASS). The optical observations presented in this paper are summarised in Table~\ref{tab:opticalobs}. The observational log is given in Table~\ref{tab:obswatcher} of the online appendix.

\begin{table}[h]
\caption{Summary of optical/NIR observations of GRB\,120711A}
\label{tab:opticalobs}
\centering
\begin{tabular}{lcccc}
\noalign{\smallskip}
\hline
\hline
\noalign{\smallskip}
Telescope & Start time & Time span & Filter & Number of \\
\noalign{\smallskip}
& (s since T$_{0}$) & (ks) & & detections\\
            \hline
\noalign{\smallskip}
\emph{Watcher} & 60 & 2 &Clear & 15\\
PROMPT3 & 67.4 & 161 & $B$ & 14 \\
PROMPT1 & 38.9 & 74 &$V$ & 21\\
PROMPT4 & 68.3 & 187 & $R$ & 25 \\
PROMPT5 & 7154.8 & 179 & $I$ & 14\\
REM & 279 & 7 & $H$ & 11\\
GROND & 21160 & 350 & g$^\prime$ & 25\\
GROND & 21160& 350 &  r$^\prime$ & 25\\
GROND & 21160& 350 & i$^\prime$ &  25\\
GROND & 21160& 350 & z$^\prime$ & 25\\
GROND & 21160& 350 & $J$ & 25\\
GROND & 21160& 350 & $H$ & 24\\
GROND & 21160& 350 & $K$ & 23\\
\noalign{\smallskip}
            \hline
\end{tabular}
\end{table}

\subsection{Skynet/PROMPT observations}
\label{sec:skynet}
\emph{Skynet}/PROMPT \citep{2005NCimC..28..767R} consists of twelve Ritchey-Chretien telescopes of diameters ranging from 0.41\,m to 0.80\,m located at the Cerro Tololo Inter-American Observatory (CTIO) and Siding Spring Observatory. These telescopes are designed to respond quickly to GRB alerts and perform quasi-simultaneous multi-filter observations in the optical/NIR band. Four PROMPT telescopes (see Table~\ref{tab:opticalobs}) were used to monitor the optical afterglow of GRB\,120711A from T$_{0}$+39\,s. The photometry in the $BVRI$ bands was calibrated using four stars from the APASS DR7 catalogue \citep{2011AAS...21812601H}. Since APASS was created using the $BV$g$^\prime$r$^\prime$i$^\prime$ filters, transformations to convert the r$^\prime$ and i$^\prime$ magnitudes into the $R$ and $I$ filters are required. The observational log is given in Table~\ref{tab:obsprompt} of the online appendix.

\subsection{GROND observations}
GRB\,120711A was observed in three sets of observations spanning 350\,ks in total (see Table~\ref{tab:opticalobs}) with the seven-channel Gamma-Ray burst Optical \& Near-infrared Detector, GROND, mounted on the 2.2m MPG/ESO telescope stationed in La Silla, Chile \citep{Greiner2008PASP}. The first epoch \citep[reported by][]{Elliott2012GCN13438} was delayed by several hours because the position of the GRB was very southerly and lay close to the Sun, so that it only rose above the pointing limit of the telescope near morning twilight. Even so, all observations were obtained at a high airmass and under mediocre to poor seeing conditions.

The GROND optical and NIR image reduction and photometry was performed by calling on standard IRAF tasks \citep{Tody1993ASPC} using the custom GROND pipeline \citep{Yoldas2008AIPC}, similar to the procedure described in \cite{Kruehler2008ApJ}. Optical photometric calibration was performed relative to the magnitudes of over forty secondary standards in the GRB field. During photometric conditions (more than a month after the GRB occurred, to allow observations at lower airmass), an SDSS field \citep{Aihara2011ApJS} at R.A. (J2000) = 06:59:33.6, Dec. (J2000) = -17:27:00 was observed within a few minutes of the observations of the GRB field. The obtained zero-points were corrected for atmospheric extinction and used to calibrate stars in the GRB field. The apparent magnitudes of the afterglow were measured with respect to these secondary standards. The absolute calibration of $JHK_S$ bands was obtained with respect to magnitudes of the Two Micron All Sky Survey (2MASS) catalogue using stars within the GRB field \citep{Skrutskie2006AJ} and converted to AB magnitudes. The observational log is given in Table~\ref{tab:obslog} of the online appendix.

\subsection{REM observations}
Early-time NIR data starting 129\,s after the GRB trigger (see Table~\ref{tab:opticalobs}) were also collected using the 60\,cm robotic telescope REM \citep{{2001AN....322..275Z},{2004SPIE.5492.1613C}} located at the European Southern Observatory (ESO) La Silla observatory (Chile). The data were reduced by using standard procedures and calibrated by isolated unsaturated 2MASS stars in the field. The observational log is given in Table~\ref{tab:obsrem} of the online appendix.

\section{Results}
\label{sec:results}

\subsection{Prompt emission}
\begin{figure}[ht!]
\centering
\resizebox{\hsize}{!}{\includegraphics[angle=0,width=12cm]{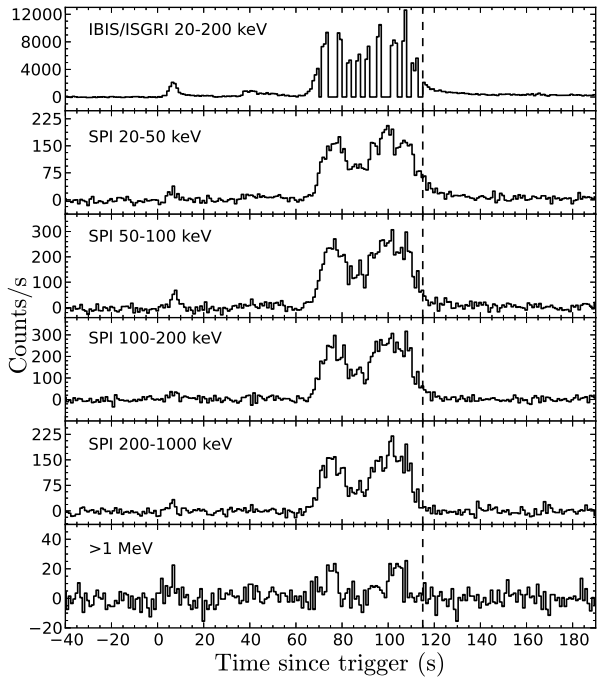}}
\caption{Energy-resolved light curves of GRB\,120711A. The top panel shows the 20\,--\,200\,keV IBIS light curve that is severely affected by telemetry gaps. The lower panels show the SPI light curves in five energy bands. In all cases, the light curves are binned over 1\,s. The dashed vertical line at T$_{0}$+115\,s represents the end of the T$_{90}$ duration.}
\label{grblcs}
\end{figure}

As shown in Fig.~\ref{grblcs}, GRB\,120711A was a long and bright GRB with a T$_{90}$ of $\sim$\,115\,s in the SPI 20\,--\,200\,keV energy range. It had a peak flux of 32\,ph\,cm$^{-2}$\,s$^{-1}$ in the 20\,keV\,--\,8\,MeV band and 27\,ph\,cm$^{-2}$\,s$^{-1}$ in the 20\,--\,200\,keV band. The burst consisted of a hard precursor followed by a soft flare at $\sim$\,T$_{0}$+40\,s, mostly visible below 50\,keV, and then $\sim$\,60\,s of long multi-peaked and overlapping pulses with emission above 1\,MeV (Fig.~\ref{grblcs}). The prompt emission was analysed using only SPI data because of significant telemetry gaps in IBIS due to buffer saturation.

The time-averaged spectrum over the T$_{90}$ emission is best fit using a power-law with exponential cutoff ($\chi^{2}$/dof\,=35/30) with photon index $\Gamma$, of 1.05$\pm$0.02 and a peak energy of 1130$^{+141}_{-27}$\,keV, making it the hardest GRB (in terms of peak energy) triggered by \emph{INTEGRAL} \citep{{Foley:2008p553},{Vianello:2009p722}}. The total fluence measured over T$_{90}$ is (2.8\,$\pm$\,0.4)$\times$10$^{-4}$\,erg\,cm$^{-2}$ in the 20\,keV\,--\,8\,MeV energy band, and (4.4\,$\pm$\,0.5)$\times$10$^{-5}$\,erg\,cm$^{-2}$ in the 20\,--\,200\,keV band.

\subsection{Post-GRB emission properties}
\subsubsection{Temporal analysis}
\label{sec:temporal}
\begin{figure}[ht!]
\centering
\resizebox{\hsize}{!}{\includegraphics[angle=0,width=12cm]{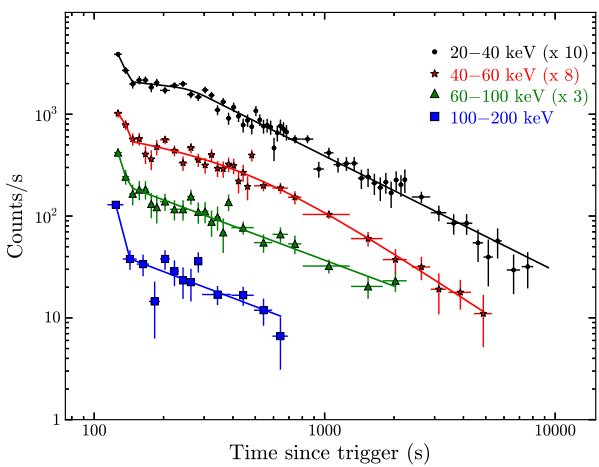}}
\caption[]{Energy-resolved IBIS light curves of GRB\,120711A with a maximum bin size of 1\,ks. The solid lines represent the best-fit model to the data. In all cases, the light curves show the temporal behaviour in each band up to the time of the last significant detection (3$\sigma$ above background). The upper three curves are re-normalised for display purposes.}
\label{ibislonglived}
\end{figure}

The background-subtracted IBIS light curve in four energy bands (20\,--\,40\,keV, 40\,--\,60\,keV, 60\,--\,100\,keV and 100\,--\,200\,keV) is shown in Fig.~\ref{ibislonglived} with a bin size that increases with time up to a highest value of 1\,ks. The emission is detected above 3$\sigma$ in each time bin for $\sim$\,10\,ks in the 20\,--\,40\,keV band and up to $\sim$\,2\,ks in the 60\,--\,100\,keV band. Emission at the 5$\sigma$ level in the 20\,--\,40\,keV energy band is found at even later times by combining IBIS data from three science windows from T$_{0}$+10\,ks to T$_{0}$+27.8\,ks. 

The data in all four energy bands can be fit using a series of smoothly connected power-laws (see Appendix\,A of \citealt{2011A&A...526A..23S}), where the simplest model (single power-law) is initially used and additional power-law segments are added when necessary \citep{2013MNRAS.428..729M}. The best-fit temporal parameters for the four energy bands are given in Table~\ref{tab:ibislc}. All energy bands show a short steep decay that ends at $\sim$\,T$_{0}$+140\,s (Fig.~\ref{ibislonglived}). This steep decay phase is followed by a slower decay, which may then break again depending on the energy band considered (Fig.~\ref{ibislonglived}). 

In the softest energy band, 20\,--\,40\,keV, the best fit consists of three power-laws ($\chi^{2}$/dof\,=\,63/52). The F-test probability of a chance improvement of this model is 3\,$\times$\,10$^{-6}$ ($>$\,4$\sigma$) over one with a single temporal break ($\chi^{2}$/dof\,=\,123/58) and 10$^{-4}$ ($>$\,3$\sigma$) over a model with two breaks ($\chi^{2}$/dof\,=\,98/56). The 40\,--\,60\,keV energy range is best fit using a two-break model ($\chi^{2}$/dof\,=\,27/23), with an F-test probability of chance improvement of 2.5\,$\times$\,10$^{-3}$ ($\sim$\,3$\sigma$) over a single-break model ($\chi^{2}$/dof\,=\,51/26). Although not required statistically, the model derived from the IBIS 20\,--\,40\,keV band can also adequately describe the higher-energy light curves. The observed differences may be caused by the reduced temporal sensitivity above 40\,keV.

\begin{table}[ht!]
\caption{Best-fit temporal parameters for the long-lasting emission from GRB\,120711A observed with IBIS.}
\label{tab:ibislc}
\centering
\begin{tabular}{lcccc}
\noalign{\smallskip}
\hline
\hline
\noalign{\smallskip}
Param.& 20\,--\,40 & 40\,--\,60 & 60\,--\,100 & 100\,--\,200 \\
& (keV) & (keV) & (keV) & (keV)\\
\noalign{\smallskip}
            \hline
\noalign{\smallskip}
$\alpha_{\gamma, 1}$      & 10.0$\pm$0.6 & 6$\pm$1 & 10$\pm$1 & 16$\pm$5 \\
$t_{\rm break,1}$\,(s) & 140$\pm$5 & 142$\pm$14 & 133$\pm$10 & 141$\pm$15 \\
$\alpha_{\gamma, 2}$ & 0.26$\pm$0.30 & 1.00$\pm$0.05 & 0.86$\pm$0.05 & 0.85$\pm$0.07\\
$t_{\rm break,2}$\,(s) & 283$\pm$36 & 500$\pm$250 & -- & -- \\
$\alpha_{\gamma, 3}$ & 1.13$\pm$0.04 & 1.5$\pm$0.6 & -- & -- \\
$\chi^{2}$/dof  & 63/52 & 27/23 & 12/17 & 10/9 \\
\noalign{\smallskip}
            \hline
\end{tabular}
\end{table}

\begin{figure}[ht!]
\centering
\resizebox{\hsize}{!}{\includegraphics[angle=0,width=12cm]{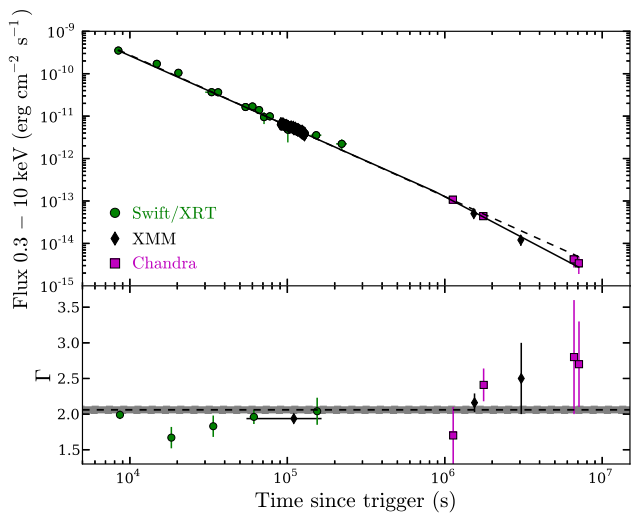}}
\caption{\textit{Top panel}: X-ray light curve of the observed flux in the 0.3\,--\,10\,keV band using \emph{Swift}, \emph{XMM-Newton}, and \emph{Chandra} observations. The solid line corresponds to the best-fit model (broken power-law), while the dashed line shows the fit when a single power-law model is considered. \textit{Bottom panel}: photon index evolution during the X-ray follow-up observations. The dashed line is the photon index ($\Gamma$\,=\,2.06) obtained by the joint spectral fit to the IBIS and \emph{Swift} data. The 1$\sigma$ error ($\pm$\,0.05) is shown as a grey area.}
\label{xraylc}
\end{figure}

\begin{figure}[ht!]
\centering
\resizebox{\hsize}{!}{\includegraphics[angle=0,width=12cm]{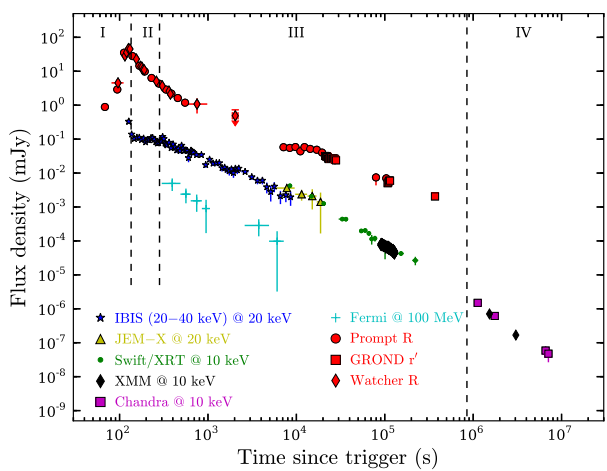}}
\caption{Afterglow light curves of GRB\,120711A as observed at soft $\gamma$-rays, X-rays, optical $R$-band and above 100\,MeV. The dashed vertical lines give the break times observed in the soft $\gamma$/X-ray light curves that form the 4 temporal segments (I to IV).}
\label{afterglow}
\end{figure}

All the X-ray data share a common energy band (0.3\,--\,10\,keV), and thus are fit as a single data set as shown in Fig.~\ref{xraylc}. The spectrum from the whole \emph{Swift}/XRT-PC data set (see Table~\ref{tab:xrayspec}) is used to convert count rates to flux units assuming no spectral evolution during the afterglow phase. This results in a best fit consisting of a broken power-law ($\chi^{2}$/dof of 116/96) with break time t$_{\rm break, 3}$\,=\,860\,$\pm$\,500\,ks and temporal indices $\alpha_{\rm X,1}$\,=\,1.65\,$\pm$\,0.04 and $\alpha_{\rm X,2}$\,=\,1.96\,$\pm$\,0.22. The F-test probability for this model compared with a single power-law decay ($\chi^{2}$/dof of 129/98) is 6\,$\times$\,10$^{-3}$ ($\sim$\,3$\sigma$). It should be noted that the large error on the break time is mainly due to the lack of X-ray data from $\sim$\,T$_{0}$+200\,ks to $\sim$\,T$_{0}$+1\,Ms. Fig.~\ref{afterglow} shows the flux density light curve post-GRB using data from IBIS in the 20\,--\,40\,keV energy band, \emph{Swift}/XRT, \emph{XMM-Newton}, \emph{Chandra}, \emph{Fermi}, and optical data from \emph{Watcher}, PROMPT (both using the $R$-filter band), and GROND in the r$^{\prime}$ band. The \emph{Fermi}/LAT data is best fit using a single power-law decay with index $\alpha_{\rm Fermi}$\,=\,1.30\,$\pm$\,0.13, consistent with temporal decays for long-lasting emission observed by \emph{Fermi}/LAT \citep{collaboration:2013tp}. 

Thus, the post-GRB emission from X-rays to $\gamma$-rays shows four distinct temporal segments as shown in Fig.~\ref{afterglow}, determined by the break times obtained from the two high-energy temporal fits.

\begin{figure}[ht!]
\centering
\resizebox{\hsize}{!}{\includegraphics[angle=0,width=12cm]{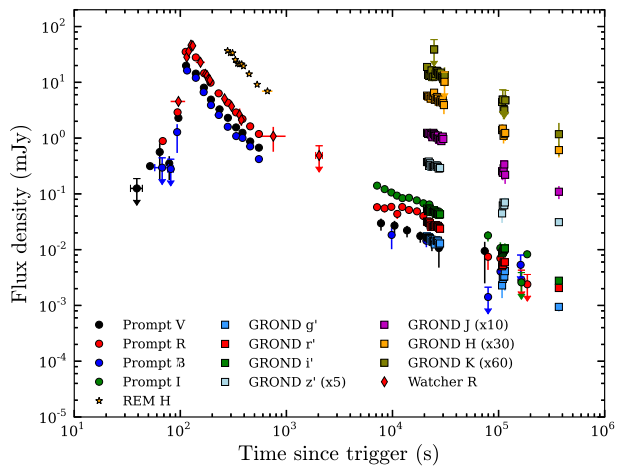}}
\caption{Multi-band optical/NIR light curve observed with \emph{Watcher}, \emph{Skynet}/PROMPT, GROND, and REM. The $z'$$J$$H$$K$ bands are scaled for display purposes.}
\label{opticallc}
\end{figure}

The multi-band optical/NIR light curve plotted in Fig.~\ref{opticallc} shows a fast-rising optical flash peaking at an $R$ brightness of $\sim$\,11.5\,mag $\sim$\,T$_{0}$\,+126\,s. The peak is followed by a steep decay that seems to break at later times. The behaviour of the optical/NIR light curve after $\sim$\,T$_{0}$+10\,ks is complex, with small fluctuations especially evident in the $R$-filter. In fact, during the second epoch of GROND observations, the optical/NIR afterglow seems to re-brighten in all filters, indicating late activity on top of the standard afterglow decay.

Little or no colour change is seen throughout the entire optical/NIR observing campaign. Therefore we fit the optical/NIR data using the $R$-band only where the light curve is best sampled by combining \emph{Watcher} and PROMPT data. For times $<$\,T$_{0}$\,+1\,ks, the best-fit model ($\chi^{2}$/dof of 30/24) consists of a fast-rising power-law of index $\alpha_{\rm opt, 1}$\,=\,-9\,$\pm$\,2 that peaks at T$_{0}$\,+126\,$\pm$\,5\,s and decays as a broken power-law with $\alpha_{\rm opt, 2}$\,=\,3.30\,$\pm$\,0.20, $\alpha_{\rm opt, 3}$\,=\,1.80\,$\pm$\,0.13 and $t_{\rm break, opt}$\,=\,215\,$\pm$\,14\,s. The F-test probability of this test over a model with a single power-law decay after the optical peak ($\chi^{2}$/dof of 74/26) is 2\,$\times$\,10$^{-5}$ ($\sim$\,4$\sigma$).

\begin{figure}[ht!]
\centering
\resizebox{\hsize}{!}{\includegraphics[angle=0,width=12cm]{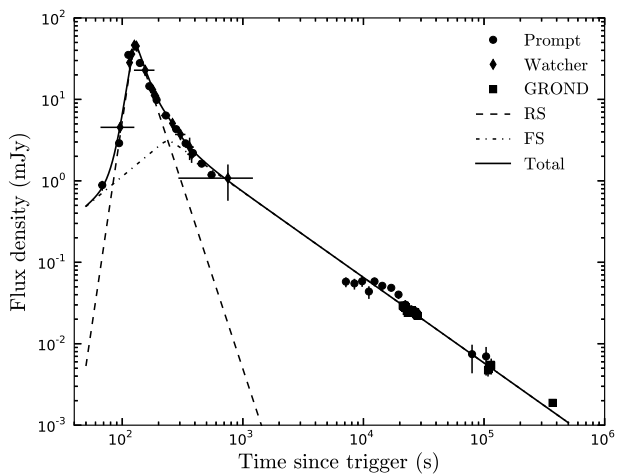}}
\caption{$R$-band optical light curve of GRB\,120711A. The best-fit model (solid line) consists of two components: a reverse shock (RS, dashed line) and a forward shock (FS, dashed-dotted line).}
\label{opticalfit}
\end{figure}

The late-time optical/NIR data ($>$\,T$_{0}$\,+1\,ks) can be fit by considering a single power-law decay of index $\alpha_{\rm opt, 4}$\,=\,1.00\,$\pm$\,0.05 using the simultaneous multi-band data from GROND. This implies a temporal break in the optical/NIR somewhere around $\sim$\,T$_{0}$\,+1\,ks. As shown in Fig.~\ref{opticalfit}, the decay between T$_{0}$\,+\,215\,s and $\sim$\,T$_{0}$\,+1\,ks ($\alpha_{\rm opt, 3}$\,$\sim$\,1.80) can be explained by the contribution from the decay of two broken power-laws, one with temporal decay $\alpha_{\rm opt, 2}$ and peaking at T$_{0}$\,+\,126\,s, consistent with reverse shock (RS) emission, and the second with the late decay, $\alpha_{\rm opt, 4}$, consistent with forward shock (FS) emission. This FS component is best fit with a rising power-law index of $\alpha_{\rm opt, FS,1}$\,=\,-1.23\,$\pm$\,0.5 and peaking at T$_{0}$\,+\,238$^{+20}_{-50}$\,s.

The unusually fast rising power-law index of the optical flash is obtained with the data referring to the trigger time, T$_{0}$, which corresponds to the time at which the precursor was detected. As discussed by \cite{Zhang:2006p369}, for multi-peaked GRBs, it seems more reasonable to refer to the beginning of the main emission than to the trigger time. This results in a power-law index $\alpha_{\rm opt, 1}$\,$\sim$\,5.4, more consistent with those expected from a reverse shock in a homogeneous environment \citep{2003ApJ...597..455K}.

\subsubsection{Spectral analysis}
\label{sec:spectral}
Time-averaged spectral analysis was performed for the four segments of the decaying  high-energy light curve (Fig.~\ref{afterglow}). For the soft X-ray observations, two absorption components were considered, consisting of an intrinsic absorption fixed at \textit{z}\,=\,1.405 and the Galactic absorption of $N_{\rm H}=7.9\times10^{20}$\,cm$^{-2}$ at the GRB sky coordinates \citep{Kalberla:2005p1538}. For \emph{INTEGRAL} and \emph{Fermi} data, where the absorption is no longer relevant, a single power-law model was adopted. The parameters obtained from the spectral analysis for the four segments are given in Table~\ref{tab:xrayspec}. For segments where data from more than one instrument are available, the spectral results are shown independently for each instrument. For the data in segment IV, the intrinsic absorption cannot be constrained and the data were fitted assuming only the Galactic contribution. This model is favoured over a model with a fixed intrinsic absorption, especially for the \emph{XMM-Newton} observation at T$_{0}$\,+\,1.5\,Ms (see Table~\ref{tab:xrayspec}). This lack of intrinsic absorption at late times could indicate evolution of $N_{\rm H, intr}$ with time. However, in this case it is more likely to be due to the poorly constrained fit caused by the low number of counts during this segment.

\begin{table*}[ht]
\caption{Spectral analysis of GRB 120711A per segment and instrument. The models considered are power-law (PL), blackbody+powerlaw (BBPL), and broken power-law (BKNPL).}
\label{tab:xrayspec}
\centering
\begin{tabular}{clcccccccc}
\noalign{\smallskip}
\hline
\hline
\noalign{\smallskip}
Segment & Inst.$^{a}$ & Time interval & Model & $N_{\rm H, intr}$ & \textit{z} & $\Gamma$ & kT/$E_{\rm break}^{b}$ & $\chi^{2}$/dof & F-test$^{c}$ \\
\noalign{\smallskip}
& & (ks since trigger) & & ($10^{22}$\,cm$^{-2}$) & & & (keV) &\\
\noalign{\smallskip}
            \hline
\noalign{\smallskip}
I & IBIS & 0.115--0.13 & PL &---& --- & 1.69$\pm$0.06 & --- & 35/28 & --- \\
 &  & & BBPL & --- & --- &1.36$\pm$0.12 & 4.8$\pm$0.9 & 19/26 & 4\,$\times$\,10$^{-4}$ \\
 &  & & BKNPL & --- & --- &1.38$\pm$0.10 & 44$\pm$6 & 18/26 & 2\,$\times$\,10$^{-4}$ \\
  &  & &       &     &     &2.24$\pm$0.18 &   &   &   \\
\noalign{\smallskip}
II & IBIS & 0.13--0.28 & PL & --- & --- &2.22$\pm$0.05 & --- & 60/28 & --- \\
 &  & & BBPL & --- & --- & 1.93$\pm$0.15 & 3.6$\pm$1.3 & 47/26 & 4\,$\times$\,10$^{-2}$ \\
 &  & & BKNPL & --- & --- & 1.0$\pm$0.5 & 86$^{+10}_{-18}$ & 45/26 & 2\,$\times$\,10$^{-2}$ \\
 &  & &       &     &     &2.35$\pm$0.08 &   &   &   \\
\noalign{\smallskip}
III & IBIS & 0.28--2.5 & PL & --- & --- & 2.02$\pm$0.06 & --- & 27/25 & --- \\
& \emph{Fermi}/LAT & 0.3--1.05 & PL & --- & --- & 2.00$\pm$0.3 & --- & 0.17/1 & --- \\
\noalign{\smallskip}
 & IBIS & 2.5-10 & PL & --- & --- & 2.5$\pm$0.5 & --- & 1.43/5 & --- \\
 & JEM-X & 7--20 & PL & --- & --- & 1.77$\pm$0.27 & --- & 7/5 & --- \\
 & \emph{Swift}/XRT-wt & 8.3--9 & PL & 1.79$\pm$0.22 & 1.405$^{d}$ & 1.99$\pm$0.06 & --- & 111/93 & --- \\
 & \emph{Swift}/XRT-pc & 14.7--239 & PL & 1.17$\pm$0.11 & 1.405$^{d}$ & 1.87$\pm$0.04 & --- & 153/152 & --- \\
& \emph{XMM}/pn & 90--130 & PL & 0.87$\pm$0.03 &  1.405$^{d}$ & 1.950$\pm$0.014 & --- & 163/156 & --- \\
\noalign{\smallskip}
IV & \emph{Chandra}$^{e}$ & 1126--1136  & PL & --- & --- & 1.7$\pm$0.4 & --- & 11/9 & --- \\
 &  &   & PL & 1.17$^{d}$ & 1.405$^{d}$ & 1.84$\pm$0.22 & --- & 12/9 & --- \\
& \emph{XMM}/pn$^{e}$ & 1530--1560  & PL & --- &  --- & 2.16$\pm$0.13 & --- & 16/16 & --- \\
&  &   & PL & 1.17$^{d}$ &  1.405$^{d}$ & 2.81$\pm$0.20 & --- & 27/16 & --- \\
& \emph{Chandra}$^{e}$ & 1755--1775  & PL & --- &  --- & 2.41$\pm$0.23 & --- & 5/9 & --- \\
&  &   & PL & 1.17$^{d}$ &  1.405$^{d}$ & 2.8$\pm$0.3 & --- & 8/9 & --- \\
& \emph{XMM}/pn$^{e}$ & 3040--3080 & PL & --- &  --- &2.5$\pm$0.5 & --- & 11/10 & --- \\
&  &  & PL & 1.17$^{d}$ &  1.405$^{d}$ & 4.4$\pm$0.5 & --- & 16/10 & --- \\
 & \emph{Chandra}$^{e}$ & 6600--6640  & PL & --- &  --- & 2.8$\pm$0.8 & --- & 4/5 & --- \\
 &  &   & PL & 1.17$^{d}$ &  1.405$^{d}$ & 3.7$\pm$0.8 & --- & 2/5 & --- \\
& \emph{Chandra}$^{e}$ & 7100--7140  & PL & --- &  --- & 2.7$\pm$0.6 & --- & 3/3 & --- \\
&  &   & PL & 1.17$^{d}$ &  1.405$^{d}$ & 3.3$\pm$0.8 & --- & 3/3 & --- \\
\noalign{\smallskip}
            \hline
\end{tabular}
\tablefoot{$^{a}$ The energy range of the instruments is 20\,--\,200\,keV for IBIS, 0.1\,--\,10\,GeV for \emph{Fermi}/LAT, 3\,--\,35\,keV for JEM-X, and 0.3\,--\,10\,keV for \emph{Swift}/XRT, \emph{XMM-Newton}-pn and \emph{Chandra}/ACIS-S. $^{b}$ The parameter kT corresponds to the temperature of the BBPL model and $E_{\rm break}$ to the break energy of the BKNPL. $^{c}$ Throughout, the F-test probability corresponds to the improvement with respect to the simplest model (PL). $^{d}$ Parameter fixed in the spectral model. $^{e}$ Only the Galactic absorption component is included in these observations.}
\end{table*}

\begin{figure}[ht!]
\centering
\resizebox{\hsize}{!}{\includegraphics[angle=0,width=12cm]{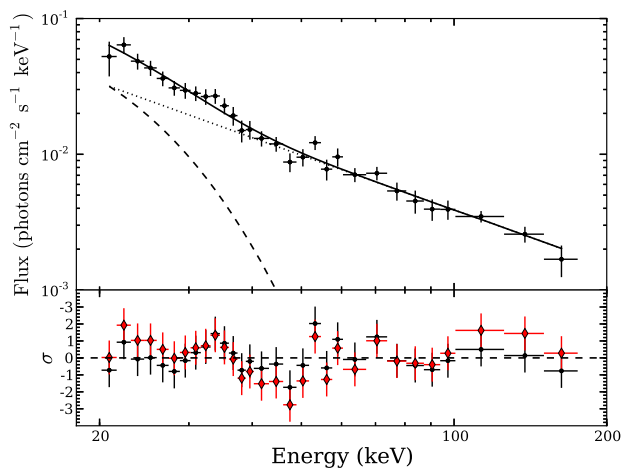}}
\caption{IBIS spectrum from 20\,keV to 200\,keV of GRB\,120711A during segment I (T$_{0}$+115\,s to T$_{0}$+130\,s). The best-fit model is shown as a solid line. The model includes blackbody and power-law components that are shown as dashed and dotted lines. The bottom panel shows the residuals from the blackbody+power-law fit (dots) and the single power-law fit (diamonds).}
\label{specseg1}
\end{figure}

The IBIS spectrum for segment I (T$_{0}$+115\,s to T$_{0}$+130\,s) is shown in Fig.~\ref{specseg1}. The null-hypothesis probability that the single power-law model is not the best fit is 0.8 ($\chi^{2}$/dof\,=\,35/28). As seen in Fig.~\ref{specseg1} (bottom panel), the distribution of the residuals suggests that an additional soft component is required. The F-test probability of a chance improvement of a two-component model (blackbody plus power-law or broken power-law) with respect to the single power-law model is $\sim$\,10$^{-4}$ (see Table~\ref{tab:xrayspec}). There is no significant difference between a fit using a blackbody+power-law and a broken power-law. A similar result is obtained for segment II (Table~\ref{tab:xrayspec}). In this case, the F-test probability of a chance improvement is $\sim$\,10$^{-2}$, and we note that all three models tested result in a poor fit for segment II (Table~\ref{tab:xrayspec}). The spectrum of the forward shock emission consists of a series of power-law segments with breaks caused by self-absorption, characteristic and cooling frequencies when crossing the observed band. During segments I and II, the observed change of the spectral index is $\Delta\beta$\,$\sim$\,1, which is higher than the value expected for the spectral breaks predicted in the forward shock model ($\Delta\beta$\,$\sim$\,0.5). A thermal component is not expected either in the forward shock spectrum. Therefore the spectral evolution during segments I and II may be caused by either late activity from the central engine or by the reverse shock.

\begin{figure}[ht!]
\centering
\resizebox{\hsize}{!}{\includegraphics[angle=0,width=12cm]{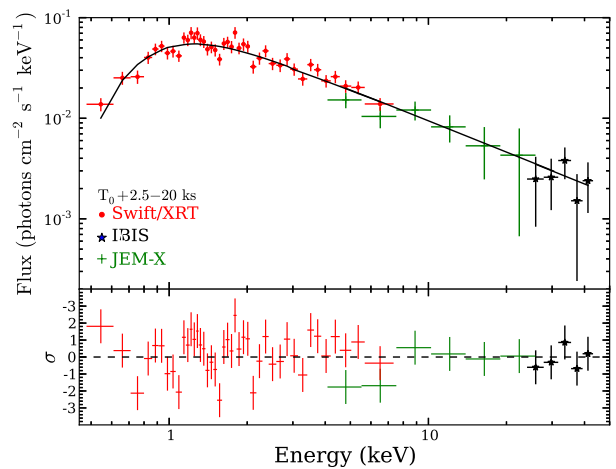}}
\caption{Broadband spectral fit from 0.3\,keV to 45\,keV during the time interval between T$_{0}$+2.5\,ks and T$_{0}$+20\,ks using data from \emph{Swift}/XRT, JEM-X, and IBIS. The data are well fit by a single power-law of $\Gamma$\,=2.06\,$\pm$\,0.05 that includes soft X-ray absorption from our galaxy and an intrinsic absorption of $N_{\rm H, intr}=1.87^{+0.14}_{-0.13}\times10^{22}$\,cm$^{-2}$. The residuals to the fit are shown in the bottom panel.}
\label{jointspec}
\end{figure}

The joint spectral fit from T$_{0}$+2.5\,ks to T$_{0}$+20\,ks (segment III) using \emph{Swift}/XRT, JEM-X, and IBIS is shown in Fig.~\ref{jointspec}. It is well fit by a single power-law of photon index $\Gamma$\,=\,2.06\,$\pm$\,0.05 ($\chi^{2}$/dof\,=\,67/51) including two absorption components at soft X-rays, a Galactic component (fixed), and intrinsic absorption with a value of $N_{\rm H, intr}=1.87^{+0.14}_{-0.13}\times10^{22}$\,cm$^{-2}$. We note that a broken power-law model would also satisfy the data with a break energy at $\sim$\,1.3\,keV fixing $\Delta\beta$\,=\,0.5 ($\chi^{2}$/dof\,=\,64/50). However, the fit is limited by the uncertainty in the normalisation between the instruments and the degeneracy between the intrinsic X-ray absorption and the low-energy power-law component.

The evolution of the photon index during the X-ray observation campaign with \emph{Swift}, \emph{XMM-Newton} and \emph{Chandra} (see Fig.~\ref{xraylc}, bottom panel) shows little spectral evolution when compared with the resulting photon index from the joint spectral fit (Fig.~\ref{jointspec}). However, these variations of the photon index are consistent within 2$\sigma$. Therefore we consider the X-ray photon index to be constant during segment III.

\subsubsection{Spectral energy distributions}
\begin{figure}[ht!]
\centering
\resizebox{\hsize}{!}{\includegraphics[angle=0,width=12cm]{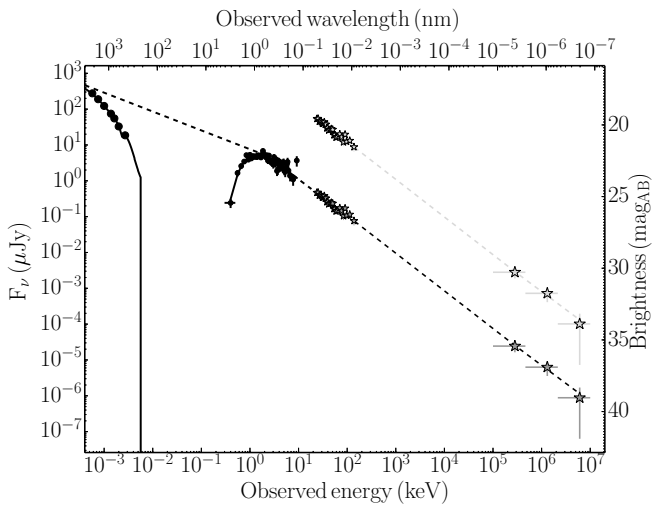}}
\caption{Spectral energy distributions at two different time intervals. The SED for IBIS and \emph{Fermi}/LAT data from T$_{0}$+300\,s to T$_{0}$+1050\,s (light-grey stars), and the SED for GROND and \emph{Swift}/XRT at T$_{0}$+26\,ks (black circles). The IBIS and \emph{Fermi}/LAT data are also shown rescaled (dark-grey stars) to the GROND-\emph{Swift}/XRT SED to illustrate the single-spectra component from X-rays to GeV.}
\label{sed}
\end{figure}

Fig.~\ref{sed} shows that the spectral energy distribution (SED) determined by GROND is very red, showing clear signs of rest-frame dust extinction. A joint fit with the X-ray data shows that the best fit is given by a broken power-law, with the cooling frequency $\nu_c$ between the optical and the X-ray bands ($\nu_c$\,=\,2.37$^{+0.97}_{-0.48}$\,keV). The data were corrected for the Galactic foreground extinction of $E_{(B-V)}=0.08$ \citep{Schlegel1998} and Galactic hydrogen column density $N_{\rm H}=7.9\times10^{20}$\,cm$^{-2}$ \citep{Kalberla:2005p1538} for this fit. The SED was determined at 26\,ks, using direct GROND observations and X-ray data for which the SED was determined over a broader time-frame (no spectral evolution is seen), and then scaled to the same time as the GROND data. Table~\ref{tab:sed} shows the different models considered under the three common extinction curves (Milky Way, SMC, and LMC). The best fit is found using the LMC dust ($\chi^2$/dof=436/490), with $\beta_{\rm opt}=0.53\pm0.02$ ($\beta_{X}=\beta_{\rm opt}+0.5$), $A_{\rm V}=0.85\pm0.06$, and $N_{\rm H, intr}=1.04^{+0.24}_{-0.21}\times10^{22}$\,cm$^{-2}$. The soft X-ray intrinsic absorption found in this fit is consistent with that obtained using X-ray data alone (see Table~\ref{tab:xrayspec}). The 2175\,{\AA} bump lies between the $g^\prime$ and the $r^\prime$ filter, so there is no direct detection of this bump in the data set. However, in the SMC fit, the $r^\prime$-band flux is significantly overestimated, whereas the $g^\prime$-band flux is similarly underestimated. These two bands are fit excellently in the LMC model.

\begin{table*}[ht!]
\caption{Resulting parameters from the SED fit between optical/NIR and X-ray data at T$_{0}$\,+\,26\,ks using a broken power-law model. In all cases $\beta_X=\beta_{opt}+0.5$.}
\label{tab:sed}
\centering
\begin{tabular}{lcccccc}
\noalign{\smallskip}
\hline
\hline
\noalign{\smallskip}
Extinction & R$_{\rm V}$$^{a}$ & A$_{\rm V}$$^{b}$ & $N_{\rm H, intr}$ & $\beta_{\rm opt}$ & $E_{\rm break}$ & $\chi^{2}$/dof \\
   model & & & (10$^{22}$\,cm$^{-2}$) & & (keV) & \\
\noalign{\smallskip}
            \hline
\noalign{\smallskip}
Milky Way & 3.08 & 0.91\,$\pm$\,0.06 & 0.86$_{-0.17}^{+0.18}$ &0.55\,$\pm$\,0.01 & 21.75 & 445/490\\
LMC & 3.16 & 0.85\,$\pm$\,0.06 & 1.04$_{-0.21}^{+0.24}$ &0.53\,$\pm$\,0.02 & 2.37$^{+0.97}_{-0.48}$ & 434/490\\
SMC & 2.93 & 0.76\,$\pm$\,0.05 & 1.03$_{-0.21}^{+0.23}$ &0.51\,$\pm$\,0.02 & 2.28$^{+0.97}_{-0.48}$ & 443/490\\
\noalign{\smallskip}
            \hline
\end{tabular}
\tablefoot{$^{a}$: frozen parameter. $^{b}$: value calculated at the GRB rest-frame}
\end{table*}

Fig.~\ref{sed} also shows the simultaneous IBIS and \emph{Fermi} data between T$_{0}$+300\,s and T$_{0}$+1050\,s, during segment III. The data are well fit by a single power-law with photon index of $\sim$\,2, which is consistent with synchrotron radiation. However, the error on the \emph{Fermi}/LAT photon index is quite large (Table~\ref{tab:xrayspec}). An analysis where the normalisation between both instruments was allowed to vary showed that the results from both instruments are consistent with each other. The lack of spectral data over more than two decades in energy between \emph{INTEGRAL} and \emph{Fermi} makes it difficult to completely rule out inverse Compton or hadronic components that peak at energies below 100\,MeV. 

\section{Interpretation of the post-GRB emission}
\label{sec:model}
The soft $\gamma$/X-ray and optical light curves for GRB\,120711A exhibit several interesting features including an optical flash, a short plateau phase at soft $\gamma$-rays; a high-energy light curve consisting of four segments; a spectral break between the optical and the X-ray emission due to $\nu_c$ (Fig.~\ref{sed}); and a single power-law spectral model between \emph{INTEGRAL} and \emph{Fermi}. The shape of the high-energy light curve in Fig.~\ref{afterglow} resembles those usually observed in GRB afterglows. However, for GRB\,120711A the emission is observed above 20\,keV during the first 10\,ks. The 20\,--\,40\,keV energy band provides the most sensitive light curve, and the temporal parameters used to model the IBIS light curve were obtained from this energy band. The properties of the emission are studied in the following sections in the context of the standard afterglow model \citep[e.g.][]{{Sari:1998p1670},{2000ApJ...536..195C},{2002ApJ...568..820G},{Zhang:2006p369}}.

\subsection{Segments I and II}

Segments I and II span the time interval from $\sim$\,T$_{0}$\,+115\,--\,283\,s, which includes the high-energy early steep decay and plateau phases and the optical flash.

The spectral index during the soft $\gamma$-ray fast decay is $\beta_{\gamma,\rm 1}$\,=\,0.38\,$\pm$\,0.10 with an additional softer component in the form of a second power-law of index, $\beta'_{\gamma,\rm 1}$\,$\sim$\,1.2 (Table~\ref{tab:xrayspec}). This additional component seems to preclude the simple relation between spectral and temporal indices that is typically used to identify high-latitude emission \citep[$\alpha=2+\beta$,][]{{1997asxo.proc..167F},{2000ApJ...541L..51K},{Zhang:2006p369}}, which assumes a single power-law. However, as reported by \cite{2011A&A...529A.142R}, this relationship can be recovered assuming that the flux and the peak energy of the high-latitude emission depend on the viewing angle (i.e. the electron population is not constant for all angles). For GRB\,120711A, the main pulse consists of two broad peaks that contribute to the high-latitude emission. This multi-peaked behaviour can lead to temporal indices that do not follow the simple correlation between the spectral and temporal indices \citep{2009MNRAS.399.1328G}. Steeper decays than the one expected from the simple approximation have also been reported for several GRBs in the \emph{Swift} sample \citep{2013MNRAS.428..729M}. 

The high-energy plateau phase in GRB\,120711A is unusually short when compared with the \emph{Swift} GRB sample \citep{2013MNRAS.428..729M}. Assuming late activity of the central engine \citep[e.g.][]{{2007ChJAA...7....1Z}}, the energy injection parameter in the case of fast or slow cooling is \textit{q}\,=\,0.3\,$\pm$\,0.3, where $\alpha_{\gamma,2}$\,=\,0.26\,$\pm$\,0.30 and is either $\beta_{\gamma,2}$\,=\,0.93\,$\pm$\,0.15 or $\beta_{\gamma,2}$\,=\,1.35\,$\pm$\,0.08, when the cooling frequency is lower than that of the IBIS band. Nevertheless, the X-ray plateau can also be linked with reverse shocks in the optical band \citep{{2007ApJ...665L..93U},{2007MNRAS.381..732G}}. For GRB\,120711A, a powerful optical flash is observed in the optical/NIR light curve (Fig.~\ref{opticallc}) that may be responsible for this high-energy plateau.

The optical/NIR light curve during these temporal segments shows a very steep increase of flux that peaks at $\sim$\,T$_{0}$\,+\,126\,s and then decays with a temporal index $\alpha_{\rm, opt, 2}$\,$\sim$\,3.3. This behaviour resembles that of a type-II reverse shock \citep{{Zhang:2003p365},{2007MNRAS.378.1043J}}. This hypothesis is also supported by the lack of chromatic evolution around the peak (Fig.~\ref{opticallc}). In this type of reverse shock, the forward and reverse shocks are expected to peak at similar times with the reverse shock outshining the forward shock. The observed increases of the reverse and forward shock emission are steeper than those expected in a wind environment, $t_{\rm RS}^{5/2}$ and $t_{\rm FS}^{1/2}$ \citep{{2000ApJ...536..195C},{2003ApJ...597..455K}}, which could suggest an ISM environment. However, this steep increase in the flux can also be achieved in a wind environment if self-absorption cannot be ignored \citep{2000ApJ...536..195C}. Additionally, the rising slope of the forward shock model is not well constrained for GRB\,120711A.

Regardless of the environment, in this type of reverse-forward shock emission, a flattening of the light curve is expected at later times when the forward shock dominates, with a temporal index $\sim$\,1.1. For GRB\,120711A, a break in the decay is seen by the end of segment II ($\sim$\,T$_{0}$\,+\,215\,s). However, the new temporal decay index is steeper than that expected for pure forward shock emission ($\alpha_{\rm, opt, 3}$\,$\sim$\,1.80). The interpretation of this part of the optical/NIR light curve is discussed in Sect.~\ref{sec:s3}. 

In the on-axis GRB model, the time at which the reverse shock finishes crossing the ejecta can be interpreted as the deceleration time of the jet, $t_{\rm dec}$. For GRB\,120711A this time ($t_{\rm peak}$\,$\sim$\,T$_{0}$\,+\,126\,s) is similar to the duration of the burst ($\sim$\,115\,s). Therefore, GRB\,120711A seems to be in an intermediate state between the so-called thin-shell case for which t$_{\rm peak}$\,$>$\,T$_{90}$, and the thick-shell case for which t$_{\rm peak}$\,$<$\,T$_{90}$ \citep[e.g.][]{{Sari:1999p1259},{2005MNRAS.363...93Z},{Meszaros:2006p261}}. The time of the peak of the forward shock (t$_{\rm peak}$\,$\sim$\,T$_{0}$\,+\,240\,s) can be interpreted as the onset of the afterglow emission and thus it can also be used to estimate $t_{\rm dec}$ \citep[e.g.][]{{Meszaros:2006p261},{2007A&A...469L..13M}}.

\subsection{Segment III: multi-wavelength afterglow emission}
\label{sec:s3}
Segment III spans most of the afterglow emission from $\sim$\,T$_{0}$\,+283\,s to $\sim$\,T$_{0}$\,+900\,ks with data available from near-infrared, optical, X-rays, soft $\gamma$-rays, and photons with energies $>$\,100\,MeV.

The SED shown in Fig.~\ref{sed}, which combines optical/NIR and X-ray data, indicates that a break is needed between the two energy bands. Considering that the peak of the forward shock occurred at $\sim$\,T$_{0}$\,+\,126\,s, $\nu_{\rm m}$\,$<$\,$\nu_{\rm opt}$\,$<$\,$\nu_{\rm c}$\,$\sim$\,$\nu_{\rm X}$\,$<$\,$\nu_{\gamma}$ during most of segment III, where $\nu_{\rm m}$ is the characteristic frequency and $\nu_{\rm c}$ the cooling frequency. In an ISM environment, $\nu_{\rm c}$ is expected to decrease with time as $\nu_{\rm c}$\,$\propto$\,t$^{-1/2}$, while in a wind environment $\nu_{\rm c}$ is expected to increase with time as $\nu_{\rm c}$\,$\propto$\,t$^{1/2}$ \citep[e.g.][]{{2000ApJ...536..195C},{2002ApJ...568..820G},{Zhang:2006p369}}.

As mentioned, there are two possible environments depending on the density profile $\rho=Ar^{-k}$, homogeneous or ISM when $k$\,=\,0 and wind when $k$\,=\,2. An estimate of the real $k$-index can be obtained if the observed frequency is below $\nu_{\rm c}$  \citep{2008ApJ...672..433S}. At T$_{0}$\,+\,26\,ks, the optical temporal and spectral indices are $\alpha_{\rm opt, 4}$\,=\,1.00\,$\pm$\,0.05 and $\beta_{\rm opt}$\,=\,0.53\,$\pm$\,0.02, respectively. Thus, the index of the density profile, $k$ given by Eq.~\ref{eq:density} is

\begin{equation}
\label{eq:density}
\begin{array}{l}
k=\dfrac{4(3\beta-2\alpha)}{3\beta-2\alpha-1}\,=\,1.2\,\pm\,0.3,
\end{array}
\end{equation}
which corresponds to an intermediate case between ISM and wind environments at a $\sim$\,4\,$\sigma$ confidence level.

In the IBIS data, $\nu_{\rm c}$\,$<$\,$\nu_{\gamma}$, which means that, the ISM and wind environments cannot be distinguished. The relationship between the spectral and temporal index in the slow-cooling regime is $\alpha_{\rm th,\gamma}$\footnote{The underscript $th$ is used to indicate expected values from the closure relations.}=(3$\beta_{\gamma}$-1)/2\,=\,1.03\,$\pm$\,0.09. This value is consistent with the observed temporal decay of the IBIS data ($\alpha_{\gamma, 3}$\,=\,1.13\,$\pm$\,0.04). Thus, the IBIS soft $\gamma$-ray data during this segment are consistent with forward shock emission. Similarly, the \emph{Fermi}/LAT data are also consistent with forward shock emission. The marginal difference between the IBIS and \emph{Fermi}/LAT temporal decays could be explained by the reduced efficiency of the shock to accelerate electrons that emit at energies above 100\,MeV \citep{2012ApJ...749...80S}.

If the optical/NIR data simultaneous with IBIS are considered as pure forward shock emission, then $\alpha_{\rm opt, 3}$\,$>$\,$\alpha_{\gamma, 3}$, favouring the wind environment. However, the closure relation in slow cooling for the wind environment is not compatible with the observed decay. As mentioned in Sect.~\ref{sec:temporal} (see Fig.~\ref{opticalfit}), the optical decay index of $\alpha_{\rm opt, 3}$\,$\sim$\,1.80 can be obtained from a combination of the fast-decaying reverse shock and the forward shock with temporal index $\alpha_{\rm opt, 4}$, which is similar to $\alpha_{\gamma, 3}$. This lack of a temporal break between the optical and soft $\gamma$-ray bands is incompatible with the ISM or wind environments when $\nu_{\rm opt}$\,$<$\,$\nu_{\rm c}$\,$<$\,$\nu_{\gamma}$ ($\Delta\alpha_{\rm th}$\,$\sim$\,0.25). However, if a more complete set of closure relations is considered where $k$\,=\,1.2 \citep{2008ApJ...672..433S}, the expected temporal index for the optical band becomes $\alpha_{\rm th, opt}$\,=\,[6$\beta_{\rm opt}$(4-$k$+2$k$)]/4(4-$k$)\,$\sim$\,1.01, which is consistent with $\alpha_{\rm opt, 4}$. As previously mentioned, $\alpha_{\rm th,\gamma}$ does not depend on $k$ and is consistent with the negligible difference between the optical and $\gamma$-ray temporal indices found for GRB\,120711A. Therefore, the afterglow emission of GRB\,120711A is consistent with a mildly stratified wind-like environment. Note that the  late-time variability in the optical emission reported in Sect.~\ref{sec:temporal} may imply that $\alpha_{\rm opt, 4}$ is actually steeper than measured, which could lead to an environment more like the classic wind profile (i.e. $k$\,=\,2).

The electron spectral index, $p$, can be derived from the IBIS and optical data for a wind-like environment of $k$\,$\sim$\,1.2 as $p$\,=\,2\,$\beta_{\gamma}$\,=\,2.04\,$\pm$\,0.12 ($\beta_{\gamma}$\,=\,1.02\,$\pm$\,0.06) and $p$\,=\,2\,$\beta_{\rm opt}$\,+\,1\,=\,2.06\,$\pm$\,0.04 ($\beta_{\rm opt}$\,=\,0.53\,$\pm$\,0.02), respectively.

The X-ray data in segment III show a much steeper temporal decay than the IBIS data with no significant spectral change. The location of $\nu_c$ within the X-ray band precludes the closure relations between the spectral and temporal indices, since they are defined for cases above and below the break frequencies. 

The smoothing parameter, $s$ of the spectral break caused by $\nu_c$ is determined by the electron spectral index as $s$\,=\,(0.80-0.03\,$p$)\,$\sim$\,0.7 \citep{2002ApJ...568..820G}. Such a smooth break (the lower $|s|$, the smoother the break) is extremely difficult to identify in the narrow X-ray band of 0.3\,--\,10\,keV or even in the joint spectral fit shown in Fig.~\ref{jointspec} from 0.3\,--\,40\,keV. The fact that a broken power-law was fitted using a single power-law component when using only X-ray data might explain the 2\,$\sigma$ variations of the photon index shown in Fig.~\ref{xraylc} (bottom panel) and the relative softening seen at $>$\,T$_{0}$\,+\,1\,Ms.

\cite{2002ApJ...568..820G} described the flux density near $\nu_c$ as
\begin{equation}
\label{eq:flux}
\begin{array}{l}
F_{\nu}=F_{\nu_{c}} \left[\left(\dfrac{\nu}{\nu_{c}}\right)^{-s\beta_1} + \left(\dfrac{\nu}{\nu_{c}}\right)^{-s\beta_2} \right]^{-1/s}.
\end{array}
\end{equation}

This means that near the spectral break caused by $\nu_c$, the temporal evolution of the observed flux is $F_{\nu}$\,$\propto$\,$F_{\nu_{c}}$\,$\propto$\,$t^{1/2-p}$\,$\sim$\,$t^{-1.6}$, which is different from the temporal evolution predicted by the closure relations when $\nu_{\rm obs}$\,$>$\,$\nu_c$, $t^{(2-3p)/4}$\,$\sim$\,$t^{-1.1}$. This steeper temporal decay is consistent with the observed X-ray decay of $\alpha_{\rm X, 1}$\,=\,1.65\,$\pm$\,0.04. The lack of X-ray spectral evolution can be explained by the mildly stratified environment with $k$\,$\sim$\,1.2 in GRB\,120711A. As mentioned earlier, for a wind environment the cooling frequency is expected to evolve as $\nu_{\rm c}$\,$\propto$\,t$^{1/2}$. However, a non-evolving $\nu_{\rm c}$ can be found when $k$\,$\sim$\,1.5, suggesting that for GRB\,120711A, $\nu_{\rm c}$ remains within the X-ray band for the entire observation \citep[see also GRB\,130427A, ][]{2013arXiv1307.4401P}. The time evolution of $\nu_{\rm c}$ might also be modified if the microphysical parameters of the afterglow emission are time dependent \citep{{2006MNRAS.369.2059P},{2011A&A...535A..57F}}.

The closure relations when $\nu_{\rm obs}$\,$<$\,$\nu_c$ in an ISM environment predict a temporal index consistent with that found using only the \emph{Swift} X-ray data ($\alpha_{\rm th, opt}$\,=\,(3$p$-2)/4\,$\sim$\,1.5), which leads to $p$\,$\sim$\,3. However, this model is rejected when combining the X-ray data with all the multi-wavelength data obtained for GRB\,120711A. Hence, it is possible that GRBs with limited optical/NIR observations might be described with an ISM model when in reality they occur in a wind-like environment.

\subsection{Segments IV: late X-ray afterglow}
\label{sec:s4}
The X-ray light curve steepens in this last segment. The values of the temporal and spectral indices are $\alpha_{\rm X,2}$\,=\,1.96\,$\pm$\,0.22 and $\beta_{\rm X,2}$\,=\,1.5\,$\pm$\,0.5. No significant spectral change is seen during the break at T$_{0}\sim$\,1\,Ms, which suggests the occurrence of a jet break. The change in the temporal index with the previous segment is $\Delta\alpha_{\rm X}$\,=\,0.31\,$\pm$\,0.22 when compared with the X-ray decay. This value is consistent with the expected change due to a jet break for the wind model \citep[$\sim$\,0.4][]{Kumar:2000p2152}. For a non-spreading jet in a wind environment with $\nu_{\rm X}$\,$>$\,$\nu_{\rm c}$, the expected temporal index is $\alpha_{\rm th}=3\beta_{\rm X}/2$\,$\sim$\,2.3\,$\pm$\,1.5, consistent with the observed decay after the break. As expected for a jet break, the observed temporal decay after the break is consistent with the electron spectral index \citep{{Kumar:2000p2152}}.

Even though there is good agreement between the model and the data in the case of a jet break, the tendency towards a softer spectrum after the break could also be attributed to the inefficiency of the shock to accelerate X-ray emitting electrons \citep{2012ApJ...749...80S}.

\section{Emission parameters of GRB\,120711A in an wind environment}
\label{sec:parameters}
The standard afterglow model in a stratified wind-like environment contains four free parameters \citep[e.g.][]{{2000ApJ...536..195C},{2002ApJ...568..820G}}: the kinetic energy, $E_{\rm k}$, the energy equipartition fractions of electrons and magnetic field, $\epsilon_{\rm e}$ and $\epsilon_{\rm B}$, and the wind parameter, $A_*$. $A_*$ is defined by $A_*$\,=\,($\dot{M}_{\rm W}$\,/4\,$\pi$\,$V_{\rm W}$)/(5\,$\times$\,10$^{11}$)\,g\,cm$^{-1}$, where $\dot{M}_{\rm W}$ is the mass-loss rate, and $V_{\rm W}$ is the wind velocity of a typical Wolf-Rayet star ($V_{\rm W}$\,=\,1000\,km\,s$^{-1}$). Although studies have shown that the microphysical parameters can be time dependent \citep{2006MNRAS.369.2059P}, for GRB\,120711A they are considered as constants. From Eq.~\ref{eq:density}, the best density profile in this case is $k$\,=\,1.2\,$\pm$\,0.3. The formulation by \cite{2002ApJ...568..820G} considers the two cases k\,=\,0 and 2. A complete derivation for $k$\,=\,1.2 is beyond the scope of this work. The parameter most affected by the $k$ value is the initial bulk Lorentz factor, which varies by a factor of $\sim$\,2 when comparing the two extreme cases of $k$\,=\,0 and $k$\,=\,2. Therefore, to allow calculation of the microphysical parameters in this case, we assumed $k$\,=\,2 and an electron spectral index of $\sim$\,2.1.

Considering the formulation from \cite{2002ApJ...568..820G} to describe the forward shock emission, three constraints on the fireball parameters can be obtained as follows: at T$_{0}$\,+\,26\,ks, $\nu_c$\,$\sim$\,2.37\,keV with an unabsorbed flux of $F_{\nu_c}$\,$\sim$\,5.33\,$\mu$Jy; and at T$_{0}$\,+\,238\,s the forward shock peaks and the location of $\nu_m$ can be established at $\nu_m$\,$\sim$\,4.68\,$\times$\,10$^{14}$\,Hz. These constraints result in the following three conditions:

\begin{equation}
\label{eq:param}
\begin{array}{lll}
\epsilon_{\rm e}^{2}\epsilon_{\rm B}^{1/2}E_{k,52,\rm iso}^{1/2}=9.3\times10^{-4},\\
\\
A_{*}^{-2}\epsilon_{\rm B}^{-3/2}E_{k,52,\rm iso}^{1/2}=2.5\times10^{7},\\
\\
\epsilon_{\rm e}^{1.1}\epsilon_{\rm B}^{1.6}A_{*}^{2.1}E_{k,52,\rm iso}^{0.5}=2.3\times10^{-7},
\end{array}
\end{equation}
where the luminosity distance \textit{d$_{\rm L,28}$}\,=\,3.18\,cm calculated for a redshift of 1.405 \citep{2012GCN..13441...1T} has been taken into account.

A fourth condition must be imposed to find all unknown parameters in Eq.~\ref{eq:param}. The lack of a constraint on the location of the self-absorption frequency ($\nu_{sa}$) requires that one parameter must be fixed. The radiative efficiency, $\eta_{\gamma}$, to convert total energy, $E_{\rm total}$, into radiation is defined as $\eta_{\gamma}=E_{\gamma}/E_{\rm total}$, where $E_{\rm total}=E_{\rm \gamma}+E_{\rm k}$. Most long GRBs are well fitted with $\eta_{\gamma}$\,$<$\,0.2 \citep{2007ApJ...655..989Z}. However, \cite{2011ApJ...738..138R} reported that while that statement is true for most \emph{Swift}/BAT GRBs, for \emph{Fermi}/LAT GRBs it might not be valid because higher efficiencies are required ($\eta_{\gamma}$\,$\sim$\,0.5 or even higher in a few cases). In these studies all other microphysical parameters were assumed to be the same for all GRBs in the sample. Considering the ambiguity on $\eta_{\gamma}$ and that GRB\,120711A is also a \emph{Fermi}/LAT GRB, all physical parameters were calculated for two cases, $\eta_{\gamma}$\,$\sim$\,0.2 and $\eta_{\gamma}$\,$\sim$\,0.5.

Using the data from the prompt emission, the rest-frame k-corrected isotropic energy radiated during the T$_{90}$ in the 1\,keV\,--\,10\,MeV energy band is $E_{\gamma, \rm iso}$\,=\,1.65\,$\times\,10^{54}$\,erg. Thus, assuming $\eta_{\gamma}$\,$\sim$\,0.2 (0.5), the isotropic kinetic energy is $E_{\rm k,iso}$\,=\,6.6\,$\times\,10^{54}$\,erg (1.65\,$\times\,10^{54}$\,erg). Applying this result to Eq.~\ref{eq:param} gives $\epsilon_{\rm B}$\,=\,2\,$\times$\,10$^{-3}$ (4\,$\times$\,10$^{-5}$), $\epsilon_{\rm e}$\,=\,0.03 (0.12), and $A_{*}$\,=\,0.11 (1.6). These values are consistent with those reported for other GRBs \citep[e.g.][]{{2002ApJ...571..779P},{2007ApJ...655..989Z}}. The estimated wind parameter implies a mass-loss rate for the progenitor of GRB 120711A of $\dot{M}_{\rm W}$\,$\sim$\,10$^{-6}$\,$M_{\odot}$\,yr$^{-1}$ (10$^{-5}$\,$M_{\odot}$\,yr$^{-1}$), which is consistent with the expected mass-loss rate of an evolved Wolf-Rayet star at the end of its life \citep{{1989A&A...220..135L},{Chevalier:1999p2185}}. This model also predicts the location of the self-absorption frequency, $\nu_{\rm sa}$, at the time of the ATCA radio observation (T$_{0}$\,+\,3.78\,days) at 51\,GHz (241 GHz), which is higher than the frequency used for the observation \citep[34 GHz, ][]{2012GCN..13485...1H}. Accodingly, the lack of a detection of the radio afterglow emission from GRB\,120711A might be due to self-absorption.

The peak emission from the reverse shock corresponds to the time when the shock has finished crossing the ejecta and therefore is a good indicator of the deceleration time. From \cite{2005MNRAS.363...93Z}, the Lorentz factor at this time can be expressed as 

\begin{equation}
\label{eq:Lorentz}
\begin{array}{l}
\gamma_{\rm dec}(t_{\rm dec})= 165 \left( \dfrac{(1+z)E_{\rm k,iso,53}}{A_{*,-1}t_{\rm dec,s}} \right)^{1/4}\sim\,170\,(62).
\end{array}
\end{equation}

At the deceleration time, the initial bulk Lorentz factor, $\gamma_{0}$, is expected to be twice the Lorentz factor at $t_{\rm dec}$ \citep[e.g.][]{{Sari:1999p1259},{Meszaros:2006p261}}, which results in $\gamma_{0}$\,=\,2\,$\gamma_{\rm dec}$\,$\sim$\,340 (124). 

The initial bulk Lorentz factor can also be constrained by considering that the peak of the forward shock ($\sim$\,T$_{0}$\,+238\,s) is the onset of the afterglow emission and marks the deceleration of the fireball. Using the expressions from \cite{2007A&A...469L..13M} for the thin-shell case (t$_{\rm peak}$\,$>$\,T$_{90}$), the initial Lorentz factor is $\gamma_{0}$\,$\sim$\,402 (260). These values are slightly higher (more than $\times$\,2 in the case of $\eta_{\gamma}$\,$\sim$\,0.5) than those measured using the peak of the reverse shock. However, considering the large uncertainty on the time of the forward shock peak, they can be considered to be compatible.

The deceleration radius, $R_{\rm dec}$, in a wind environment expressed as a function of the initial Lorentz factor \citep{2000ApJ...536..195C} can be estimated as

\begin{equation}
\label{eq:radius}
\begin{array}{l}
R(t)=1.1\times\,10^{17} \left( \dfrac{2t_{\rm days}E_{\rm k,iso,52}}{(1+z)A_{*}} \right)^{1/2}\,\,\textrm{cm,}
\end{array}
\end{equation}
which results in $R_{\rm dec}$\,=3\,$\times$\,10$^{17}$\,cm (3\,$\times$\,10$^{16}$\,cm), consistent with that reported for other GRBs \citep[e.g.][]{2006cosp...36...77Z}.

In Sect.~\ref{sec:s4}, the last temporal segment in the X-ray light curve can be explained by a jet break. At the time of the jet break ($t_{\rm jet}$\,$\sim$\,T$_{0}$\,+\,0.9\,Ms), the jet half-opening angle can be assumed to be $\theta_{\rm jet}$\,$\sim$\,$\gamma^{-1}(t_{\rm jet})$. Therefore, the jet half-opening angle using the wind model \citep{{2000ApJ...536..195C},{2006A&A...452..839G}} can be expressed as

\begin{equation}
\label{eq:gamma2}
\begin{array}{l}
\theta_{\rm jet}=0.2016 \left( \dfrac{A_{*}t_{\rm days}}{(1+z)E_{\rm k,iso,52}} \right)^{1/4},
\end{array}
\end{equation}
which results in $\theta_{\rm jet}$\,$\sim$\,0.03\,rad (2$^{\circ}$) for $\eta_{\gamma}$\,$\sim$\,0.2 and $\theta_{\rm jet}$\,$\sim$\,0.09\,rad (5$^{\circ}$) for $\eta_{\gamma}$\,$\sim$\,0.5. This corresponds to a Lorentz factor at the time of the jet break of $\gamma_{\rm jet}$\,=\,30 (11). The high Lorentz factor obtained at the jet break for $\eta_{\gamma}$\,$\sim$\,0.2 could indicate that $\eta_{\gamma}$\,$>$\,0.2 may be preferred for GRB\,120711A.

The collimation factor, $f_{\rm b}$, is then $f_{\rm b}$\,=\,$1-\rm cos\,\theta_{\rm jet}$\,=\,4\,$\times$\,10$^{-4}$ (4\,$\times$\,10$^{-3}$). Therefore, the corrected radiative and kinetic energies are $E_{\gamma}$\,=\,$E_{\gamma, \rm iso}f_{\rm b}$\,=\,7\,$\times$\,10$^{50}$\,erg (7\,$\times$\,10$^{51}$\,erg) and $E_{\rm k}$\,=\,$E_{\rm k,iso}f_{\rm b}$\,=\,3\,$\times$\,10$^{51}$\,erg (7\,$\times$\,10$^{51}$\,erg). The total energy of the fireball is then $E_{\rm total}$\,=\,$E_{\gamma}$\,+\,$E_{\rm k}$\,$\sim$\,4\,$\times$\,10$^{51}$\,erg (1.4\,$\times$\,10$^{52}$\,erg). As a result of these values, either $\eta_{\gamma}$\,$\sim$\,0.2 is preferred or GRB\,120711A approaches the class of hyper-energetic
GRBs suggested by \cite{2011ApJ...732...29C}.

The baryon load $M_{\rm fb}$ is described as a function of the corrected kinetic energy of the blast-wave and the initial Lorentz factor by

\begin{equation}
\begin{array}{l}
M_{\rm fb}=\dfrac{E_{\rm k}}{\gamma_{0}c^{2}},
\end{array}
\end{equation}
and gives a value of $M_{\rm fb}$\,=\,5\,$\times$\,10$^{-6}$\,$M_{\odot}$ (3\,$\times$\,10$^{-5}$\,$M_{\odot}$) for GRB\,120711A, using the beam-corrected $E_{\rm k}$. $M_{\rm fb}$ is consistent with the value expected from the fireball model \citep[e.g.][and references therein]{Piran:2005p2117}.

The emission parameters obtained for GRB\,120711A are given in Table~\ref{tab:physics}.

\begin{table*}[ht!]
\caption{Emission parameters obtained for GRB\,120711A for two cases with different radiation efficiency, $\eta_{\gamma}$.}
\label{tab:physics}
\centering
\begin{tabular}{llccc}
\noalign{\smallskip}
\hline
\hline
\noalign{\smallskip}
Parameter & Symbol & Approx. values & & Approx. values \\
          &        & for $\eta_{\gamma}$\,=\,0.2 & & for $\eta_{\gamma}$\,=\,0.5\\
\noalign{\smallskip}
            \hline
\noalign{\smallskip}
Electron spectral index &$p$ & 2.1&  & 2.1\\
Redshift &$z$ & 1.405 & & 1.405\\
Luminosity distance &$d_{\rm L}$ & 3.18\,$\times\,10^{28}$\,cm & & 3.18\,$\times\,10^{28}$\,cm\\
Deceleration time &$t_{\rm dec}$ & T$_{0}$\,+\,126\,s & & T$_{0}$\,+\,126\,s\\
Time of the jet &$t_{\rm jet}$ & T$_{0}$\,+\,8.5\,Ms& & T$_{0}$\,+\,8.5\,Ms\\
Peak energy &$E_{\rm peak}$ & 1130\,keV & & 1130\,keV\\
Rest-frame peak energy &$E_{\rm peak,z}$ & 2260\,keV& & 2260\,keV\\
Isotropic energy &$E_{\gamma,\rm iso}$ & 1.65\,$\times\,10^{54}$\,erg & & 1.65\,$\times\,10^{54}$\,erg\\
Isotropic kinetic energy &$E_{\rm k,iso}$ & 6.6\,$\times\,10^{54}$\,erg& & 1.65\,$\times\,10^{54}$\,erg\\
Wind density parameter & $A_{*}$ & 0.11 && 1.6 \\
Electron energy fraction  &$\epsilon_{\rm e}$ & 0.03 && 0.12\\
Magnetic energy fraction & $\epsilon_{\rm B}$ & 2\,$\times$\,10$^{-3}$ && 4\,$\times$\,10$^{-5}$\\
Mass-loss rate & $\dot{M}_{\rm W}$ & 10$^{-6}$\,$M_{\odot}$\,yr$^{-1}$ && 10$^{-5}$\,$M_{\odot}$\,yr$^{-1}$\\
Initial bulk Lorentz factor &$\gamma_{0}$ & 340 && 124\\
Jet half-opening angle &$\theta_{\rm jet}$ & 0.03\,rad (2$^{\circ}$)&& 0.09\,rad (5$^{\circ}$)\\
Collimation factor &$f_{b}$ & 4\,$\times$\,10$^{-4}$&& 4\,$\times$\,10$^{-3}$\\
Radiated energy &$E_{\gamma}$ & 7\,$\times$\,10$^{50}$\,erg&& 7\,$\times$\,10$^{51}$\,erg\\
Kinetic energy &$E_{\rm k}$ & 3\,$\times$\,10$^{51}$\,erg&& 7\,$\times$\,10$^{51}$\,erg\\
Total energy &$E_{\rm total}$ & 4\,$\times$\,10$^{51}$\,erg&& 1.4\,$\times$\,10$^{52}$\,erg\\
Deceleration radius &$R_{\rm dec}$ & 10$^{16}$\,cm && 10$^{15}$\,cm\\
Baryon load &$M_{\rm fb}$ & 2\,$\times$\,10$^{-5}$\,$M_{\odot}$&& 1\,$\times$\,10$^{-4}$\,$M_{\odot}$\\
\noalign{\smallskip}
            \hline
\end{tabular}
\end{table*}

\section{Discussion}
\label{sec:disc}
\subsection{GRB\,120711A as a member of the long GRB population}
The large number of GRBs triggered by BATSE provide one of the best samples to study the distribution of observed $E_{\rm peak}$ for long GRBs \citep{Kaneko:2006p243}. Despite possible observational biases, the BATSE $E_{\rm peak}$ distribution peaked around 200\,--\,400\,keV, with a high-energy tail extending up to $\sim$\,3\,MeV. Similar results have recently been published using \emph{Fermi}/GBM data \citep{2012arXiv1201.2981G}. In these distributions GRB\,120711A is one of the hardest GRBs observed to date. GRB\,120711A is the hardest GRB triggered by \emph{INTEGRAL} \citep{{Foley:2008p553},{Vianello:2009p722}}. In the rest frame, GRB\,120711A belongs to the top 1\% of the hardest GRBs with known redshift and also to the top 1\% of the brightest GRBs in terms of $E_{\rm \gamma, iso}$ \citep[e.g.][]{2012MNRAS.420..483G}. It should be noted that some of the most powerful GRBs including GRB\,120711A are also detected by \emph{Fermi}/LAT \citep{{2010A&A...516A..71M},{2011ApJ...738..138R},{collaboration:2013tp}}.

The emission from GRB\,120711A is well modelled using a density profile of $k$\,$\sim$\,1.2, which corresponds to an intermediate case between the ISM and wind environments. A stratified environment seems to be the preferred model for most \emph{Fermi}/LAT GRBs \citep{2011ApJ...732...29C}. This is in contrast to the vast majority of GRBs that are consistent with a homogeneous environment \citep{2011A&A...526A..23S}. A similar intermediate environment has been found in the powerful GRB\,130427A \citep{2013arXiv1307.4401P}. Recent analysis of the \emph{Swift} GRB sample also indicates that the wind environment seems to be preferred over a homogeneous density medium for highly energetic bursts \citep{2013EAS....61..217D}. 

The analysis of several GRBs with known redshifts \citep{2012MNRAS.420..483G} indicates that the initial bulk Lorentz factor seems to depend strongly on the environment, with an average value of $\gamma_{0}$ of 138 and 66 for the ISM and wind environments. \cite{2012MNRAS.420..483G} also found a trend towards higher $\gamma_{0}$ in \emph{Fermi}/LAT GRBs with the average values increasing to $\sim$\,299 in the wind model. The results presented here indicate that GRB\,120711A falls in this category with high $\gamma_{0}$. The result found for GRB\,120711A is also consistent with the $\gamma_{0}$ distribution presented in the first \emph{Fermi}/LAT catalogue \citep{collaboration:2013tp}.

The typical jet half-opening angle of GRBs is $\sim$\,5$^{\circ}$ with some values as high as 24$^{\circ}$ \citep{2009ApJ...698...43R}, which is consistent with the values of 2$^{\circ}$ and 5$^{\circ}$ derived for GRB\,120711A. In a large sample of \emph{Swift} GRBs, \cite{2009ApJ...698...43R} reported an average time of the jet break of 1\,--\,2\,days post trigger, with a paucity of GRBs with jet breaks at times $>$\,10\,days. The lack of GRBs with breaks at late times is probably a bias caused by the lack of observations at times $>$\,2\,Ms. Currently, only 5\% of GRBs monitored by \emph{Swift} have observations at these times. In fact, in GRB\,120711A, the candidate jet break found at $\sim$\,10\,days was detected using late observations performed by \emph{XMM-Newton} and \emph{Chandra}.

GRB\,120711A satisfies the so-called Ghirlanda correlation \citep{2007A&A...466..127G} with an expected radiative energy, $E_{\gamma}$ consistent with our measurements. Interestingly, the $E_{\rm peak}$--$E_{\gamma,\rm iso}$--$t_{\rm jet}$ correlation \citep{{2005ApJ...633..611L},{2007A&A...466..127G}} predicts a jet break at much earlier times ($\sim$\,5\,days after trigger in the observer's frame).

The product of the jet half-opening angle and the initial bulk Lorentz factor, $\theta_{\rm jet}\gamma_{0}$, provides insight into the jet geometry. \cite{2012MNRAS.420..483G} found two distributions of $\theta_{\rm jet}\gamma_{0}$ depending on the environment, with a wind environment resulting in a smaller product ($\sim$\,$\times$\,6) than the ISM environment ($\sim$\,$\times$\,20). For GRB\,120711A, the product $\theta_{\rm jet}\gamma_{0}$ is $\sim$\,11 regardless of $\eta_{\gamma}$. This value falls in the tail of the $\theta_{\rm jet}\gamma_{0}$ distribution in the wind environment reported by \cite{2012MNRAS.420..483G}. However, the $\theta_{\rm jet}\gamma_{0}$ product obtained for GRB\,120711A is not as high as the values reported for the hyper-energetic population \citep{2011ApJ...732...29C}.

The isotropic energy radiated in the afterglow emission is $E_{\rm s-\gamma,iso}$\,=\,3.7\,$\times\,10^{52}$\,erg in the 20\,--\,200\,keV band (rest frame) during the IBIS data in segment III, and $E_{\rm X,iso}$\,=\,1.51\,$\times\,10^{51}$\,erg in the 0.3\,--\,30\,keV band (rest frame) during the entire \emph{Swift} campaign. Therefore, the energy in the afterglow of GRB\,120711A during the long-lived soft $\gamma$-ray emission is $\sim$\,2\,\% of the total bolometric energy radiated during the prompt phase and $\sim$\,0.1\,\% during the X-ray afterglow. GRB\,120711A is consistent with the $E_{\rm X,iso}$--$E_{\gamma,\rm iso}$--$E_{\rm peak}$ correlation \citep{{2012MNRAS.425.1199B},{2013MNRAS.428..729M}} when the IBIS data are extrapolated to the 0.3\,--\,30\,keV band (rest frame) and taking into account the absorption at soft X-rays. Thus, the observed X-ray isotropic energy becomes $E_{\rm X,iso}$\,=\,5.4\,$\times\,10^{52}$\,erg ($\sim$\,3\% of the total bolometric energy) consistent with the expected value of $E_{\rm X,iso, th}$\,$\sim$\,6\,$\times\,10^{52}$\,erg.

The optical absorption at the source rest-frame obtained for GRB\,120711A is one of the highest observed when compared with the large GRB samples presented by \cite{2010ApJ...720.1513K} and \cite{Greiner:2011p1952}. The absorption value of $A_{V}$\,$\sim$\,0.85 is caused by dust extinction in the host galaxy at $z$\,$\sim$\,1.4, which is close to the expected peak of the star formation rate \citep{2013MNRAS.432.1231C}. Following the method of \cite{Kann2006}, it can be shown that the magnitude shift, $dRc$, between what is observed and how bright the afterglow would be if the GRB had occurred at $z$=1 and with no dust extinction is $dRc$\,=-2.70\,$\pm$\,0.13\,mag for GRB\,120711A. This implies that the optical flash would have peaked at an unobscured value of $\approx$\,9th magnitude. This is fainter than the peak magnitudes of some other GRBs with reverse-shock flashes such as GRB\,990123, GRB\,050904, and GRB\,080319B, but much brighter than GRB\,060729, GRB\,061121, and GRB\,070802 \citep{2010ApJ...720.1513K}. Although the optical/NIR absorption is high, the intrinsic soft X-ray absorption from the host galaxy is not particularly large when compared with a large GRB sample \citep[e.g.][]{2007ApJ...660L.101W}.

\subsection{Long-lasting soft $\gamma$-ray emission in GRBs}
An upper limit of 2.7\,$\times$\,10$^{-5}$\,erg\,cm$^{-2}$, during the first 5400\,s post-GRB, was obtained with BATSE in the 20\,--\,300\,keV energy band for the burst GRB\,940217 with long-lasting GeV emission \citep{Hurley:1994p2071}. A previous study by \cite{2009essu.confE..48T} placed upper limits on bright \emph{INTEGRAL} GRBs of $\sim$\,5\,$\times$\,10$^{-6}$\,erg\,cm$^{-2}$ in the 20\,--\,400\,keV energy band at 1\,hour after the burst. For GRB\,041219A, the brightest GRB detected by \emph{INTEGRAL}, \cite{McBreen:2006p2017} reported a 3$\sigma$ upper limit of $\sim$\,10$^{-5}$\,erg\,cm$^{-2}$ in the 20\,--\,200\,keV range between $\sim$\,1\,--\,2\,ks after the burst using SPI data. In a re-analysis of the IBIS data from GRB\,041219A, we improved this upper limit to $\sim$\,10$^{-6}$\,erg\,cm$^{-2}$ during the same time interval. For GRB\,120711A, the detected fluence, measured in segment III for 1.2\,ks, is $\sim$\,10$^{-5}$\,erg\,cm$^{-2}$ in the 20\,--\,200\,keV energy band. Therefore, GRB\,041219A did not have long-lasting soft $\gamma$-ray emission at a comparable level to GRB\,120711A. There were no observations of GRB\,041219A above 100\,MeV. 

Since the launch of \emph{Fermi}, only GRB\,080723B was observed with similar brightness by \emph{INTEGRAL} and \emph{Fermi} simultaneously. No LAT emission was detected from this GRB and no soft $\gamma$-ray late emission was observed in the IBIS data with a 3$\sigma$ upper limit of $\sim$\,10$^{-6}$\,erg\,cm$^{-2}$ in the 20\,--\,200\,keV range between $\sim$\,1\,--\,2\,ks after the burst. The lower initial bulk Lorentz factor of $\sim$\,200, predicted for ISM environments with the $\gamma_{0}$--$E_{\rm \gamma, iso}$ correlation \citep{2010ApJ...725.2209L}, seems consistent with the concept that a high value of $\gamma_{0}$ is required for both LAT detection and long-lived soft $\gamma$-ray afterglow emission.

\begin{figure}[ht!]
\centering
\resizebox{\hsize}{!}{\includegraphics[angle=0,width=12cm]{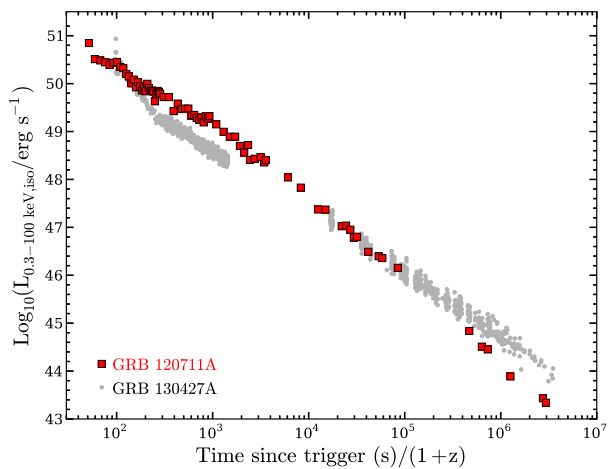}}
\caption{Isotropic luminosity light curves for the afterglow emission of GRB\,120711A and GRB\,130427A in the rest-frame energy band from 0.3\,--\,100\,keV. The light curve for GRB\,120711A consists of data from \emph{INTEGRAL}, \emph{Swift}, and \emph{XMM-Newton}. All data points for GRB\,130427A were measured with the \emph{Swift}/XRT instrument using data obtained from \texttt{http://www.swift.ac.uk}.}
\label{lccomp}
\end{figure}

GRB\,120711A and GRB\,130427A \citep[e.g.][]{{2013ApJ...776..119L},{2013arXiv1307.4401P},{Preece03012014},{Vestrand03012014},{Ackermann03012014}} are two powerful bursts with $E_{\gamma, \rm iso}$\,$\sim$\,10$^{54}$\,erg in the 1\,keV\,--\,10\,MeV energy range (about one order of magnitude higher than that estimated for GRB\,041219A \citep{{McBreen:2006p2017},{2011MNRAS.413.2173G}}) and long-lived soft $\gamma$-ray afterglow emission ($\sim$\,10\,ks for GRB\,120711A and $\sim$\,1\,ks for GRB\,130427A). The soft $\gamma$-ray detection from GRB\,130427A is most likely limited by the \emph{Swift}/BAT sensitivity considering that the LAT emission was detectable for longer than a day. In both GRBs the LAT emission seems to be consistent with synchrotron emission \citep{2013ApJ...779L...1K}. Fig.~\ref{lccomp} shows the k-corrected isotropic luminosity light curves of GRB\,120711A and GRB\,130427A in the rest-frame energy band from 0.3\,--\,100\,keV. Both afterglows are among the most luminous detected to date and show similar afterglow luminosities, with GRB\,120711A being $\sim$\,5 times brighter than GRB\,130427A at 1\,ks. At T$_{0}$\,+\,1\,ks GRB\,120711A still had significant emission above 60\,keV (Fig.~\ref{ibislonglived}) while most of the emission from GRB\,130427A was detected in the 0.3\,--\,10\,keV band. At later times, both GRBs become comparable, with GRB\,120711A steepening at $\sim$\,0.9\,Ms because of a jet break, which is not evident in the light curve of GRB\,130427A (Fig.~\ref{lccomp}). However, a jet break has been claimed for GRB\,130427A at T$_{0}$\,+37\,ks corresponding to a jet half-opening angle of $\sim$\,3.4$^{\circ}$ \citep{Maselli:2013te}, which is compatible with the jet half-opening angle found for GRB\,120711A.

\section{Conclusions}
\label{sec:conclusions}
GRB\,120711A triggered by \emph{INTEGRAL} was observed as a bright (32\,ph\,cm$^{-2}$\,s$^{-1}$ in the 20\,keV\,--\,8\,MeV band) and hard ($E_{\rm peak}$\,$\sim$\,1\,MeV) GRB with a duration of $\sim$\,115\,s and showed unprecedented long-lasting soft $\gamma$-ray emission above 60\,keV until 3\,ks post-trigger and above 20\,keV for at least 10\,ks. The burst was also detected by \emph{Fermi}/LAT in observations that started 300\,s post-trigger. A powerful optical flash was observed peaking at an R magnitude of $\sim$\,11.5 at $\sim$\,T$_{0}$+126\,s that is consistent with a reverse shock with 
contribution from the masked forward shock peak. GRB\,120711A was found to be highly absorbed with $A_{V}$\,$\sim$\,0.85 in the GRB rest-frame. The X-ray afterglow was monitored using \emph{Swift}, \emph{XMM-Newton}, and \emph{Chandra} with observations spanning the range from $\sim$\,8\,ks to $\sim$\,7\,Ms. Spectroscopic observations of the optical afterglow located the GRB at a redshift of 1.405, yielding an isotropic energy of $\sim$\,10$^{54}$\,erg in the rest-frame energy band from 1\,keV to 10\,MeV.

The extended $\gamma$-ray emission is explained by forward shocks, except for the first $\sim$\,300\,s when late activity from the central engine seems to be present. Therefore, GRB\,120711A does not appear to be part of the newly proposed ultra-long GRB population \citep[e.g.][]{2013arXiv1302.2352L}. This hypothesis is also supported by the resemblance of the light curve seen by \emph{Swift}/XRT in the softer energy band of 0.3\,--\,10\,keV and the consistence of the estimated $E_{\rm X,iso}$ from the IBIS data to the $E_{\rm X,iso}$--$E_{\gamma,\rm iso}$--$E_{\rm peak}$ correlation.

The combined optical/NIR, X-ray and $\gamma$-ray afterglow light curve was adequately modelled as synchrotron emission in an intermediate ISM-wind environment with a density profile of $k$\,$\sim$\,1.2. At $\sim$\,0.9\,Ms after the trigger, the afterglow light curve shows evidence of a jet break corresponding to a jet half-opening angle of 2$^{\circ}$\,--\,5$^{\circ}$.

The detection of GRB\,120711A by \emph{Fermi}/LAT is consistent with the observed trend that LAT GRBs have high Lorentz factors and stratified wind-like environments. In particular, GRB\,120711A has a Lorentz factor between 120\,--\,340, with a mass-loss rate between $\sim$\,10$^{-5}$\,--\,10$^{-6}$\,$M_{\odot}$\,yr$^{-1}$ depending on the radiation efficiency range considered. The baryon load is $\sim$\,10$^{-5}$\,--\,10$^{-6}$\,$M_{\odot}$, consistent with that expected in the fireball model when the emission is highly relativistic.

At this time GRB\,120711A is the only GRB with soft $\gamma$-ray emission ($>$\,20\,keV) detected up to 10\,ks after the trigger. The recently detected GRB\,130427A also has long-lasting emission above 20\,keV, although only for $\sim$\,1\,ks. The lack of long-lasting soft $\gamma$-ray emission from \emph{INTEGRAL} GRBs with higher peak fluxes than GRB\,120711A such as GRB\,041219A indicates that brightness is not the primary factor in the presence of such long-lived soft $\gamma$-ray emission. Both GRB\,120711A and GRB\,130427A have high isotropic energies and high Lorentz factors. GRB\,080723B is the only other bright GRB detected simultaneously with \emph{INTEGRAL}--\emph{Fermi}/LAT, and it does not have the properties associated with GRB\,120711A and GRB\,130427A. The existence of LAT emission and high Lorentz factors seems to be a requirement to produce soft $\gamma$-ray afterglow emission. Additionally, the fact that GRB\,120711A and GRB\,130427A had reverse shocks also suggests that these properties can be important in the production of harder afterglow emission.

\begin{acknowledgements}
We would like to thank the referee for helpful comments and suggestions. We would also like to thank N. Tanvir (Leicester) for his useful comments on the redshift measurement. This research is based on observations with \emph{INTEGRAL}, an ESA project with instruments and science data centre funded by ESA member states (especially the PI countries: Denmark, France, Germany, Italy, Switzerland, Spain), Poland and with the participation of Russia and the USA. The \emph{XMM-Newton} project is an ESA Science Mission with instruments and contributions directly funded by ESA Member States and the USA (NASA). This work made use of data supplied by the UK \emph{Swift} Science Data Centre at the University of Leicester. This research has made use of the XRT Data Analysis Software (XRTDAS) developed under the responsibility of the ASI Science Data Centre (ASDC), Italy; and of \emph{Fermi} data, a NASA project with international collaboration from France, Japan, Italy, and Sweden. The scientific results reported in this article are partly based on data obtained from the Chandra Data Archive. We would like to thank N. Schartel and the anonymous TAC member who carefully studied our ToO proposal to follow this GRB. AMC and LH acknowledge support from Science Foundation Ireland under grant 09/RFP/AST/2400 and LH and CW from 11/RFP/AST/3188. LH and MT acknowledge support from the EU FP7 under grant number 283783. TK acknowledges support by the European Commission under the Marie Curie Intra-European Fellowship Programme. The Dark Cosmology Centre is funded by the Danish National Research Foundation. Part of the funding for GROND (both hardware and personnel) was generously granted from the Leibniz-Prize to G. Hasinger (DFG grant HA 1850/28-1). AR acknowledges support by the Thueringer Landessternwarte Tautenburg. PS acknowledges support through the Sofja Kovalevskaja Award from 
the Alexander von Humboldt Foundation of Germany. SK acknowledge support by DFG grant Kl 766/16-1. DAK acknowledges support by the DFG cluster of excellence 'Origin and Structure of the Universe'. We acknowledge the excellent support of \emph{INTEGRAL} project scientist Chris Winkler.
\end{acknowledgements}
\bibliographystyle{aa}

\begin{thebibliography}{}
\bibitem[{Ackermann {et~al.}(2014)Ackermann, Ajello, Asano, Atwood, Axelsson,
  Baldini, Ballet, Barbiellini, Baring, Bastieri, Bechtol, Bellazzini,
  Bissaldi, Bonamente, Bregeon, Brigida, Bruel, Buehler, Burgess, Buson,
  Caliandro, Cameron, Caraveo, Cecchi, Chaplin, Charles, Chekhtman, Cheung,
  Chiang, Chiaro, Ciprini, Claus, Cleveland, Cohen-Tanugi, Collazzi, Cominsky,
  Connaughton, Conrad, Cutini, D’Ammando, de~Angelis, DeKlotz, de~Palma,
  Dermer, Desiante, Diekmann, Di~Venere, Drell, Drlica-Wagner, Favuzzi, Fegan,
  Ferrara, Finke, Fitzpatrick, Focke, Franckowiak, Fukazawa, Funk, \&
  Fusco}]{Ackermann03012014}
Ackermann, M., Ajello, M., Asano, K., {et~al.} 2014, 343, 42
\bibitem[{{Ackermann} {et~al.}(2013){Ackermann}, {Ajello}, {Asano}, {Axelsson},
  {Baldini}, {Ballet}, {Barbiellini}, {Bastieri}, {Bechtol}, {Bellazzini},
  {Bhat}, {Bissaldi}, {Bloom}, {Bonamente}, {Bonnell}, {Bouvier}, {Brandt},
  {Bregeon}, {Brigida}, {Bruel}, {Buehler}, {Burgess}, {Buson}, {Byrne},
  {Caliandro}, {Cameron}, {Caraveo}, {Cecchi}, {Charles}, {Chaves},
  {Chekhtman}, {Chiang}, {Chiaro}, {Ciprini}, {Claus}, {Cohen-Tanugi},
  {Connaughton}, {Conrad}, {Cutini}, {D'Ammando}, {de Angelis}, {de Palma},
  {Dermer}, {Desiante}, {Digel}, {Dingus}, {Di Venere}, {Drell},
  {Drlica-Wagner}, {Dubois}, {Favuzzi}, {Ferrara}, {Fitzpatrick}, {Foley},
  {Franckowiak}, {Fukazawa}, {Fusco}, {Gargano}, {Gasparrini}, {Gehrels},
  {Germani}, {Giglietto}, {Giommi}, {Giordano}, {Giroletti}, {Glanzman},
  {Godfrey}, {Goldstein}, {Granot}, {Grenier}, {Grove}, {Gruber}, {Guiriec},
  {Hadasch}, {Hanabata}, {Hayashida}, {Horan}, {Hou}, {Hughes}, {Inoue},
  {Jackson}, {Jogler}, {J{\'o}hannesson}, {Johnson}, {Johnson}, {Kamae},
  {Kataoka}, {Kawano}, {Kippen}, {Kn{\"o}dlseder}, {Kocevski}, {Kouveliotou},
  {Kuss}, {Lande}, {Larsson}, {Latronico}, {Lee}, {Longo}, {Loparco},
  {Lovellette}, {Lubrano}, {Massaro}, {Mayer}, {Mazziotta}, {McBreen},
  {McEnery}, {McGlynn}, {Michelson}, {Mizuno}, {Moiseev}, {Monte}, {Monzani},
  {Moretti}, {Morselli}, {Murgia}, {Nemmen}, {Nuss}, {Nymark}, {Ohno},
  {Ohsugi}, {Omodei}, {Orienti}, {Orlando}, {Paciesas}, {Paneque}, {Panetta},
  {Pelassa}, {Perkins}, {Pesce-Rollins}, {Piron}, {Pivato}, {Porter}, {Preece},
  {Racusin}, {Rain{\`o}}, {Rando}, {Rau}, {Razzano}, {Razzaque}, {Reimer},
  {Reimer}, {Reposeur}, {Ritz}, {Romoli}, {Roth}, {Ryde}, {Saz Parkinson},
  {Schalk}, {Sgr{\`o}}, {Siskind}, {Sonbas}, {Spandre}, {Spinelli}, {Suson},
  {Tajima}, {Takahashi}, {Takeuchi}, {Tanaka}, {Thayer}, {Thayer}, {Thompson},
  {Tibaldo}, {Tierney}, {Tinivella}, {Torres}, {Tosti}, {Troja}, {Tronconi},
  {Usher}, {Vandenbroucke}, {van der Horst}, {Vasileiou}, {Vianello}, {Vitale},
  {von Kienlin}, {Winer}, {Wood}, {Wood}, {Xiong}, \&
  {Yang}}]{collaboration:2013tp}
{Ackermann}, M., {Ajello}, M., {Asano}, K., {et~al.} 2013, \apjs, 209, 11
\bibitem[{{Aihara} {et~al.}(2011){Aihara}, {Allende Prieto}, {An}, {Anderson},
  {Aubourg}, {Balbinot}, {Beers}, {Berlind}, {Bickerton}, {Bizyaev}, {Blanton},
  {Bochanski}, {Bolton}, {Bovy}, {Brandt}, {Brinkmann}, {Brown}, {Brownstein},
  {Busca}, {Campbell}, {Carr}, {Chen}, {Chiappini}, {Comparat}, {Connolly},
  {Cortes}, {Croft}, {Cuesta}, {da Costa}, {Davenport}, {Dawson}, {Dhital},
  {Ealet}, {Ebelke}, {Edmondson}, {Eisenstein}, {Escoffier}, {Esposito},
  {Evans}, {Fan}, {Femen{\'{\i}}a Castell{\'a}}, {Font-Ribera}, {Frinchaboy},
  {Ge}, {Gillespie}, {Gilmore}, {Gonz{\'a}lez Hern{\'a}ndez}, {Gott}, {Gould},
  {Grebel}, {Gunn}, {Hamilton}, {Harding}, {Harris}, {Hawley}, {Hearty}, {Ho},
  {Hogg}, {Holtzman}, {Honscheid}, {Inada}, {Ivans}, {Jiang}, {Johnson},
  {Jordan}, {Jordan}, {Kazin}, {Kirkby}, {Klaene}, {Knapp}, {Kneib},
  {Kochanek}, {Koesterke}, {Kollmeier}, {Kron}, {Lampeitl}, {Lang}, {Le Goff},
  {Lee}, {Lin}, {Long}, {Loomis}, {Lucatello}, {Lundgren}, {Lupton}, {Ma},
  {MacDonald}, {Mahadevan}, {Maia}, {Makler}, {Malanushenko}, {Malanushenko},
  {Mandelbaum}, {Maraston}, {Margala}, {Masters}, {McBride}, {McGehee},
  {McGreer}, {M{\'e}nard}, {Miralda-Escud{\'e}}, {Morrison}, {Mullally},
  {Muna}, {Munn}, {Murayama}, {Myers}, {Naugle}, {Neto}, {Nguyen}, {Nichol},
  {O'Connell}, {Ogando}, {Olmstead}, {Oravetz}, {Padmanabhan},
  {Palanque-Delabrouille}, {Pan}, {Pandey}, {P{\^a}ris}, {Percival},
  {Petitjean}, {Pfaffenberger}, {Pforr}, {Phleps}, {Pichon}, {Pieri}, {Prada},
  {Price-Whelan}, {Raddick}, {Ramos}, {Reyl{\'e}}, {Rich}, {Richards}, {Rix},
  {Robin}, {Rocha-Pinto}, {Rockosi}, {Roe}, {Rollinde}, {Ross}, {Ross},
  {Rossetto}, {S{\'a}nchez}, {Sayres}, {Schlegel}, {Schlesinger}, {Schmidt},
  {Schneider}, {Sheldon}, {Shu}, {Simmerer}, {Simmons}, {Sivarani}, {Snedden},
  {Sobeck}, {Steinmetz}, {Strauss}, {Szalay}, {Tanaka}, {Thakar}, {Thomas},
  {Tinker}, {Tofflemire}, {Tojeiro}, {Tremonti}, {Vandenberg}, {Vargas
  Maga{\~n}a}, {Verde}, {Vogt}, {Wake}, {Wang}, {Weaver}, {Weinberg}, {White},
  {White}, {Yanny}, {Yasuda}, {Yeche}, \& {Zehavi}}]{Aihara2011ApJS}
{Aihara}, H., {Allende Prieto}, C., {An}, D., {et~al.} 2011, \apjs, 193, 29
\bibitem[{Atwood {et~al.}(2009)Atwood, Abdo, Ackermann, Althouse, Anderson,
  Axelsson, Baldini, Ballet, Band, Barbiellini, Bartelt, Bastieri, Baughman,
  Bechtol, B{\'e}d{\'e}r{\`e}de, Bellardi, Bellazzini, Berenji, Bignami,
  Bisello, Bissaldi, Blandford, Bloom, Bogart, Bonamente, Bonnell, Borgland,
  Bouvier, Bregeon, Brez, Brigida, Bruel, Burnett, Busetto, Caliandro, Cameron,
  Caraveo, Carius, Carlson, Casandjian, Cavazzuti, Ceccanti, Cecchi, Charles,
  Chekhtman, Cheung, Chiang, Chipaux, Cillis, Ciprini, Claus, Cohen-Tanugi,
  Condamoor, Conrad, Corbet, Corucci, Costamante, Cutini, Davis, Decotigny,
  DeKlotz, Dermer, De~Angelis, Digel, Do~Couto E~Silva, Drell, Dubois, Dumora,
  Edmonds, Fabiani, Farnier, Favuzzi, Flath, Fleury, Focke, Funk, Fusco,
  Gargano, Gasparrini, Gehrels, Gentit, Germani, Giebels, Giglietto, Giommi,
  Giordano, Glanzman, Godfrey, Grenier, Grondin, Grove, Guillemot, Guiriec,
  Haller, Harding, Hart, Hays, Healey, Hirayama, Hjalmarsdotter, Horn, Hughes,
  Johannesson, Johansson, Johnson, Johnson, Johnson, Johnson, Kamae, Katagiri,
  Kataoka, Kavelaars, Kawai, Kelly, Kerr, Klamra, Knodlseder, Kocian, Komin,
  Kuehn, Kuss, Landriu, Latronico, Lee, Lee, Lemoine-Goumard, Lionetto, Longo,
  Loparco, Lott, Lovellette, Lubrano, Madejski, Makeev, Marangelli, Massai,
  Mazziotta, Mcenery, Menon, Meurer, Michelson, Minuti, Mirizzi, Mitthumsiri,
  Mizuno, Moiseev, Monte, Monzani, Moretti, Morselli, Moskalenko, Murgia,
  Nakamori, Nishino, Nolan, Norris, Nuss, Ohno, Ohsugi, Omodei, Orlando, Ormes,
  Paccagnella, Paneque, Panetta, Parent, Pearce, Pepe, Perazzo, Pesce-Rollins,
  Picozza, Pieri, Pinchera, Piron, Porter, Poupard, Raino, Rando, Rapposelli,
  Razzano, Reimer, Reimer, Reposeur, Reyes, Ritz, Rochester, Rodriguez, Romani,
  Roth, Russell, Ryde, Sabatini, Sadrozinski, Sanchez, Sander, Sapozhnikov,
  Parkinson, Scargle, Schalk, Scolieri, Sgro, Share, Shaw, Shimokawabe,
  Shrader, Sierpowska-Bartosik, Siskind, Smith, Smith, Spandre, Spinelli,
  Starck, Stephens, Strickman, Strong, Suson, Tajima, Takahashi, Takahashi,
  Tanaka, Tenze, Tether, Thayer, Thayer, Thompson, Tibaldo, Tibolla, Torres,
  Tosti, Tramacere, Turri, Usher, Vilchez, Vitale, Wang, Watters, Winer, Wood,
  Ylinen, \& Ziegler}]{2009ApJ...697.1071A}
Atwood, W.~B., Abdo, A.~A., Ackermann, M., {et~al.} 2009, ApJ, 697, 1071
\bibitem[{Bernardini {et~al.}(2011)Bernardini, Margutti, Chincarini, Guidorzi,
  \& Mao}]{2011A&A...526A..27B}
Bernardini, M.~G., Margutti, R., Chincarini, G., Guidorzi, C., \& Mao, J. 2011,
  A\&A, 526, 27
\bibitem[{Bernardini {et~al.}(2012)Bernardini, Margutti, Zaninoni, \&
  Chincarini}]{2012MNRAS.425.1199B}
Bernardini, M.~G., Margutti, R., Zaninoni, E., \& Chincarini, G. 2012, MNRAS,
  425, 1199
\bibitem[{Bozzo {et~al.}(2012)Bozzo, G{\"o}tz, Mereghetti, Ferrigno, Gibaud, \&
  Borkowski}]{2012GCN..13435...1B}
Bozzo, E., G{\"o}tz, D., Mereghetti, S., {et~al.} 2012, GCN 13435
\bibitem[{Burenin {et~al.}(1999{\natexlab{a}})Burenin, Vikhlinin, Gilfanov,
  Terekhov, Tkachenko, Sazonov, Churazov, Sunyaev, Goldoni, Claret, Goldwurm,
  Paul, Roques, Jourdain, Pelaez, \& Vedrenne}]{1999A&A...344L..53B}
Burenin, R.~A., Vikhlinin, A.~A., Gilfanov, M.~R., {et~al.} 1999{\natexlab{a}},
  A\&A, 344, L53
\bibitem[{Burenin {et~al.}(1999{\natexlab{b}})Burenin, Vikhlinin, Terekhov,
  Tkachenko, Sazonov, Gilfanov, Churazov, Sunyaev, Goldoni, Claret, Goldwurm,
  Paul, Roques, Jourdain, Vedrenne, \& Mandrou}]{1999A&AS..138..443B}
Burenin, R.~A., Vikhlinin, A.~A., Terekhov, O.~V., {et~al.} 1999{\natexlab{b}},
  A\&AS, 138, 443
\bibitem[{Burrows(2005)}]{Burrows:2005p2248}
Burrows, D.~N. 2005, Science, 309, 1833
\bibitem[{Burrows {et~al.}(2005)Burrows, Hill, Nousek, Kennea, Wells, Osborne,
  Abbey, Beardmore, Mukerjee, Short, Chincarini, Campana, Citterio, Moretti,
  Pagani, Tagliaferri, Giommi, Capalbi, Tamburelli, Angelini, Cusumano,
  Br{\"a}uninger, Burkert, \& Hartner}]{Burrows:2005p1620}
Burrows, D.~N., Hill, J.~E., Nousek, J.~A., {et~al.} 2005, SSR, 120, 165
\bibitem[{Cenko {et~al.}(2011)Cenko, Frail, Harrison, Haislip, Reichart,
  Butler, Cobb, Cucchiara, Berger, Bloom, Chandra, Fox, Perley, Prochaska,
  Filippenko, Glazebrook, Ivarsen, Kasliwal, Kulkarni, LaCluyze, Lopez, Morgan,
  Pettini, \& Rana}]{2011ApJ...732...29C}
Cenko, S.~B., Frail, D.~A., Harrison, F.~A., {et~al.} 2011, ApJ, 732, 29
\bibitem[{Chevalier \& Li(1999)}]{Chevalier:1999p2185}
Chevalier, R.~A. \& Li, Z.-Y. 1999, ApJ, 520, L29
\bibitem[{Chevalier \& Li(2000)}]{2000ApJ...536..195C}
Chevalier, R.~A. \& Li, Z.-Y. 2000, ApJ, 536, 195
\bibitem[{Chincarini {et~al.}(2010)Chincarini, Mao, Margutti, Bernardini,
  Guidorzi, Pasotti, Giannios, Della~Valle, Moretti, Romano, D'Avanzo,
  Cusumano, \& Giommi}]{2010MNRAS.406.2113C}
Chincarini, G., Mao, J., Margutti, R., {et~al.} 2010, MNRAS, 406, 2113
\bibitem[{Connaughton(2002)}]{2002ApJ...567.1028C}
Connaughton, V. 2002, ApJ, 567, 1028
\bibitem[{{Covino} {et~al.}(2013){Covino}, {Melandri}, {Salvaterra}, {Campana},
  {Vergani}, {Bernardini}, {D'Avanzo}, {D'Elia}, {Fugazza}, {Ghirlanda},
  {Ghisellini}, {Gomboc}, {Jin}, {Kr{\"u}hler}, {Malesani}, {Nava},
  {Sbarufatti}, \& {Tagliaferri}}]{2013MNRAS.432.1231C}
{Covino}, S., {Melandri}, A., {Salvaterra}, R., {et~al.} 2013, \mnras, 432,
  1231
\bibitem[{{Covino} {et~al.}(2004){Covino}, {Stefanon}, {Sciuto},
  {Fernandez-Soto}, {Tosti}, {Zerbi}, {Chincarini}, {Antonelli}, {Conconi},
  {Cutispoto}, {Molinari}, {Nicastro}, \& {Rodono}}]{2004SPIE.5492.1613C}
{Covino}, S., {Stefanon}, M., {Sciuto}, G., {et~al.} 2004, in Society of
  Photo-Optical Instrumentation Engineers (SPIE) Conference Series, Vol. 5492,
  Ground-based Instrumentation for Astronomy, ed. A.~F.~M. {Moorwood} \&
  M.~{Iye}, 1613--1622
\bibitem[{De~Angelis(2001)}]{DeAngelis:2001p1637}
De~Angelis, A. 2001, New worlds in astroparticle physics, 140
\bibitem[{De~Pasquale {et~al.}(2013)De~Pasquale, Schulze, Kann, Oates, \&
  Zhang}]{2013EAS....61..217D}
De~Pasquale, M., Schulze, S., Kann, D.~A., Oates, S., \& Zhang, B. 2013, EAS
  Publications Series, 61, 217
\bibitem[{{Elliott} {et~al.}(2012){Elliott}, {Klose}, \&
  {Greiner}}]{Elliott2012GCN13438}
{Elliott}, J., {Klose}, S., \& {Greiner}, J. 2012, GCN, 13438
\bibitem[{Evans {et~al.}(2009)Evans, Beardmore, Page, Osborne, O'Brien,
  Willingale, Starling, Burrows, Godet, Vetere, Racusin, Goad, Wiersema,
  Angelini, Capalbi, Chincarini, Gehrels, Kennea, Margutti, Morris, Mountford,
  Pagani, Perri, Romano, \& Tanvir}]{Evans:2009p4}
Evans, P.~A., Beardmore, A.~P., Page, K.~L., {et~al.} 2009, MNRAS, 397, 1177
\bibitem[{Evans {et~al.}(2007)Evans, Beardmore, Page, Tyler, Osborne, Goad,
  O'Brien, Vetere, Racusin, Morris, Burrows, Capalbi, Perri, Gehrels, \&
  Romano}]{Evans:2007p208}
Evans, P.~A., Beardmore, A.~P., Page, K.~L., {et~al.} 2007, A\&A, 469, 379
\bibitem[{{Evans} {et~al.}(2014){Evans}, {Willingale}, {Osborne}, {O'Brien},
  {Tanvir}, {Frederiks}, {Pal'shin}, {Svinkin}, {Lien}, {Cummings}, {Xiong},
  {Zhang}, {G{\"o}tz}, {Savchenko}, {Negoro}, {Nakahira}, {Suzuki}, {Wiersema},
  {Starling}, {Castro-Tirado}, {Beardmore}, {S{\'a}nchez-Ram{\'{\i}}rez},
  {Gorosabel}, {Jeong}, {Kennea}, {Burrows}, \&
  {Gehrels}}]{2014arXiv1403.4079E}
{Evans}, P.~A., {Willingale}, R., {Osborne}, J.~P., {et~al.} 2014, astro-ph:
  1403.4079
\bibitem[{Fenimore \& Sumner(1997)}]{1997asxo.proc..167F}
Fenimore, E. \& Sumner, M.~C. 1997, "All-Sky X-Ray Observations in the Next
  Decade", 167
\bibitem[{{Ferrero} {et~al.}(2010){Ferrero}, {Hanlon}, {Felletti}, {French},
  {Melady}, {McBreen}, {Kub{\'a}nek}, {Jel{\'{\i}}nek}, {McBreen}, {Meintjes},
  {Calitz}, \& {Hoffman}}]{2010AdAst2010E..36F}
{Ferrero}, A., {Hanlon}, L., {Felletti}, R., {et~al.} 2010, Advances in
  Astronomy, 2010
\bibitem[{Filgas {et~al.}(2011)Filgas, Greiner, Schady, Kruhler, Updike, Klose,
  Nardini, Kann, Rossi, Sudilovsky, Afonso, Clemens, Elliott,
  Nicuesa~Guelbenzu, Olivares~E, \& Rau}]{2011A&A...535A..57F}
Filgas, R., Greiner, J., Schady, P., {et~al.} 2011, A\&A, 535, 57
\bibitem[{Fitzpatrick(2012)}]{2012grb..confE..24F}
Fitzpatrick, G. 2012, in Proceedings of the Gamma-ray Bursts 2012 Conference,
  Munich (Germany), Proceedings of Science, Vol. 152, 24
\bibitem[{Foley {et~al.}(2008)Foley, McGlynn, Hanlon, McBreen, \&
  McBreen}]{Foley:2008p553}
Foley, S., McGlynn, S., Hanlon, L., McBreen, S., \& McBreen, B. 2008, A\&A,
  484, 143
\bibitem[{Frail {et~al.}(2001)Frail, Kulkarni, Sari, Djorgovski, Bloom, Galama,
  Reichart, Berger, Harrison, Price, Yost, Diercks, Goodrich, \&
  Chaffee}]{Frail:2001p1785}
Frail, D.~A., Kulkarni, S.~R., Sari, R., {et~al.} 2001, ApJ, 562, L55
\bibitem[{{Frederiks} {et~al.}(2013){Frederiks}, {Hurley}, {Svinkin},
  {Pal'shin}, {Mangano}, {Oates}, {Aptekar}, {Golenetskii}, {Mazets},
  {Oleynik}, {Tsvetkova}, {Ulanov}, {Kokomov}, {Cline}, {Burrows}, {Krimm},
  {Pagani}, {Sbarufatti}, {Siegel}, {Mitrofanov}, {Golovin}, {Litvak}, {Sanin},
  {Boynton}, {Fellows}, {Harshman}, {Enos}, {Starr}, {von Kienlin}, {Rau},
  {Zhang}, \& {Goldstein}}]{2013ApJ...779..151F}
{Frederiks}, D.~D., {Hurley}, K., {Svinkin}, D.~S., {et~al.} 2013, \apj, 779,
  151
\bibitem[{French {et~al.}(2004)French, Hanlon, McBreen, McBreen, Moran, Smith,
  Giltinan, Meintjes, \& Hoffman}]{2004AIPC..727..741F}
French, J., Hanlon, L., McBreen, B., {et~al.} 2004, American Institute of
  Physics Conference Series, 727, 741
\bibitem[{Garmire {et~al.}(2003)Garmire, Bautz, Ford, Nousek, \&
  Ricker}]{Garmire:2003p1650}
Garmire, G.~P., Bautz, M.~W., Ford, P.~G., Nousek, J.~A., \& Ricker, G.~R.
  2003, X-Ray and Gamma-Ray Telescopes and Instruments for Astronomy. Edited by
  Joachim E. Truemper, 4851, 28
\bibitem[{Gehrels {et~al.}(2004)Gehrels, Chincarini, Giommi, Mason, Nousek,
  Wells, White, Barthelmy, Burrows, Cominsky, Hurley, Marshall,
  M{\'e}sz{\'a}ros, Roming, Angelini, Barbier, Belloni, Campana, Caraveo,
  Chester, Citterio, Cline, Cropper, Cummings, Dean, Feigelson, Fenimore,
  Frail, Fruchter, Garmire, Gendreau, Ghisellini, Greiner, Hill, Hunsberger,
  Krimm, Kulkarni, Kumar, Lebrun, Lloyd-Ronning, Markwardt, Mattson, Mushotzky,
  Norris, Osborne, Paczynski, Palmer, Park, Parsons, Paul, Rees, Reynolds,
  Rhoads, Sasseen, Schaefer, Short, Smale, Smith, Stella, Tagliaferri,
  Takahashi, Tashiro, Townsley, Tueller, Turner, Vietri, Voges, Ward,
  Willingale, Zerbi, \& Zhang}]{Gehrels:2004p1618}
Gehrels, N., Chincarini, G., Giommi, P., {et~al.} 2004, ApJ, 611, 1005
\bibitem[{{Gendre} {et~al.}(2013){Gendre}, {Stratta}, {Atteia}, {Basa},
  {Bo{\"e}r}, {Coward}, {Cutini}, {D'Elia}, {Howell}, {Klotz}, \&
  {Piro}}]{2012arXiv1212.2392G}
{Gendre}, B., {Stratta}, G., {Atteia}, J.~L., {et~al.} 2013, \apj, 766, 30
\bibitem[{Genet {et~al.}(2007)Genet, Daigne, \&
  Mochkovitch}]{2007MNRAS.381..732G}
Genet, F., Daigne, F., \& Mochkovitch, R. 2007, MNRAS, 381, 732
\bibitem[{Genet \& Granot(2009)}]{2009MNRAS.399.1328G}
Genet, F. \& Granot, J. 2009, MNRAS, 399, 1328
\bibitem[{Ghirlanda {et~al.}(2006)Ghirlanda, Ghisellini, Firmani, Nava,
  Tavecchio, \& Lazzati}]{2006A&A...452..839G}
Ghirlanda, G., Ghisellini, G., Firmani, C., {et~al.} 2006, A\&A, 452, 839
\bibitem[{Ghirlanda {et~al.}(2012)Ghirlanda, Nava, Ghisellini, Celotti, Burlon,
  Covino, \& Melandri}]{2012MNRAS.420..483G}
Ghirlanda, G., Nava, L., Ghisellini, G., {et~al.} 2012, MNRAS, 420, 483
\bibitem[{Ghirlanda {et~al.}(2007)Ghirlanda, Nava, Ghisellini, \&
  Firmani}]{2007A&A...466..127G}
Ghirlanda, G., Nava, L., Ghisellini, G., \& Firmani, C. 2007, A\&A, 466, 127
\bibitem[{Ghisellini {et~al.}(2010)Ghisellini, Ghirlanda, Nava, \&
  Celotti}]{2010MNRAS.403..926G}
Ghisellini, G., Ghirlanda, G., Nava, L., \& Celotti, A. 2010, MNRAS, 403, 926
\bibitem[{Giblin {et~al.}(2002)Giblin, Connaughton, van Paradijs, Preece,
  Briggs, Kouveliotou, Wijers, \& Fishman}]{2002ApJ...570..573G}
Giblin, T.~W., Connaughton, V., van Paradijs, J., {et~al.} 2002, ApJ, 570, 573
\bibitem[{{Giblin} {et~al.}(1999){Giblin}, {van Paradijs}, {Kouveliotou},
  {Connaughton}, {Wijers}, {Briggs}, {Preece}, \&
  {Fishman}}]{1999ApJ...524L..47G}
{Giblin}, T.~W., {van Paradijs}, J., {Kouveliotou}, C., {et~al.} 1999, \apjl,
  524, L47
\bibitem[{{Giuliani} \& {Mereghetti}(2014)}]{2013arXiv1307.8345G}
{Giuliani}, A. \& {Mereghetti}, S. 2014, \aap, 563, A6
\bibitem[{{Goldstein} {et~al.}(2012){Goldstein}, {Burgess}, {Preece}, {Briggs},
  {Guiriec}, {van der Horst}, {Connaughton}, {Wilson-Hodge}, {Paciesas},
  {Meegan}, {von Kienlin}, {Bhat}, {Bissaldi}, {Chaplin}, {Diehl}, {Fishman},
  {Fitzpatrick}, {Foley}, {Gibby}, {Giles}, {Greiner}, {Gruber}, {Kippen},
  {Kouveliotou}, {McBreen}, {McGlynn}, {Rau}, \&
  {Tierney}}]{2012arXiv1201.2981G}
{Goldstein}, A., {Burgess}, J.~M., {Preece}, R.~D., {et~al.} 2012, \apjs, 199,
  19
\bibitem[{G{\"o}tz {et~al.}(2011)G{\"o}tz, Covino, Hasco{\"e}t,
  Fern{\'a}ndez-Soto, Daigne, Mochkovitch, \& Esposito}]{2011MNRAS.413.2173G}
G{\"o}tz, D., Covino, S., Hasco{\"e}t, R., {et~al.} 2011, MNRAS, 413, 2173
\bibitem[{G{\"o}tz {et~al.}(2012)G{\"o}tz, Mereghetti, Bozzo, Ferrigno, Gibaud,
  \& Borkowski}]{2012GCN..13434...1G}
G{\"o}tz, D., Mereghetti, S., Bozzo, E., {et~al.} 2012, GCN 13434
\bibitem[{Granot \& Sari(2002)}]{2002ApJ...568..820G}
Granot, J. \& Sari, R. 2002, ApJ, 820
\bibitem[{Grebenev \& Chelovekov(2007)}]{Grebenev:2007p2363}
Grebenev, S.~A. \& Chelovekov, I.~V. 2007, AL, 33, 789
\bibitem[{{Greiner} {et~al.}(2008){Greiner}, {Bornemann}, {Clemens}, {Deuter},
  {Hasinger}, {Honsberg}, {Huber}, {Huber}, {Krauss}, {Kr{\"u}hler},
  {K{\"u}pc{\"u} Yolda{\c s}}, {Mayer-Hasselwander}, {Mican}, {Primak},
  {Schrey}, {Steiner}, {Szokoly}, {Th{\"o}ne}, {Yolda{\c s}}, {Klose}, {Laux},
  \& {Winkler}}]{Greiner2008PASP}
{Greiner}, J., {Bornemann}, W., {Clemens}, C., {et~al.} 2008, \pasp, 120, 405
\bibitem[{Greiner {et~al.}(2011)Greiner, Kr{\"u}hler, Klose, Afonso, Clemens,
  Filgas, Hartmann, K{\"u}pc{\"u}~Yolda{\c s}, Nardini, Olivares~E, Rau, Rossi,
  Schady, \& Updike}]{Greiner:2011p1952}
Greiner, J., Kr{\"u}hler, T., Klose, S., {et~al.} 2011, A\&A, 526, 30
\bibitem[{{Grupe} {et~al.}(2010){Grupe}, {Burrows}, {Wu}, {Wang}, {Zhang},
  {Liang}, {Garmire}, {Nousek}, {Gehrels}, {Ricker}, \&
  {Bautz}}]{2010ApJ...711.1008G}
{Grupe}, D., {Burrows}, D.~N., {Wu}, X.-F., {et~al.} 2010, \apj, 711, 1008
\bibitem[{{Grupe} {et~al.}(2007){Grupe}, {Gronwall}, {Wang}, {Roming},
  {Cummings}, {Zhang}, {M{\'e}sz{\'a}ros}, {Trigo}, {O'Brien}, {Page},
  {Beardmore}, {Godet}, {vanden Berk}, {Brown}, {Koch}, {Morris}, {Stroh},
  {Burrows}, {Nousek}, {McMath Chester}, {Immler}, {Mangano}, {Romano},
  {Chincarini}, {Osborne}, {Sakamoto}, \& {Gehrels}}]{2007ApJ...662..443G}
{Grupe}, D., {Gronwall}, C., {Wang}, X.-Y., {et~al.} 2007, \apj, 662, 443
\bibitem[{{Hancock} {et~al.}(2012){Hancock}, {Murphy}, {Gaensler}, {Bell},
  {Burlon}, \& {de Ugarte Postigo}}]{2012GCN..13485...1H}
{Hancock}, P., {Murphy}, T., {Gaensler}, B., {et~al.} 2012, GCN, 13485, 1
\bibitem[{Hanlon {et~al.}(2012)Hanlon, Martin-Carrillo, Zhang, \& von
  Kienlin}]{2012GCN..13468...1H}
Hanlon, L., Martin-Carrillo, A., Zhang, X.~L., \& von Kienlin, A. 2012, GCN
  13468
\bibitem[{{Henden} {et~al.}(2011){Henden}, {Levine}, {Terrell}, {Smith}, \&
  {Welch}}]{2011AAS...21812601H}
{Henden}, A.~A., {Levine}, S.~E., {Terrell}, D., {Smith}, T.~C., \& {Welch},
  D.~L. 2011, in American Astronomical Society Meeting Abstracts \#218, 126.01
\bibitem[{Hurley {et~al.}(1994)Hurley, Dingus, Mukherjee, Sreekumar,
  Kouveliotou, Meegan, Fishman, Band, Ford, Bertsch, Cline, Fichtel, Hartman,
  Hunter, Thompson, Kanbach, Mayer-Hasselwander, von Montigny, Sommer, Lin,
  Nolan, Michelson, Kniffen, Mattox, Schneid, Boer, \& Niel}]{Hurley:1994p2071}
Hurley, K., Dingus, B.~L., Mukherjee, R., {et~al.} 1994, Nature, 372, 652
\bibitem[{Jansen {et~al.}(2001)Jansen, Lumb, Altieri, Clavel, Ehle, Erd,
  Gabriel, Guainazzi, Gondoin, Much, Munoz, Santos, Schartel, Texier, \&
  Vacanti}]{Jansen:2001p1572}
Jansen, F., Lumb, D., Altieri, B., {et~al.} 2001, A\&A, 365, L1
\bibitem[{Jin \& Fan(2007)}]{2007MNRAS.378.1043J}
Jin, Z.~P. \& Fan, Y.~Z. 2007, MNRAS, 378, 1043
\bibitem[{Kalberla {et~al.}(2005)Kalberla, Burton, Hartmann, Arnal, Bajaja,
  Morras, \& P{\"o}ppel}]{Kalberla:2005p1538}
Kalberla, P. M.~W., Burton, W.~B., Hartmann, D., {et~al.} 2005, A\&A, 440, 775
\bibitem[{Kaneko {et~al.}(2006)Kaneko, Preece, Briggs, Paciesas, Meegan, \&
  Band}]{Kaneko:2006p243}
Kaneko, Y., Preece, R.~D., Briggs, M.~S., {et~al.} 2006, ApJSS, 166, 298
\bibitem[{{Kann} {et~al.}(2006){Kann}, {Klose}, \& {Zeh}}]{Kann2006}
{Kann}, D.~A., {Klose}, S., \& {Zeh}, A. 2006, \apj, 641, 993
\bibitem[{{Kann} {et~al.}(2010){Kann}, {Klose}, {Zhang}, {Malesani}, {Nakar},
  {Pozanenko}, {Wilson}, {Butler}, {Jakobsson}, {Schulze}, {Andreev},
  {Antonelli}, {Bikmaev}, {Biryukov}, {B{\"o}ttcher}, {Burenin}, {Castro
  Cer{\'o}n}, {Castro-Tirado}, {Chincarini}, {Cobb}, {Covino}, {D'Avanzo},
  {D'Elia}, {Della Valle}, {de Ugarte Postigo}, {Efimov}, {Ferrero}, {Fugazza},
  {Fynbo}, {G{\aa}lfalk}, {Grundahl}, {Gorosabel}, {Gupta}, {Guziy}, {Hafizov},
  {Hjorth}, {Holhjem}, {Ibrahimov}, {Im}, {Israel}, {Je{\'l}inek}, {Jensen},
  {Karimov}, {Khamitov}, {Kizilo{\v g}lu}, {Klunko}, {Kub{\'a}nek}, {Kutyrev},
  {Laursen}, {Levan}, {Mannucci}, {Martin}, {Mescheryakov}, {Mirabal},
  {Norris}, {Ovaldsen}, {Paraficz}, {Pavlenko}, {Piranomonte}, {Rossi},
  {Rumyantsev}, {Salinas}, {Sergeev}, {Sharapov}, {Sollerman}, {Stecklum},
  {Stella}, {Tagliaferri}, {Tanvir}, {Telting}, {Testa}, {Updike}, {Volnova},
  {Watson}, {Wiersema}, \& {Xu}}]{2010ApJ...720.1513K}
{Kann}, D.~A., {Klose}, S., {Zhang}, B., {et~al.} 2010, \apj, 720, 1513
\bibitem[{Kobayashi \& Zhang(2003)}]{2003ApJ...597..455K}
Kobayashi, S. \& Zhang, B. 2003, ApJ, 597, 455
\bibitem[{Kocevski {et~al.}(2012)Kocevski, Vianello, Omodei, \&
  Digel}]{2012GCN..13452...1K}
Kocevski, D., Vianello, G., Omodei, N., \& Digel, S. 2012, GCN 13452
\bibitem[{Kouveliotou {et~al.}(2013)Kouveliotou, Granot, Racusin, Bellm,
  Vianello, Oates, Fryer, Boggs, Christensen, Craig, Dermer, Gehrels, Hailey,
  Harrison, Melandri, Mcenery, Mundell, Stern, Tagliaferri, \&
  Zhang}]{2013ApJ...779L...1K}
Kouveliotou, C., Granot, J., Racusin, J.~L., {et~al.} 2013, ApJL, 779, L1
\bibitem[{{Kr{\"u}hler} {et~al.}(2008){Kr{\"u}hler}, {K{\"u}pc{\"u} Yolda{\c
  s}}, {Greiner}, {Clemens}, {McBreen}, {Primak}, {Savaglio}, {Yolda{\c s}},
  {Szokoly}, \& {Klose}}]{Kruehler2008ApJ}
{Kr{\"u}hler}, T., {K{\"u}pc{\"u} Yolda{\c s}}, A., {Greiner}, J., {et~al.}
  2008, \apj, 685, 376
\bibitem[{Kumar \& Barniol~Duran(2009)}]{2009MNRAS.400L..75K}
Kumar, P. \& Barniol~Duran, R. 2009, MNRASL, 400, L75
\bibitem[{Kumar \& Panaitescu(2000{\natexlab{a}})}]{2000ApJ...541L..51K}
Kumar, P. \& Panaitescu, A. 2000{\natexlab{a}}, ApJ, 541, L51
\bibitem[{Kumar \& Panaitescu(2000{\natexlab{b}})}]{Kumar:2000p2152}
Kumar, P. \& Panaitescu, A. 2000{\natexlab{b}}, ApJ, 541, L9
\bibitem[{{K\"upc\"u Yolda{\c s}} {et~al.}(2008){K\"upc\"u Yolda{\c s}},
  {Kr{\"u}hler}, {Greiner}, {Yolda{\c s}}, {Clemens}, {Szokoly}, {Primak}, \&
  {Klose}}]{Yoldas2008AIPC}
{K\"upc\"u Yolda{\c s}}, A., {Kr{\"u}hler}, T., {Greiner}, J., {et~al.} 2008,
  in American Institute of Physics Conference Series, Vol. 1000, American
  Institute of Physics Conference Series, ed. {M.~Galassi, D.~Palmer, \&
  E.~Fenimore}, 227--231
\bibitem[{LaCluyz\'e {et~al.}(2012)LaCluyz\'e, Haislip, Ivarsen, Reichart,
  Moore, Cromartie, Egger, Foster, Frank, Nysewander, Oza, Speckhard, Trotter,
  \& Crain}]{2012GCN..13430...1L}
LaCluyz\'e, A., Haislip, J., Ivarsen, K., {et~al.} 2012, GCN 13430
\bibitem[{Langer(1989)}]{1989A&A...220..135L}
Langer, N. 1989, A\&A, 220, 135
\bibitem[{{Laskar} {et~al.}(2013){Laskar}, {Berger}, {Zauderer}, {Margutti},
  {Soderberg}, {Chakraborti}, {Lunnan}, {Chornock}, {Chandra}, \&
  {Ray}}]{2013ApJ...776..119L}
{Laskar}, T., {Berger}, E., {Zauderer}, B.~A., {et~al.} 2013, \apj, 776, 119
\bibitem[{Lebrun {et~al.}(2003)Lebrun, Leray, Lavocat, Cr{\'e}tolle,
  Arqu{\`e}s, Blondel, Bonnin, Bou{\`e}re, Cara, Chaleil, Daly, Desages,
  Dzitko, Horeau, Laurent, Limousin, Mathy, Mauguen, Meignier, Molini{\'e},
  Poindron, Rouger, Sauvageon, \& Tourrette}]{Lebrun:2003p249}
Lebrun, F., Leray, J.~P., Lavocat, P., {et~al.} 2003, A\&A, 411, L141
\bibitem[{{Levan} {et~al.}(2014){Levan}, {Tanvir}, {Starling}, {Wiersema},
  {Page}, {Perley}, {Schulze}, {Wynn}, {Chornock}, {Hjorth}, {Cenko},
  {Fruchter}, {O'Brien}, {Brown}, {Tunnicliffe}, {Malesani}, {Jakobsson},
  {Watson}, {Berger}, {Bersier}, {Cobb}, {Covino}, {Cucchiara}, {de Ugarte
  Postigo}, {Fox}, {Gal-Yam}, {Goldoni}, {Gorosabel}, {Kaper}, {Kr{\"u}hler},
  {Karjalainen}, {Osborne}, {Pian}, {S{\'a}nchez-Ram{\'{\i}}rez}, {Schmidt},
  {Skillen}, {Tagliaferri}, {Th{\"o}ne}, {Vaduvescu}, {Wijers}, \&
  {Zauderer}}]{2013arXiv1302.2352L}
{Levan}, A.~J., {Tanvir}, N.~R., {Starling}, R.~L.~C., {et~al.} 2014, \apj,
  781, 13
\bibitem[{Leventis {et~al.}(2014)Leventis, Wijers, \& van~der
  Horst}]{Leventis:2014kx}
Leventis, K., Wijers, R. A. M.~J., \& van~der Horst, A.~J. 2014, MNRAS, 437,
  2448
\bibitem[{{Liang} \& {Zhang}(2005)}]{2005ApJ...633..611L}
{Liang}, E. \& {Zhang}, B. 2005, \apj, 633, 611
\bibitem[{Liang {et~al.}(2010)Liang, Yi, Zhang, L{\"u}, Zhang, \&
  Zhang}]{2010ApJ...725.2209L}
Liang, E.-W., Yi, S.-X., Zhang, J., {et~al.} 2010, ApJ, 725, 2209
\bibitem[{Lund {et~al.}(2003)Lund, Budtz-J{\o}rgensen, Westergaard, Brandt,
  Rasmussen, Hornstrup, Oxborrow, Chenevez, Jensen, Laursen, Andersen,
  Mogensen, Rasmussen, Om{\o}, Pedersen, Polny, Andersson, Andersson,
  K{\"a}m{\"a}r{\"a}inen, Vilhu, Huovelin, Maisala, Morawski, Juchnikowski,
  Costa, Feroci, Rubini, Rapisarda, Morelli, Carassiti, Frontera, Pelliciari,
  Loffredo, Mart{\'\i}nez~N{\'u}{\~n}ez, Reglero, Velasco, Larsson, Svensson,
  Zdziarski, Castro-Tirado, Attina, Goria, Giulianelli, Cordero, Rezazad,
  Schmidt, Carli, Gomez, Jensen, Sarri, Tiemon, Orr, Much, Kretschmar, \&
  Schnopper}]{Lund:2003p250}
Lund, N., Budtz-J{\o}rgensen, C., Westergaard, N.~J., {et~al.} 2003, A\&A, 411,
  L231
\bibitem[{Maiorano {et~al.}(2005)Maiorano, Masetti, Palazzi, Frontera, Grandi,
  Pian, Amati, Nicastro, Soffitta, Guidorzi, Landi, Montanari, Orlandini,
  Corsi, Piro, Antonelli, Costa, Feroci, Heise, Kuulkers, \& in't
  Zand}]{2005A&A...438..821M}
Maiorano, E., Masetti, N., Palazzi, E., {et~al.} 2005, A\&A, 438, 821
\bibitem[{Margutti {et~al.}(2011)Margutti, Chincarini, Granot, Guidorzi,
  Berger, Bernardini, Gehrels, Soderberg, Stamatikos, \&
  Zaninoni}]{2011MNRAS.417.2144M}
Margutti, R., Chincarini, G., Granot, J., {et~al.} 2011, MNRAS, astro-ph.HE,
  2144
\bibitem[{Margutti {et~al.}(2010)Margutti, Guidorzi, Chincarini, Bernardini,
  Genet, Mao, \& Pasotti}]{2010MNRAS.406.2149M}
Margutti, R., Guidorzi, C., Chincarini, G., {et~al.} 2010, MNRAS, 406, 2149
\bibitem[{Margutti {et~al.}(2013)Margutti, Zaninoni, Bernardini, Chincarini,
  Pasotti, Guidorzi, Angelini, Burrows, Capalbi, Evans, Gehrels, Kennea,
  Mangano, Moretti, Nousek, Osborne, Page, Perri, Racusin, Romano, Sbarufatti,
  Stafford, \& Stamatikos}]{2013MNRAS.428..729M}
Margutti, R., Zaninoni, E., Bernardini, M.~G., {et~al.} 2013, MNRAS, 428, 729
\bibitem[{Martin-Carrillo \& Hanlon(2013)}]{2013arXiv1302.0560M}
Martin-Carrillo, A. \& Hanlon, L. 2013, Proceedings of the 9th INTEGRAL
  Workshop, Paris (France), Proceedings of Science, Vol 176, 121
\bibitem[{{Maselli} {et~al.}(2014){Maselli}, {Melandri}, {Nava}, {Mundell},
  {Kawai}, {Campana}, {Covino}, {Cummings}, {Cusumano}, {Evans}, {Ghirlanda},
  {Ghisellini}, {Guidorzi}, {Kobayashi}, {Kuin}, {La Parola}, {Mangano},
  {Oates}, {Sakamoto}, {Serino}, {Virgili}, {Zhang}, {Barthelmy}, {Beardmore},
  {Bernardini}, {Bersier}, {Burrows}, {Calderone}, {Capalbi}, {Chiang},
  {D'Avanzo}, {D'Elia}, {De Pasquale}, {Fugazza}, {Gehrels}, {Gomboc},
  {Harrison}, {Hanayama}, {Japelj}, {Kennea}, {Kopac}, {Kouveliotou}, {Kuroda},
  {Levan}, {Malesani}, {Marshall}, {Nousek}, {O'Brien}, {Osborne}, {Pagani},
  {Page}, {Page}, {Perri}, {Pritchard}, {Romano}, {Saito}, {Sbarufatti},
  {Salvaterra}, {Steele}, {Tanvir}, {Vianello}, {Weigand}, {Wiersema}, {Yatsu},
  {Yoshii}, \& {Tagliaferri}}]{Maselli:2013te}
{Maselli}, A., {Melandri}, A., {Nava}, L., {et~al.} 2014, Science, 343, 48
\bibitem[{McBreen {et~al.}(2006)McBreen, Hanlon, McGlynn, McBreen, Foley,
  Preece, von Kienlin, \& Williams}]{McBreen:2006p2017}
McBreen, S., Hanlon, L., McGlynn, S., {et~al.} 2006, A\&A, 455, 433
\bibitem[{{McBreen} {et~al.}(2010){McBreen}, {Kr{\"u}hler}, {Rau}, {Greiner},
  {Kann}, {Savaglio}, {Afonso}, {Clemens}, {Filgas}, {Klose}, {K{\"u}pc{\"u}
  Yolda{\c s}}, {Olivares E.}, {Rossi}, {Szokoly}, {Updike}, \& {Yolda{\c
  s}}}]{2010A&A...516A..71M}
{McBreen}, S., {Kr{\"u}hler}, T., {Rau}, A., {et~al.} 2010, \aap, 516, A71
\bibitem[{M{\'e}sz{\'a}ros(2006)}]{Meszaros:2006p261}
M{\'e}sz{\'a}ros, P. 2006, Reports on Progress in Physics, 69, 2259
\bibitem[{Molinari {et~al.}(2007)Molinari, Vergani, Malesani, Covino, D'Avanzo,
  Chincarini, Zerbi, Antonelli, Conconi, Testa, Tosti, Vitali, D'Alessio,
  Malaspina, Nicastro, Palazzi, Guetta, Campana, Goldoni, Masetti, Meurs,
  Monfardini, Norci, Pian, Piranomonte, Rizzuto, Stefanon, Stella, Tagliaferri,
  Ward, Ihle, Gonzalez, Pizarro, Sinclaire, \&
  Valenzuela}]{2007A&A...469L..13M}
Molinari, E., Vergani, S.~D., Malesani, D., {et~al.} 2007, A\&A, 469, L13
\bibitem[{Nousek {et~al.}(2006)Nousek, Kouveliotou, Grupe, Page, Granot,
  Ramirez-Ruiz, Patel, Burrows, Mangano, Barthelmy, Beardmore, Campana,
  Capalbi, Chincarini, Cusumano, Falcone, Gehrels, Giommi, Goad, Godet,
  Hurkett, Kennea, Moretti, O'Brien, Osborne, Romano, Tagliaferri, \&
  Wells}]{Nousek:2006p277}
Nousek, J.~A., Kouveliotou, C., Grupe, D., {et~al.} 2006, ApJ, 642, 389
\bibitem[{Panaitescu \& Kumar(2002)}]{2002ApJ...571..779P}
Panaitescu, A. \& Kumar, P. 2002, ApJ, 571, 779
\bibitem[{Panaitescu {et~al.}(2006)Panaitescu, Meszaros, Burrows, Nousek,
  Gehrels, O'Brien, \& Willingale}]{2006MNRAS.369.2059P}
Panaitescu, A., Meszaros, P., Burrows, D., {et~al.} 2006, MNRAS, 369, 2059
\bibitem[{Peng {et~al.}(2013)Peng, Hu, Xi, Wang, Lu, Liang, \&
  Zhang}]{2013arXiv1302.4876P}
Peng, F.-K., Hu, Y.-D., Xi, S.-Q., {et~al.} 2013, astro-ph: 1302.4876
\bibitem[{{Perley} {et~al.}(2014){Perley}, {Cenko}, {Corsi}, {Tanvir}, {Levan},
  {Kann}, {Sonbas}, {Wiersema}, {Zheng}, {Zhao}, {Bai}, {Bremer},
  {Castro-Tirado}, {Chang}, {Clubb}, {Frail}, {Fruchter}, {G{\"o}{\u g}{\"u}{\c
  s}}, {Greiner}, {G{\"u}ver}, {Horesh}, {Filippenko}, {Klose}, {Mao},
  {Morgan}, {Pozanenko}, {Schmidl}, {Stecklum}, {Tanga}, {Volnova}, {Volvach},
  {Wang}, {Winters}, \& {Xin}}]{2013arXiv1307.4401P}
{Perley}, D.~A., {Cenko}, S.~B., {Corsi}, A., {et~al.} 2014, \apj, 781, 37
\bibitem[{Piran(2005)}]{Piran:2005p2117}
Piran, T. 2005, Reviews of Modern Physics, 76, 1
\bibitem[{Preece {et~al.}(2014)Preece, Burgess, von Kienlin, Bhat, Briggs,
  Byrne, Chaplin, Cleveland, Collazzi, Connaughton, Diekmann, Fitzpatrick,
  Foley, Gibby, Giles, Goldstein, Greiner, Gruber, Jenke, \&
  Kippen}]{Preece03012014}
Preece, R., Burgess, J.~M., von Kienlin, A., {et~al.} 2014, 343, 51
\bibitem[{Racusin {et~al.}(2009)Racusin, Liang, Burrows, Falcone, Sakamoto,
  Zhang, Zhang, Evans, \& Osborne}]{2009ApJ...698...43R}
Racusin, J.~L., Liang, E.~W., Burrows, D.~N., {et~al.} 2009, ApJ, 698, 43
\bibitem[{Racusin {et~al.}(2011)Racusin, Oates, Schady, Burrows, De~Pasquale,
  Donato, Gehrels, Koch, McEnery, Piran, Roming, Sakamoto, Swenson, Troja,
  Vasileiou, Virgili, Wanderman, \& Zhang}]{2011ApJ...738..138R}
Racusin, J.~L., Oates, S.~R., Schady, P., {et~al.} 2011, ApJ, 738, 138
\bibitem[{{Reichart} {et~al.}(2005){Reichart}, {Nysewander}, {Moran},
  {Bartelme}, {Bayliss}, {Foster}, {Clemens}, {Price}, {Evans}, {Salmonson},
  {Trammell}, {Carney}, {Keohane}, \& {Gotwals}}]{2005NCimC..28..767R}
{Reichart}, D., {Nysewander}, M., {Moran}, J., {et~al.} 2005, Nuovo Cimento C
  Geophysics Space Physics C, 28, 767
\bibitem[{Rossi {et~al.}(2011)Rossi, Schulze, Klose, Kann, Rau, Krimm,
  Johannesson, Panaitescu, Yuan, Ferrero, Kruhler, Greiner, Schady, Pandey,
  Amati, Afonso, Akerlof, Arnold, Clemens, Filgas, Hartmann,
  K{\"u}pc{\"u}~Yolda{\c s}, McBreen, McKay, Nicuesa~Guelbenzu, Olivares,
  Paciesas, Rykoff, Szokoly, Updike, \& Yoldas}]{2011A&A...529A.142R}
Rossi, A., Schulze, S., Klose, S., {et~al.} 2011, A\&A, 529, 142
\bibitem[{Sagi \& Nakar(2012)}]{2012ApJ...749...80S}
Sagi, E. \& Nakar, E. 2012, ApJ, 749, 80
\bibitem[{Sari \& Piran(1999)}]{Sari:1999p1259}
Sari, R. \& Piran, T. 1999, ApJ, 520, 641
\bibitem[{Sari {et~al.}(1998)Sari, Piran, \& Narayan}]{Sari:1998p1670}
Sari, R., Piran, T., \& Narayan, R. 1998, ApJL, 497, L17
\bibitem[{{Schlegel} {et~al.}(1998){Schlegel}, {Finkbeiner}, \&
  {Davis}}]{Schlegel1998}
{Schlegel}, D.~J., {Finkbeiner}, D.~P., \& {Davis}, M. 1998, \apj, 500, 525
\bibitem[{Schulze {et~al.}(2011)Schulze, Klose, Bj{\"o}rnsson, Jakobsson, Kann,
  Rossi, Kruhler, Greiner, \& Ferrero}]{2011A&A...526A..23S}
Schulze, S., Klose, S., Bj{\"o}rnsson, G., {et~al.} 2011, A\&A, 526, 23
\bibitem[{{Skrutskie} {et~al.}(2006){Skrutskie}, {Cutri}, {Stiening},
  {Weinberg}, {Schneider}, {Carpenter}, {Beichman}, {Capps}, {Chester},
  {Elias}, {Huchra}, {Liebert}, {Lonsdale}, {Monet}, {Price}, {Seitzer},
  {Jarrett}, {Kirkpatrick}, {Gizis}, {Howard}, {Evans}, {Fowler}, {Fullmer},
  {Hurt}, {Light}, {Kopan}, {Marsh}, {McCallon}, {Tam}, {Van Dyk}, \&
  {Wheelock}}]{Skrutskie2006AJ}
{Skrutskie}, M.~F., {Cutri}, R.~M., {Stiening}, R., {et~al.} 2006, \aj, 131,
  1163
\bibitem[{Starling {et~al.}(2008)Starling, van~der Horst, Rol, Wijers,
  Kouveliotou, Wiersema, Curran, \& Weltevrede}]{2008ApJ...672..433S}
Starling, R. L.~C., van~der Horst, A.~J., Rol, E., {et~al.} 2008, ApJ, 672, 433
\bibitem[{Str{\"u}der {et~al.}(2001)Str{\"u}der, Briel, Dennerl, Hartmann,
  Kendziorra, Meidinger, Pfeffermann, Reppin, Aschenbach, Bornemann,
  Br{\"a}uninger, Burkert, Elender, Freyberg, Haberl, Hartner, Heuschmann,
  Hippmann, Kastelic, Kemmer, Kettenring, Kink, Krause, M{\"u}ller, Oppitz,
  Pietsch, Popp, Predehl, Read, Stephan, St{\"o}tter, Tr{\"u}mper, Holl,
  Kemmer, Soltau, St{\"o}tter, Weber, Weichert, von Zanthier, Carathanassis,
  Lutz, Richter, Solc, B{\"o}ttcher, Kuster, Staubert, Abbey, Holland, Turner,
  Balasini, Bignami, La~Palombara, Villa, Buttler, Gianini, Lain{\'e}, Lumb, \&
  Dhez}]{Struder:2001p1573}
Str{\"u}der, L., Briel, U., Dennerl, K., {et~al.} 2001, A\&A, 365, L18
\bibitem[{Tam {et~al.}(2012)Tam, Li, \& Kong}]{2012GCN..13444...1T}
Tam, P. H.~T., Li, K.~L., \& Kong, A. K.~H. 2012, GCN 13444
\bibitem[{Tanvir {et~al.}(2012)Tanvir, Wiersema, Levan, Fox, Fruchter, \&
  Krogsrud}]{2012GCN..13441...1T}
Tanvir, N.~R., Wiersema, K., Levan, A.~J., {et~al.} 2012, GCN 13441
\bibitem[{Th{\"o}ne {et~al.}(2011)Th{\"o}ne, de~Ugarte~Postigo, Fryer, Page,
  Gorosabel, Aloy, Perley, Kouveliotou, Janka, Mimica, Racusin, Krimm,
  Cummings, Oates, Holland, Siegel, De~Pasquale, Sonbas, Im, Park, Kann, Guziy,
  Garc{\'\i}a, Llorente, Bundy, Choi, Jeong, Korhonen, Kubanek, Lim, Moskvitin,
  Munoz-Darias, Pak, \& Parrish}]{2011Natur.480...72T}
Th{\"o}ne, C.~C., de~Ugarte~Postigo, A., Fryer, C.~L., {et~al.} 2011, Nature,
  480, 72
\bibitem[{{Tody}(1993)}]{Tody1993ASPC}
{Tody}, D. 1993, ASP Conf. Ser., 52, 173
\bibitem[{Topinka {et~al.}(2009)Topinka, Meehan, Martin-Carrillo, Hanlon,
  McBreen, \& Foley}]{2009essu.confE..48T}
Topinka, M., Meehan, S., Martin-Carrillo, A., {et~al.} 2009, in Proceedings of
  The Extreme sky: Sampling the Universe above 10 keV, Otranto (Italy),
  Proceedings of Science, Vol.~96, 48
\bibitem[{Turner {et~al.}(2001)Turner, Abbey, Arnaud, Balasini, Barbera,
  Belsole, Bennie, Bernard, Bignami, Boer, Briel, Butler, Cara, Chabaud, Cole,
  Collura, Conte, Cros, Denby, Dhez, Di~Coco, Dowson, Ferrando, Ghizzardi,
  Gianotti, Goodall, Gretton, Griffiths, Hainaut, Hochedez, Holland, Jourdain,
  Kendziorra, Lagostina, Laine, La~Palombara, Lortholary, Lumb, Marty, Molendi,
  Pigot, Poindron, Pounds, Reeves, Reppin, Rothenflug, Salvetat, Sauvageot,
  Schmitt, Sembay, Short, Spragg, Stephen, Str{\"u}der, Tiengo, Trifoglio,
  Tr{\"u}mper, Vercellone, Vigroux, Villa, Ward, Whitehead, \&
  Zonca}]{Turner:2001p1574}
Turner, M. J.~L., Abbey, A., Arnaud, M., {et~al.} 2001, A\&A, 365, L27
\bibitem[{Uhm \& Beloborodov(2007)}]{2007ApJ...665L..93U}
Uhm, Z.~L. \& Beloborodov, A.~M. 2007, ApJ, 665, L93
\bibitem[{Vedrenne {et~al.}(2003)Vedrenne, Roques, Sch{\"o}nfelder, Mandrou,
  Lichti, von Kienlin, Cordier, Schanne, Kn{\"o}dlseder, Skinner, Jean,
  Sanchez, Caraveo, Teegarden, von Ballmoos, Bouchet, Paul, Matteson, Boggs,
  Wunderer, Leleux, Weidenspointner, Durouchoux, Diehl, Strong, Cass{\'e},
  Clair, \& Andr{\'e}}]{Vedrenne:2003p343}
Vedrenne, G., Roques, J.-P., Sch{\"o}nfelder, V., {et~al.} 2003, A\&A, 411, L63
\bibitem[{Vestrand {et~al.}(2014)Vestrand, Wren, Panaitescu, Wozniak, Davis,
  Palmer, Vianello, Omodei, Xiong, Briggs, Elphick, Paciesas, \&
  Rosing}]{Vestrand03012014}
Vestrand, W.~T., Wren, J.~A., Panaitescu, A., {et~al.} 2014, Science, 343, 38
\bibitem[{Vianello {et~al.}(2009)Vianello, G{\"o}tz, \&
  Mereghetti}]{Vianello:2009p722}
Vianello, G., G{\"o}tz, D., \& Mereghetti, S. 2009, A\&A, 495, 1005
\bibitem[{Virgili {et~al.}(2013)Virgili, Mundell, Pal'shin, Guidorzi, Margutti,
  Melandri, Harrison, Kobayashi, Chornock, Henden, Updike, Cenko, Tanvir,
  Steele, Cucchiara, Gomboc, Levan, Cano, Mottram, Clay, Bersier, Kopac,
  Japelj, Filippenko, Li, Svinkin, Golenetskii, Hartmann, Milne, Williams,
  O'Brien, Fox, \& Berger}]{2013ApJ...778...54V}
Virgili, F.~J., Mundell, C.~G., Pal'shin, V., {et~al.} 2013, ApJ, 778, 54
\bibitem[{Watson {et~al.}(2007)Watson, Hjorth, Fynbo, Jakobsson, Foley,
  Sollerman, \& Wijers}]{2007ApJ...660L.101W}
Watson, D., Hjorth, J., Fynbo, J. P.~U., {et~al.} 2007, ApJ, 660, L101
\bibitem[{{Weisskopf} {et~al.}(2002){Weisskopf}, {Brinkman}, {Canizares},
  {Garmire}, {Murray}, \& {Van Speybroeck}}]{2002PASP..114....1W}
{Weisskopf}, M.~C., {Brinkman}, B., {Canizares}, C., {et~al.} 2002, \pasp, 114,
  1
\bibitem[{Winkler {et~al.}(2003)Winkler, Courvoisier, Cocco, Gehrels,
  Gim{\'e}nez, Grebenev, Hermsen, Mas-Hesse, Lebrun, Lund, Palumbo, Paul,
  Roques, Schnopper, Sch{\"o}nfelder, Sunyaev, Teegarden, Ubertini, Vedrenne,
  \& Dean}]{Winkler:2003p1629}
Winkler, C., Courvoisier, T. J.-L., Cocco, G.~D., {et~al.} 2003, A\&A, 411, L1
\bibitem[{{Wu} {et~al.}(2013){Wu}, {Hou}, \& {Lei}}]{2013arXiv1302.4878W}
{Wu}, X.-F., {Hou}, S.-J., \& {Lei}, W.-H. 2013, \apjl, 767, L36
\bibitem[{{Zerbi} {et~al.}(2001){Zerbi}, {Chincarini}, {Ghisellini},
  {Rondon{\'o}}, {Tosti}, {Antonelli}, {Conconi}, {Covino}, {Cutispoto},
  {Molinari}, {Nicastro}, {Palazzi}, {Akerlof}, {Burderi}, {Campana}, {Crimi},
  {Danzinger}, {di Paola}, {Fernandez-Soto}, {Fiore}, {Frontera}, {Fugazza},
  {Gentile}, {Goldoni}, {Israel}, {Jordan}, {Lorenzetti}, {McBreen},
  {Martinetti}, {Mazzoleni}, {Masetti}, {Messina}, {Meurs}, {Monfardini},
  {Nucciarelli}, {Orlandini}, {Paul}, {Pian}, {Saracco}, {Sardone}, {Stella},
  {Tagliaferri}, {Tavani}, {Testa}, \& {Vitali}}]{2001AN....322..275Z}
{Zerbi}, R.~M., {Chincarini}, G., {Ghisellini}, G., {et~al.} 2001,
  Astronomische Nachrichten, 322, 275
\bibitem[{{Zhang}(2006)}]{2006cosp...36...77Z}
{Zhang}, B. 2006, in COSPAR Meeting, Vol.~36, 36th COSPAR Scientific Assembly,
  77
\bibitem[{Zhang(2007)}]{2007ChJAA...7....1Z}
Zhang, B. 2007, Chinese Journal of Astronomy and Astrophysics, 7, 1
\bibitem[{Zhang {et~al.}(2006)Zhang, Fan, Dyks, Kobayashi, M{\'e}sz{\'a}ros,
  Burrows, Nousek, \& Gehrels}]{Zhang:2006p369}
Zhang, B., Fan, Y.~Z., Dyks, J., {et~al.} 2006, ApJ, 642, 354
\bibitem[{Zhang {et~al.}(2003)Zhang, Kobayashi, \&
  M{\'e}sz{\'a}ros}]{Zhang:2003p365}
Zhang, B., Kobayashi, S., \& M{\'e}sz{\'a}ros, P. 2003, ApJ, 595, 950
\bibitem[{Zhang {et~al.}(2007)Zhang, Liang, Page, Grupe, Zhang, Barthelmy,
  Burrows, Campana, Chincarini, Gehrels, Kobayashi, M{\'e}sz{\'a}ros, Moretti,
  Nousek, O'Brien, Osborne, Roming, Sakamoto, Schady, \&
  Willingale}]{2007ApJ...655..989Z}
Zhang, B., Liang, E., Page, K.~L., {et~al.} 2007, ApJ, 989
\bibitem[{Zou {et~al.}(2005)Zou, Wu, \& Dai}]{2005MNRAS.363...93Z}
Zou, Y.~C., Wu, X.~F., \& Dai, Z.~G. 2005, MNRAS, 363, 93
\end{thebibliography}

\onecolumn
\Online
\begin{appendix}
\section{Logs of the optical/NIR observations of GRB\,120711A}

\longtab{
\begin{longtable}{ccrc}
\caption{\label{tab:obswatcher} \emph{Watcher} observations of the afterglow of GRB\,120711A.}\\
\hline
\hline                
Time since trigger [s] & mag$\pm$error & exp [s] & filter\\ 
\hline
96	&	14.47	$\pm$	0.14	&	30	&	$R$	\\
115	&	12.49	$\pm$	0.16	&	5	&	$R$	\\
120	&	12.23	$\pm$	0.15	&	5	&	$R$	\\
126	&	11.95	$\pm$	0.09	&	5	&	$R$	\\
131	&	11.98	$\pm$	0.14	&	5	&	$R$	\\
155	&	12.72	$\pm$	0.15	&	30	&	$R$	\\
173	&	13.26	$\pm$	0.12	&	5	&	$R$	\\
179	&	13.32	$\pm$	0.08	&	5	&	$R$	\\
184	&	13.49	$\pm$	0.12	&	5	&	$R$	\\
190	&	13.52	$\pm$	0.11	&	5	&	$R$	\\
261	&	14.35	$\pm$	0.13	&	5	&	$R$	\\
304	&	14.69	$\pm$	0.15	&	30	&	$R$	\\
362	&	15.08	$\pm$	0.31	&	30	&	$R$	\\
376	&	15.30	$\pm$	0.22	&	30	&	$R$	\\
752	&	16.03	$\pm$	0.47	&	460	&	$R$	\\
2035	&	$>$16.90	&	300	&	$R$	\\
\hline									
\hline									
\end{longtable}
}

\longtab{
\begin{longtable}{ccrc}
\caption{\label{tab:obsprompt} PROMPT observations of the afterglow of GRB\,120711A.}\\
\hline
\hline                
Time since trigger [s] & mag$\pm$error & exp [s] & filter\\ 
\hline
\endhead
\caption{\label{tab:obsprompt} PROMPT observations of the afterglow of GRB\,120711A.}\\
\hline
\hline                
Time since trigger [s] & mag$\pm$error & exp [s] & filter\\ 
\hline
\endhead 
\endfoot
67.392	&	$>$17.90		&	10	&	$B$	\\					
81.216	&	$>$17.96		&	5	&	$B$	\\					
93.312	&	16.30	$^{+	0.58	}	_{-	0.388	}$	&	5	&	$B$	\\
115.776	&	13.54	$^{+	0.02	}	_{-	0.014	}$	&	20	&	$B$	\\
139.968	&	13.87	$^{+	0.03	}	_{-	0.028	}$	&	10	&	$B$	\\
165.888	&	14.51	$^{+	0.03	}	_{-	0.027	}$	&	20	&	$B$	\\
194.4	&	15.09	$^{+	0.05	}	_{-	0.044	}$	&	20	&	$B$	\\
230.688	&	15.54	$^{+	0.04	}	_{-	0.037	}$	&	40	&	$B$	\\
279.936	&	16.07	$^{+	0.06	}	_{-	0.054	}$	&	40	&	$B$	\\
336.096	&	16.48	$^{+	0.08	}	_{-	0.078	}$	&	40	&	$B$	\\
385.344	&	16.57	$^{+	0.09	}	_{-	0.083	}$	&	40	&	$B$	\\
455.328	&	16.95	$^{+	0.07	}	_{-	0.07	}$	&	80	&	$B$	\\
550.368	&	17.51	$^{+	0.13	}	_{-	0.118	}$	&	80	&	$B$	\\
9816.768	&	20.91	$^{+	0.45	}	_{-	0.321	}$	&	80	&	$B$	\\
20867.328	&	21.17	$^{+	0.24	}	_{-	0.194	}$	&	80	&	$B$	\\
79668.576	&	$>$23.70		&	80	&	$B$	\\					
104302.08	&	22.55	$^{+	1.84	}	_{-	0.79	}$	&	80	&	$B$	\\
161438.4	&	$>$22.25	&	80	&	$B$	\\						
\hline
38.88	&	$>$18.66		&	5	&	$V$	\\					
51.84	&	17.66	$^{+	2.12	}	_{-	0.853	}$	&	5	&	$V$	\\
63.936	&	17.03	$^{+	0.68	}	_{-	0.434	}$	&	5	&	$V$	\\
78.624	&	17.54	$^{+	0.50	}	_{-	0.348	}$	&	10	&	$V$	\\
95.904	&	15.50	$^{+	0.07	}	_{-	0.067	}$	&	10	&	$V$	\\
113.184	&	13.16	$^{+	0.01	}	_{-	0.014	}$	&	10	&	$V$	\\
139.968	&	13.51	$^{+	0.01	}	_{-	0.011	}$	&	20	&	$V$	\\
166.752	&	14.14	$^{+	0.02	}	_{-	0.015	}$	&	20	&	$V$	\\
194.4	&	14.68	$^{+	0.02	}	_{-	0.022	}$	&	20	&	$V$	\\
234.144	&	15.11	$^{+	0.02	}	_{-	0.018	}$	&	40	&	$V$	\\
280.8	&	15.49	$^{+	0.03	}	_{-	0.024	}$	&	40	&	$V$	\\
335.232	&	15.92	$^{+	0.03	}	_{-	0.033	}$	&	40	&	$V$	\\
385.344	&	16.16	$^{+	0.04	}	_{-	0.039	}$	&	40	&	$V$	\\
455.328	&	16.55	$^{+	0.04	}	_{-	0.034	}$	&	80	&	$V$	\\
550.368	&	16.83	$^{+	0.04	}	_{-	0.042	}$	&	80	&	$V$	\\
7788.96	&	20.22	$^{+	0.26	}	_{-	0.212	}$	&	80	&	$V$	\\
10445.76	&	20.33	$^{+	0.27	}	_{-	0.22	}$	&	80	&	$V$	\\
13715.136	&	20.54	$^{+	0.25	}	_{-	0.20	}$	&	80	&	$V$	\\
18290.88	&	20.79	$^{+	0.24	}	_{-	0.20	}$	&	80	&	$V$	\\
23536.224	&	21.09	$^{+	0.30	}	_{-	0.23	}$	&	80	&	$V$	\\
27376.704	&	21.33	$^{+	0.55	}	_{-	0.38	}$	&	80	&	$V$	\\
74052.576	&	21.46	$^{+	0.74	}	_{-	0.46	}$	&	80	&	$V$	\\
\hline
68.256	&	16.35	$^{+	0.15	}	_{-	0.129	}$	&	10	&	$R$	\\
94.176	&	15.07	$^{+	0.04	}	_{-	0.043	}$	&	10	&	$R$	\\
112.32	&	12.36	$^{+	0.01	}	_{-	0.008	}$	&	10	&	$R$	\\
139.968	&	12.61	$^{+	0.01	}	_{-	0.006	}$	&	20	&	$R$	\\
167.616	&	13.32	$^{+	0.01	}	_{-	0.009	}$	&	20	&	$R$	\\
192.672	&	13.74	$^{+	0.02	}	_{-	0.018	}$	&	10	&	$R$	\\
229.824	&	14.22	$^{+	0.01	}	_{-	0.01	}$	&	40	&	$R$	\\
279.936	&	14.63	$^{+	0.01	}	_{-	0.013	}$	&	40	&	$R$	\\
335.232	&	15.07	$^{+	0.02	}	_{-	0.017	}$	&	40	&	$R$	\\
385.344	&	15.37	$^{+	0.02	}	_{-	0.022	}$	&	40	&	$R$	\\
455.328	&	15.69	$^{+	0.02	}	_{-	0.018	}$	&	80	&	$R$	\\
550.368	&	16.04	$^{+	0.02	}	_{-	0.023	}$	&	80	&	$R$	\\
7152.192	&	19.32	$^{+	0.15	}	_{-	0.128	}$	&	80	&	$R$	\\
8417.088	&	19.37	$^{+	0.16	}	_{-	0.14	}$	&	80	&	$R$	\\
9751.968	&	19.30	$^{+	0.15	}	_{-	0.13	}$	&	80	&	$R$	\\
11093.76	&	19.62	$^{+	0.19	}	_{-	0.161	}$	&	80	&	$R$	\\
12351.744	&	19.31	$^{+	0.13	}	_{-	0.11	}$	&	80	&	$R$	\\
14328.576	&	19.44	$^{+	0.10	}	_{-	0.09	}$	&	80	&	$R$	\\
17005.248	&	19.50	$^{+	0.09	}	_{-	0.09	}$	&	80	&	$R$	\\
19535.904	&	19.71	$^{+	0.12	}	_{-	0.11	}$	&	80	&	$R$	\\
22239.36	&	20.02	$^{+	0.16	}	_{-	0.14	}$	&	80	&	$R$	\\
24875.424	&	20.17	$^{+	0.14	}	_{-	0.12	}$	&	80	&	$R$	\\
27325.728	&	20.27	$^{+	0.16	}	_{-	0.14	}$	&	80	&	$R$	\\
79545.024	&	21.54	$^{+	0.42	}	_{-	0.31	}$	&	80	&	$R$	\\
103998.816	&	21.61	$^{+	0.42	}	_{-	0.31	}$	&	80	&	$R$	\\
164298.24	&	$>$22.58		&	80	&	$R$	\\					
186698.304	&	$>$22.77	&	80	&	$R$	\\		
\hline
7154.784	&	18.15	$^{+	0.09	}	_{-	0.085	}$	&	80	&	$I$	\\
8429.184	&	18.30	$^{+	0.10	}	_{-	0.091	}$	&	80	&	$I$	\\
9758.016	&	18.45	$^{+	0.11	}	_{-	0.099	}$	&	80	&	$I$	\\
11022.048	&	18.59	$^{+	0.11	}	_{-	0.101	}$	&	80	&	$I$	\\
12403.584	&	18.71	$^{+	0.11	}	_{-	0.103	}$	&	80	&	$I$	\\
14313.024	&	18.69	$^{+	0.08	}	_{-	0.072	}$	&	80	&	$I$	\\
16938.72	&	18.81	$^{+	0.07	}	_{-	0.068	}$	&	80	&	$I$	\\
19554.048	&	18.94	$^{+	0.10	}	_{-	0.089	}$	&	80	&	$I$	\\
22256.64	&	18.99	$^{+	0.10	}	_{-	0.09	}$	&	80	&	$I$	\\
24876.288	&	19.24	$^{+	0.11	}	_{-	0.096	}$	&	80	&	$I$	\\
27337.824	&	19.30	$^{+	0.11	}	_{-	0.102	}$	&	80	&	$I$	\\
79460.352	&	20.38	$^{+	0.22	}	_{-	0.183	}$	&	80	&	$I$	\\
103984.992	&	20.95	$^{+	0.35	}	_{-	0.269	}$	&	80	&	$I$	\\
164234.304	&	$>$22.48		&	80	&	$I$	\\					
185945.76	&	21.22	$^{+	1.65	}	_{-	0.751	}$	&	80	&	$I$	\\
\hline									
\hline									
\end{longtable}
}

\longtab{
\begin{longtable}{ccrcl}
\caption{\label{obslog} GROND observations of the afterglow of GRB\,120711A.}\\
\hline
\hline                
Time since trigger [ks] & mag$\pm$error & exp [s] & seeing (arcsec) & filter\\ 
\hline
\endfirsthead
\caption{\label{tab:obslog} GROND observations of the afterglow of GRB 120711A.}\\
\hline
\hline                
Time since trigger [ks] & mag$\pm$error & exp [s] & seeing (arcsec) & filter\\ 
\hline
\endhead 
\endfoot
21.16	& $	20.81	\pm	0.11	$ &	142	&	2.0	&	$g^\prime$	\\			
21.62	& $	20.85	\pm	0.14	$ &	142	&	2.1	&	$g^\prime$	\\			
21.94	& $	20.85	\pm	0.09	$ &	115	&	2.1	&	$g^\prime$	\\			
22.14	& $	20.86	\pm	0.10	$ &	115	&	2.3	&	$g^\prime$	\\			
22.34	& $	20.88	\pm	0.09	$ &	115	&	2.1	&	$g^\prime$	\\			
22.53	& $	20.91	\pm	0.09	$ &	115	&	2.0	&	$g^\prime$	\\			
22.86	& $	20.95	\pm	0.09	$ &	142	&	2.0	&	$g^\prime$	\\			
23.32	& $	20.95	\pm	0.13	$ &	142	&	2.2	&	$g^\prime$	\\			
23.78	& $	21.03	\pm	0.11	$ &	142	&	1.9	&	$g^\prime$	\\			
24.21	& $	21.01	\pm	0.09	$ &	142	&	2.0	&	$g^\prime$	\\			
24.64	& $	21.04	\pm	0.08	$ &	142	&	2.0	&	$g^\prime$	\\			
25.07	& $	21.17	\pm	0.09	$ &	142	&	2.1	&	$g^\prime$	\\			
25.50	& $	20.97	\pm	0.09	$ &	142	&	2.1	&	$g^\prime$	\\			
25.93	& $	20.99	\pm	0.08	$ &	142	&	1.8	&	$g^\prime$	\\			
26.36	& $	21.01	\pm	0.07	$ &	142	&	1.9	&	$g^\prime$	\\			
26.80	& $	20.99	\pm	0.07	$ &	142	&	1.6	&	$g^\prime$	\\			
27.25	& $	21.09	\pm	0.07	$ &	142	&	1.5	&	$g^\prime$	\\			
27.66	& $	21.17	\pm	0.09	$ &	142	&	1.6	&	$g^\prime$	\\			
28.08	& $	21.12	\pm	0.10	$ &	142	&	1.6	&	$g^\prime$	\\			
107.89	& $	23.00	\pm	0.41	$ &	919	&	3.3	&	$g^\prime$	\\			
109.58	& $	22.71	\pm	0.15	$ &	1501	&	2.4	&	$g^\prime$	\\			
111.42	& $	22.63	\pm	0.13	$ &	1501	&	2.7	&	$g^\prime$	\\			
113.25	& $	22.60	\pm	0.10	$ &	1501	&	2.4	&	$g^\prime$	\\			
114.64	& $	22.36	\pm	0.55	$ &	212	&	2.2	&	$g^\prime$	\\			
370.07	& $	23.96	\pm	0.17	$ &	6463	&	2.4	&	$g^\prime$	\\													
\hline											
21.16	& $	20.14	\pm	0.05	$ &	142	&	2.2	&	$r^\prime$	\\
21.62	& $	20.17	\pm	0.04	$ &	142	&	2.0	&	$r^\prime$	\\
21.94	& $	20.15	\pm	0.05	$ &	115	&	2.1	&	$r^\prime$	\\
22.14	& $	20.20	\pm	0.04	$ &	115	&	2.0	&	$r^\prime$	\\
22.34	& $	20.20	\pm	0.05	$ &	115	&	2.0	&	$r^\prime$	\\
22.53	& $	20.22	\pm	0.04	$ &	115	&	2.1	&	$r^\prime$	\\
22.86	& $	20.31	\pm	0.05	$ &	142	&	2.0	&	$r^\prime$	\\
23.32	& $	20.37	\pm	0.06	$ &	142	&	2.1	&	$r^\prime$	\\
23.78	& $	20.33	\pm	0.08	$ &	142	&	2.2	&	$r^\prime$	\\
24.21	& $	20.36	\pm	0.05	$ &	142	&	1.8	&	$r^\prime$	\\
24.64	& $	20.35	\pm	0.04	$ &	142	&	1.8	&	$r^\prime$	\\
25.07	& $	20.30	\pm	0.04	$ &	142	&	1.9	&	$r^\prime$	\\
25.50	& $	20.30	\pm	0.04	$ &	142	&	1.9	&	$r^\prime$	\\
25.93	& $	20.32	\pm	0.04	$ &	142	&	1.6	&	$r^\prime$	\\
26.36	& $	20.35	\pm	0.04	$ &	142	&	1.7	&	$r^\prime$	\\
26.80	& $	20.33	\pm	0.04	$ &	142	&	1.5	&	$r^\prime$	\\
27.25	& $	20.39	\pm	0.03	$ &	142	&	1.4	&	$r^\prime$	\\
27.66	& $	20.43	\pm	0.05	$ &	142	&	1.6	&	$r^\prime$	\\
28.08	& $	20.46	\pm	0.04	$ &	142	&	1.5	&	$r^\prime$	\\
107.89	& $	22.15	\pm	0.16	$ &	919	&	3.0	&	$r^\prime$	\\
109.58	& $	22.08	\pm	0.08	$ &	1501	&	2.3	&	$r^\prime$	\\
111.42	& $	21.97	\pm	0.08	$ &	1501	&	2.4	&	$r^\prime$	\\
113.25	& $	21.96	\pm	0.06	$ &	1501	&	2.1	&	$r^\prime$	\\
114.64	& $	21.96	\pm	0.22	$ &	212	&	2.2	&	$r^\prime$	\\
370.07	& $	23.12	\pm	0.08	$ &	6463	&	2.2	&	$r^\prime$	\\
\hline											
21.16	& $	19.55	\pm	0.06	$ &	142	&	2.0	&	$i^\prime$	\\
21.62	& $	19.67	\pm	0.07	$ &	142	&	1.9	&	$i^\prime$	\\
21.94	& $	19.58	\pm	0.05	$ &	115	&	1.9	&	$i^\prime$	\\
22.14	& $	19.63	\pm	0.04	$ &	115	&	1.8	&	$i^\prime$	\\
22.34	& $	19.56	\pm	0.05	$ &	115	&	1.8	&	$i^\prime$	\\
22.53	& $	19.61	\pm	0.04	$ &	115	&	1.9	&	$i^\prime$	\\
22.86	& $	19.64	\pm	0.05	$ &	142	&	1.8	&	$i^\prime$	\\
23.32	& $	19.71	\pm	0.05	$ &	142	&	1.9	&	$i^\prime$	\\
23.78	& $	19.71	\pm	0.06	$ &	142	&	2.0	&	$i^\prime$	\\
24.21	& $	19.71	\pm	0.05	$ &	142	&	1.6	&	$i^\prime$	\\
24.64	& $	19.64	\pm	0.05	$ &	142	&	1.6	&	$i^\prime$	\\
25.07	& $	19.70	\pm	0.05	$ &	142	&	1.7	&	$i^\prime$	\\
25.50	& $	19.74	\pm	0.05	$ &	142	&	1.7	&	$i^\prime$	\\
25.93	& $	19.79	\pm	0.05	$ &	142	&	1.4	&	$i^\prime$	\\
26.36	& $	19.80	\pm	0.05	$ &	142	&	1.5	&	$i^\prime$	\\
26.80	& $	19.81	\pm	0.05	$ &	142	&	1.3	&	$i^\prime$	\\
27.25	& $	19.78	\pm	0.04	$ &	142	&	1.2	&	$i^\prime$	\\
27.66	& $	19.83	\pm	0.05	$ &	142	&	1.4	&	$i^\prime$	\\
28.08	& $	19.82	\pm	0.04	$ &	142	&	1.3	&	$i^\prime$	\\
107.89	& $	21.56	\pm	0.15	$ &	919	&	2.9	&	$i^\prime$	\\
109.58	& $	21.48	\pm	0.13	$ &	1501	&	2.2	&	$i^\prime$	\\
111.42	& $	21.43	\pm	0.09	$ &	1501	&	2.2	&	$i^\prime$	\\
113.25	& $	21.37	\pm	0.09	$ &	1501	&	2.0	&	$i^\prime$	\\
114.64	& $	21.35	\pm	0.23	$ &	212	&	2.0	&	$i^\prime$	\\
370.07	& $	22.79	\pm	0.14	$ &	6463	&	2.1	&	$i^\prime$	\\
\hline											
21.16	& $	19.22	\pm	0.06	$ &	142	&	1.8	&	$z^\prime$	\\
21.62	& $	19.22	\pm	0.05	$ &	142	&	1.8	&	$z^\prime$	\\
21.94	& $	19.20	\pm	0.06	$ &	115	&	1.8	&	$z^\prime$	\\
22.14	& $	19.39	\pm	0.06	$ &	115	&	1.7	&	$z^\prime$	\\
22.34	& $	19.28	\pm	0.04	$ &	115	&	1.7	&	$z^\prime$	\\
22.53	& $	19.27	\pm	0.04	$ &	115	&	1.9	&	$z^\prime$	\\
22.86	& $	19.35	\pm	0.05	$ &	142	&	1.7	&	$z^\prime$	\\
23.32	& $	19.42	\pm	0.09	$ &	142	&	1.8	&	$z^\prime$	\\
23.78	& $	19.41	\pm	0.08	$ &	142	&	1.9	&	$z^\prime$	\\
24.21	& $	19.41	\pm	0.05	$ &	142	&	1.5	&	$z^\prime$	\\
24.64	& $	19.42	\pm	0.06	$ &	142	&	1.5	&	$z^\prime$	\\
25.07	& $	19.38	\pm	0.05	$ &	142	&	1.6	&	$z^\prime$	\\
25.50	& $	19.37	\pm	0.05	$ &	142	&	1.6	&	$z^\prime$	\\
25.93	& $	19.42	\pm	0.05	$ &	142	&	1.3	&	$z^\prime$	\\
26.36	& $	19.38	\pm	0.05	$ &	142	&	1.4	&	$z^\prime$	\\
26.80	& $	19.44	\pm	0.06	$ &	142	&	1.2	&	$z^\prime$	\\
27.25	& $	19.51	\pm	0.05	$ &	142	&	1.2	&	$z^\prime$	\\
27.66	& $	19.50	\pm	0.05	$ &	142	&	1.3	&	$z^\prime$	\\
28.08	& $	19.49	\pm	0.06	$ &	142	&	1.2	&	$z^\prime$	\\
107.89	& $	21.52	\pm	0.32	$ &	919	&	2.7	&	$z^\prime$	\\
109.58	& $	21.17	\pm	0.11	$ &	1501	&	2.1	&	$z^\prime$	\\
111.42	& $	21.03	\pm	0.17	$ &	1501	&	2.1	&	$z^\prime$	\\
113.25	& $	21.16	\pm	0.09	$ &	1501	&	1.9	&	$z^\prime$	\\
114.64	& $	21.03	\pm	0.21	$ &	212	&	1.9	&	$z^\prime$	\\
370.07	& $	21.91	\pm	0.12	$ &	6463	&	1.9	&	$z^\prime$	\\
\hline											
21.19	& $	18.66	\pm	0.11	$ &	240	&	2.0	&	$J$	\\
21.64	& $	18.72	\pm	0.10	$ &	240	&	2.0	&	$J$	\\
22.26	& $	18.68	\pm	0.09	$ &	480	&	2.0	&	$J$	\\
22.89	& $	18.72	\pm	0.10	$ &	240	&	1.9	&	$J$	\\
23.35	& $	18.76	\pm	0.11	$ &	240	&	2.0	&	$J$	\\
23.80	& $	18.86	\pm	0.11	$ &	240	&	2.0	&	$J$	\\
24.23	& $	18.66	\pm	0.09	$ &	240	&	1.8	&	$J$	\\
24.66	& $	18.71	\pm	0.10	$ &	240	&	1.8	&	$J$	\\
25.09	& $	18.77	\pm	0.11	$ &	240	&	1.9	&	$J$	\\
25.52	& $	18.80	\pm	0.10	$ &	240	&	1.9	&	$J$	\\
25.96	& $	18.86	\pm	0.10	$ &	240	&	1.6	&	$J$	\\
26.39	& $	18.87	\pm	0.10	$ &	240	&	1.7	&	$J$	\\
26.82	& $	18.79	\pm	0.10	$ &	240	&	1.6	&	$J$	\\
27.25	& $	18.92	\pm	0.10	$ &	240	&	1.5	&	$J$	\\
27.68	& $	18.97	\pm	0.12	$ &	240	&	1.6	&	$J$	\\
28.11	& $	18.97	\pm	0.11	$ &	240	&	1.6	&	$J$	\\
28.52	& $	18.99	\pm	0.13	$ &	240	&	1.7	&	$J$	\\
28.92	& $	19.04	\pm	0.13	$ &	240	&	1.5	&	$J$	\\
29.33	& $	18.85	\pm	0.10	$ &	240	&	1.5	&	$J$	\\
29.74	& $	18.83	\pm	0.12	$ &	240	&	1.6	&	$J$	\\
30.14	& $	18.94	\pm	0.17	$ &	240	&	1.7	&	$J$	\\
107.92	& $	20.39	\pm	0.29	$ &	960	&	2.7	&	$J$	\\
109.61	& $	20.45	\pm	0.26	$ &	1200	&	2.2	&	$J$	\\
111.44	& $	20.25	\pm	0.18	$ &	1200	&	2.3	&	$J$	\\
113.28	& $	20.08	\pm	0.18	$ &	1200	&	2.1	&	$J$	\\
\hline											
21.19	& $	18.19	\pm	0.13	$ &	240	&	2.1	&	$H$	\\
21.64	& $	18.23	\pm	0.12	$ &	240	&	2.1	&	$H$	\\
22.26	& $	18.23	\pm	0.12	$ &	480	&	2.1	&	$H$	\\
22.89	& $	18.32	\pm	0.15	$ &	240	&	2.1	&	$H$	\\
23.35	& $	18.21	\pm	0.15	$ &	240	&	2.2	&	$H$	\\
23.80	& $	18.20	\pm	0.14	$ &	240	&	2.2	&	$H$	\\
24.23	& $	18.21	\pm	0.13	$ &	240	&	2.0	&	$H$	\\
24.66	& $	18.06	\pm	0.12	$ &	240	&	1.9	&	$H$	\\
25.09	& $	18.38	\pm	0.14	$ &	240	&	2.1	&	$H$	\\
25.52	& $	18.31	\pm	0.12	$ &	240	&	2.0	&	$H$	\\
25.96	& $	18.36	\pm	0.13	$ &	240	&	1.7	&	$H$	\\
26.39	& $	18.25	\pm	0.13	$ &	240	&	1.9	&	$H$	\\
26.82	& $	18.32	\pm	0.14	$ &	240	&	1.9	&	$H$	\\
27.25	& $	18.46	\pm	0.14	$ &	240	&	1.7	&	$H$	\\
27.68	& $	18.46	\pm	0.15	$ &	240	&	1.9	&	$H$	\\
28.11	& $	18.49	\pm	0.15	$ &	240	&	1.9	&	$H$	\\
28.52	& $	18.38	\pm	0.15	$ &	240	&	2.0	&	$H$	\\
28.92	& $	18.47	\pm	0.15	$ &	240	&	1.8	&	$H$	\\
29.33	& $	18.46	\pm	0.17	$ &	240	&	1.8	&	$H$	\\
29.74	& $	18.36	\pm	0.14	$ &	240	&	1.9	&	$H$	\\
30.14	& $	18.52	\pm	0.12	$ &	240	&	1.9	&	$H$	\\
30.54	& $	18.61	\pm	0.31	$ &	240	&	1.6	&	$H$	\\
30.94	& $ >	17.57			$ &	240	&	2.0	&	$H$	\\
107.92	& $	19.74	\pm	0.27	$ &	960	&	2.6	&	$H$	\\
109.61	& $	19.67	\pm	0.19	$ &	1200	&	2.2	&	$H$	\\
\hline											
21.19	& $	17.66	\pm	0.14	$ &	240	&	1.8	&	$K_S$	\\
21.64	& $	18.01	\pm	0.14	$ &	240	&	1.8	&	$K_S$	\\
22.26	& $	18.06	\pm	0.15	$ &	480	&	1.8	&	$K_S$	\\
22.89	& $	18.03	\pm	0.16	$ &	240	&	1.7	&	$K_S$	\\
23.35	& $	17.81	\pm	0.12	$ &	240	&	1.8	&	$K_S$	\\
23.80	& $	18.11	\pm	0.16	$ &	240	&	1.7	&	$K_S$	\\
24.23	& $	18.09	\pm	0.17	$ &	240	&	1.6	&	$K_S$	\\
24.66	& $ >	16.87			$ &	240	&	1.9	&	$K_S$	\\
25.09	& $	18.08	\pm	0.15	$ &	240	&	1.7	&	$K_S$	\\
25.52	& $	17.96	\pm	0.20	$ &	240	&	1.6	&	$K_S$	\\
25.96	& $	17.83	\pm	0.17	$ &	240	&	1.5	&	$K_S$	\\
26.39	& $	17.84	\pm	0.15	$ &	240	&	1.5	&	$K_S$	\\
26.82	& $	17.91	\pm	0.16	$ &	240	&	1.4	&	$K_S$	\\
27.25	& $	17.94	\pm	0.18	$ &	240	&	1.4	&	$K_S$	\\
27.68	& $	18.01	\pm	0.19	$ &	240	&	1.5	&	$K_S$	\\
28.11	& $	17.99	\pm	0.17	$ &	240	&	1.4	&	$K_S$	\\
28.52	& $	17.91	\pm	0.18	$ &	240	&	1.5	&	$K_S$	\\
28.92	& $	18.06	\pm	0.21	$ &	240	&	1.4	&	$K_S$	\\
29.33	& $	18.08	\pm	0.20	$ &	240	&	1.4	&	$K_S$	\\
29.74	& $	18.13	\pm	0.20	$ &	240	&	1.4	&	$K_S$	\\
30.14	& $	18.20	\pm	0.18	$ &	240	&	1.5	&	$K_S$	\\
30.54	& $	18.17	\pm	0.19	$ &	240	&	1.5	&	$K_S$	\\
30.94	& $	18.02	\pm	0.38	$ &	240	&	1.6	&	$K_S$	\\
107.92	& $	19.23	\pm	0.31	$ &	960	&	2.4	&	$K_S$	\\
109.61	& $ >	19.12			$ &	1200	&	2.1	&	$K_S$	\\
\hline									
\hline									
\end{longtable}
\tablefoot{All data are in AB magnitudes and not corrected for Galactic foreground extinction. To obtain Vega magnitudes, it is $g^\prime_{AB}-g^\prime_{Vega}=-0.062$ mag, $r^\prime_{AB}-r^\prime_{Vega}=0.178$ mag, $i^\prime_{AB}-i^\prime_{Vega}=0.410$ mag, $z^\prime_{AB}-z^\prime_{Vega}=0.543$ mag, $J_{AB}-J_{Vega}=0.929$ mag, $H_{AB}-H_{Vega}=1.394$ mag, $K_{S,AB}-K_{S,Vega}=1.859$ mag. Corrections for Galactic extinction are, using $E_{(B-V)}=0.080$ \citep{Schlegel1998} and the Galactic extinction curve of \cite{1989ApJ...345..245C}: $A_{g^\prime}=0.311$ mag, $A_{r^\prime}=0.214$ mag, $A_{i^\prime}=0.160$ mag, $A_{z^\prime}=0.119$ mag, $A_J=0.070$ mag, $A_H=0.045$ mag, $A_{K_S}=0.029$ mag.}\\
}
\newpage
\longtab{
\begin{longtable}{ccrc}
\caption{\label{tab:obsrem} REM observations of the afterglow of GRB\,120711A.}\\
\hline
\hline                
Time since trigger [s] & mag$\pm$error & exp [s] & filter\\ 
\hline
279	&	11.17	$\pm$	0.04	&	10	&	$H$	\\
296	&	11.27	$\pm$	0.04	&	10	&	$H$	\\
311	&	11.27	$\pm$	0.04	&	10	&	$H$	\\
330	&	11.57	$\pm$	0.05	&	10	&	$H$	\\
347	&	11.74	$\pm$	0.05	&	10	&	$H$	\\
362	&	11.77	$\pm$	0.06	&	10	&	$H$	\\
379	&	11.77	$\pm$	0.06	&	10	&	$H$	\\
394	&	11.86	$\pm$	0.06	&	10	&	$H$	\\
445	&	12.21	$\pm$	0.05	&	50	&	$H$	\\
530	&	12.68	$\pm$	0.05	&	50	&	$H$	\\
663	&	12.98	$\pm$	0.05	&	150	&	$H$	\\\hline									
\hline									
\end{longtable}
}

\end{appendix}
\end{document}